\pdfoutput=1
\documentclass[11pt,american,english]{article}

\usepackage[utf8]{inputenc} %
\usepackage{babel}

\usepackage[table,xcdraw]{xcolor}
\usepackage{graphicx}
\usepackage{float}
\usepackage{rotating}
\usepackage{lscape}

\usepackage[
    font=smaller,
    labelfont=sc
]{caption}
\usepackage{subcaption}

\graphicspath{
    {tsmp_draft/figs/}
    {figs/}
}

\usepackage[
    top=1in,
    bottom=1in,
    left=1in,
    right=1in
]{geometry}

\usepackage{setspace}
\usepackage{sectsty}
\usepackage[compact]{titlesec}
\usepackage{soul}
\usepackage{ragged2e}
\usepackage{changepage} %

\sectionfont{\large}
\subsectionfont{\normalsize}

\titlespacing*{\section}
    {0pt}{1.5ex plus 1ex minus 0.2ex}{1.5ex plus 0.2ex}
\titlespacing*{\subsection}
    {0pt}{1.5ex plus 1ex minus 0.2ex}{1ex plus 0.2ex}
\titlespacing*{\paragraph}
    {0pt}{1.5ex plus 1ex minus 0.2ex}{1ex plus 0.2ex}

\usepackage{amsmath}
\usepackage{amssymb}
\usepackage{mathtools}
\usepackage{amsthm}
\usepackage{witharrows}

\allowdisplaybreaks
\numberwithin{equation}{section}

\newtheorem{proposition}{Proposition}
\newtheorem{lemma}{Lemma}
\newtheorem{corollary}{Corollary}

\usepackage{array}
\usepackage{booktabs}
\usepackage{multirow}
\usepackage{makecell}
\usepackage{threeparttable}
\usepackage{adjustbox}

\usepackage[authoryear]{natbib}
\usepackage{xr}
\usepackage[bottom]{footmisc}

\usepackage[
    unicode=true,
    pdfusetitle,
    bookmarks=true,
    bookmarksnumbered=false,
    bookmarksopen=false,
    breaklinks=false,
    pdfborder={0 0 0},
    pdfborderstyle={},
    backref=false,
    colorlinks=true,
    citecolor=blue!50!black,
    linkcolor=blue!50!black,
    urlcolor=blue!80!black
]{hyperref}

\usepackage{bookmark}
\usepackage{cleveref}

\usepackage{comment}

\DeclareTextSymbolDefault{\textquotedbl}{T1}

\makeatother
\begin{document}

\title{\textsc{\Large{}Two-Sided Market Power in Firm-to-Firm Trade}\thanks{We thank the editor Arnaud Costinot and three anonymous referees for constructive comments and suggestions.
For helpful conversations and discussions, we thank Rodrigo Adao, Pol Antr\'{a}s, Ariel Burstein, Julia Cajal-Grossi, Thomas Chaney, Keith Head, Ali Hortacsu, Michael Irlacher, Oleg Itshkoki, Amit Khandelwal, Yuriy Gorodnichenko, Guillermo Marshall, Kiminori Matsuyama, Thierry Mayer, Isabelle Mejean,  Marc Melitz, Scott Orr, Ina Simonovska,  Felix Tintelnot,  Trevor Tombe, Gustavo Ventura, and Jing Zhang. 
We also thank participants at seminars at ASU, UC Berkeley, Boston College, Boston University, CREST, CUHK, Dartmouth, Duke, FED Board, Georgetown, Harvard,  UC Irvine, ITAM, JGU Mainz,  Keio, LMU, LSE, MIT, National Taiwan, Nottingham, NYU, Osaka, Oxford, Peking, Penn State, Princeton, Prometeia, Queens, UC San Diego, Sciences Po, Seoul National, Tokyo, UBC,  UCL, USC, UToronto, Wisconsin Madison, Yokohama National, and Yonsei. 
Siying Wang provided excellent research assistance. 
Any views expressed are those of the authors and not those of the US Census Bureau. The Census Bureau has reviewed this data product to ensure appropriate access, use, and disclosure avoidance protection of the confidential source data used to produce this product. This research was performed at a Federal Statistical Research Data Center under FSRDC Project Number 2109. (CBDRB-FY25-P2109-R12520). Alviarez and Kikkawa are supported in part by funding from the Social Sciences and Humanities Research Council. 
All errors are our own. 
Email: \protect\href{mailto: valviarezr@iadb.org}{valviarezr@iadb.org} (Alviarez); \protect\href{mailto: fioretti.m@unibocconi.it}{fioretti.m@unibocconi.it} (Fioretti); \protect\href{mailto: ken.kikkawa@sauder.ubc.ca}{ken.kikkawa@sauder.ubc.ca} (Kikkawa); \protect\href{mailto:mailto:morlacco\% 40usc.edu}{morlacco@usc.edu} (Morlacco).}}
\author{Alviarez, V.\\
IADB\and Fioretti, M.\\
Bocconi and CEPR\and Kikkawa, K.\\
UBC Sauder and NBER\and Morlacco, M.\\
USC}
\date{(First draft: December 2022) \\ \today}
\maketitle
\begin{abstract}
We develop and estimate a model of two-sided market power to study tariff pass-through in firm-to-firm trade. A tariff moves the negotiated price through two channels: the bilateral markup, set by bargaining, and the supplier's marginal cost, which falls alongside quantities under decreasing returns. The same decreasing returns give concentrated buyers rents to bargain over, so prices fall with the buyer's share of its supplier's output. Using U.S. customs data, we estimate substantial importer bargaining power and steep export supply. Within continuing relationships, suppliers absorb about 30 percent of the 2018 tariffs, primarily through lower marginal costs as demand contracts.



\thispagestyle{empty} 
\end{abstract}
\pagebreak{}

\setcounter{page}{1}
\section{Introduction}

Tariffs have returned to the center of economic policy. The sharp tariff increases of 2018--19 and 2025 have generated extensive evidence on their effects on prices, trade, and welfare. Yet the mechanisms that shaped tariff incidence in these episodes remain less well understood. One particularly understudied aspect of pass-through is how prices adjust within individual trading relationships. A tariff drives a wedge between what an importer pays and what its foreign supplier receives; how their negotiated price adjusts in response to the tariff determines who bears its burden. These bilateral adjustments matter in the aggregate because trade is concentrated among large importers and exporters \citep{Bernard2018d,gaubert2020granular}, and relationship-specific prices divide rents and shape trade flows \citep{antras2015global}. Yet most theories of international prices emphasize exporters' market power while treating importers as atomistic price takers. How does market power on both sides of a trading relationship affect prices and, in turn, tariff incidence?

We answer these questions with a new theory of international pricing and new evidence from U.S. import data from 2001 to 2018. We document two facts that standard pricing theories cannot jointly explain. First, even within narrow supplier--product--year cells, bilateral prices vary systematically across buyer-supplier relationships: prices rise with the \emph{supplier share}, defined as the supplier's share of the importer's purchases, and fall with the \emph{buyer share}, defined as the importer's share of the supplier's U.S. shipments. Second, among relationships observed in both 2017 and 2018, tariff pass-through declined substantially with buyer share but showed little systematic relationship to supplier share. Our estimates also indicate that the pass-through of the 2018 tariffs was incomplete on average within continuing relationships.\footnote{These estimates describe repricing within continuing, arm's-length relationships in our baseline sample of intermediate and capital goods, not incidence across U.S. imports as a whole. They therefore do not contradict product-level studies that find near-complete pass-through using broader samples \citep{amiti2019impact,amiti2020s,fajgelbaum2020return,cavallo2020tariff}. We reconcile the two sets of findings below.} Taken together, these facts are difficult to reconcile with models in which importers are atomistic and take prices as given.

To explain these patterns, we develop a bargaining model of firm-to-firm trade with two-sided market power. We extend the network oligopoly framework of \citet{kikkawa2019imperfect} by adding bilateral price bargaining and decreasing returns to scale for exporters. In equilibrium, each bilateral price equals the supplier's marginal cost, common across buyers, multiplied by a relationship-specific markup, or price-cost ratio. This ratio is a weighted average of an \emph{oligopoly markup}, which rises with supplier share, and an \emph{oligopsony markdown}, which falls with buyer share. Greater importer bargaining power shifts weight toward the latter. Both ingredients are essential for the markdown component. Decreasing returns make marginal cost rise with output, generating quasi-rents within each relationship that grow with buyer share; bargaining allows the buyer to capture part of these rents through a lower price. Under constant returns, these quasi-rents disappear. Bargaining can still divide oligopoly rents, but buyer share no longer affects the negotiated price.

The supply curvature that underpins oligopsony power is also central to tariff incidence. An import tariff affects bilateral prices through two channels: changes in the bilateral markup and changes in the supplier's marginal cost. The markup channel depends on the allocation of bargaining power. When exporters set prices, \emph{strategic complementarities} among them dampen pass-through, as in standard oligopoly models \citep[e.g.,][]{amiti2014importers,amiti2019international}. When importers hold the bargaining power, a tariff reduces purchases and compresses the buyer's markdown, generating \emph{strategic substitutability} that amplifies pass-through. Under decreasing returns, lower purchases also reduce the supplier's output and marginal cost. This cost channel dampens pass-through more strongly at higher buyer shares. We show that this channel always dominates the opposing markdown effect, leading pass-through to decline with buyer share. As export supply becomes arbitrarily steep, the entire markup channel becomes negligible relative to the cost channel, regardless of how bargaining power is allocated.  Oligopsony power can therefore substantially affect price levels even when bargaining plays little role in bilateral tariff pass-through.

Our structural estimates show that this distinction matters quantitatively in the U.S. data. Using bilateral price levels from 2001 to 2016, we estimate high importer bargaining power and strong decreasing returns to scale. Together, these features imply substantial oligopsony power and explain why bilateral prices fall with buyer share. The same supply curvature predicts that tariff pass-through also declines with buyer share. For the 2018 tariff episode, the model closely reproduces the observed gradient: for each 10 percentage point increase in buyer share, pass-through falls by 4.1 percentage points in the data and 4.2 in the model. It also matches the average: predicted pass-through is 78 percent, within the empirical confidence interval. The model attributes both the level and the gradient almost entirely to the cost channel, with little net contribution from the markup channel. Weighting by initial import shares yields predicted pass-through of about 70 percent, implying that foreign suppliers absorb roughly 30 percent of the tariff increases, primarily through lower marginal costs as demand contracts.

The resulting incidence mirrors, within each relationship, the classic terms-of-trade effect of a tariff \citep{Johnson1953}. There, a tariff reduces aggregate import demand and moves foreign suppliers down their export-supply curve. Here, the same adjustment occurs within a relationship when a tariff contracts a buyer's purchases. Two features distinguish our pass-through result from the competitive benchmark. First, relationship-level pricing remains essential even though adjustments in the bilateral markup contribute little on net. Because buyers are not atomistic, the supply response varies across relationships, which leads pass-through to decline with the buyer share. Second, the dominance of the cost channel is an estimated outcome, not a model restriction. As decreasing returns weaken, marginal costs respond less to a demand contraction and the markup channel matters more; under constant returns, it governs pass-through entirely. Different goods, settings, or horizons may therefore imply different roles for costs and markups. 

The structural parameters are identified from price levels alone. We compare prices across buyers importing the same product from the same exporter in the same year. This removes the component of marginal cost common across buyers, leaving variation in bilateral margins determined by supplier and buyer shares, together with match-specific pricing residuals. To address unobserved match-specific differences in costs, we instrument the shares with measures of market structure that exclude the focal relationship. The exclusion restriction requires these instruments to be orthogonal to match-specific pricing residuals, conditional on fixed effects. The estimates imply an importer bargaining weight of 0.83 and a returns-to-scale parameter of 0.46, corresponding to an average residual inverse export-supply elasticity of about 0.3. The median implied bilateral price--cost ratio is 1.01, compared with 1.21 under full oligopoly and 0.95 under full oligopsony.

Pass-through moments play no role in structural estimation, so the model's fit to the 2018 tariff episode, on both the average and the buyer-share gradient, is an out-of-sample validation on two untargeted moments. This evidence supports the decreasing-returns mechanism, but it does not provide independent evidence of importer bargaining power. Average pass-through alone rejects neither the baseline nor the alternatives that eliminate importer bargaining power, decreasing returns, or both. The buyer-share gradient rejects constant returns, but it does not separate the baseline from the model without importer bargaining power: at the estimated parameters, removing it lowers predicted average pass-through from 78 to 72 percent and moves the buyer-share coefficient from $-0.42$ to $-0.37$; both differences are small relative to the standard errors of the empirical estimates. This weak separation is itself a prediction of the model. At the estimated supply curvature, most of the tariff response comes through marginal cost rather than the bilateral markup. The evidence for importer bargaining power comes instead from price levels, which fall with the buyer share only when importers hold bargaining power. Among the specifications we consider, only the baseline accounts for the buyer-share evidence in price levels and in pass-through jointly.

Our focus on market power within buyer--supplier relationships goes hand in hand with sample restrictions that are consequential for the results. Starting from the universe of U.S. import transactions at the importer--exporter--product--year level, two restrictions are essential. First, each exporter must sell a product to at least two U.S. buyers, because identification exploits price differences across its buyers. Second, each relationship must be observed in consecutive years, because bilateral price changes are otherwise undefined. We then impose additional restrictions to align the data with the model. We exclude related-party trade, whose prices may reflect transfer pricing, and energy goods, whose unit values are noisy. Finally, we interpret the supply curvature at the heart of the model as short-run capacity constraints arising from relationship-specific investments \citep[e.g.,][]{Nunn2007,boehm2022convex}. We therefore restrict the baseline sample to intermediate and capital goods, where these constraints are most binding \citep{antras2015global}.\footnote{Consistent with this interpretation, estimates for consumption goods imply returns closer to constant and lower importer bargaining power. The model therefore predicts a weaker cost channel and a flatter buyer-share gradient in pass-through. The reduced-form gradient is indeed attenuated, though imprecisely estimated. By contrast, the average estimate is too imprecise to determine whether pass-through is higher or lower than in the baseline.}

Our pass-through results therefore describe repricing within the relationships in the baseline sample, not U.S. imports as a whole. This difference in scope helps reconcile our estimates of incomplete pass-through with studies that find nearly complete pass-through of the same episode \citep{amiti2019impact,amiti2020s,fajgelbaum2020return,cavallo2020tariff}. Those studies use monthly prices at the product--source-country level and cover a broader set of transactions. To investigate the role of aggregation and sample composition for the reduced-form evidence, we vary one while holding the other fixed. Holding the baseline set of relationships fixed, moving from annual bilateral prices to monthly product-level prices does not raise pass-through and, if anything, lowers it. Sample composition matters more. Pass-through approaches one as we expand the sample to include consumption goods, single-buyer suppliers, energy and related-party trade, and relationships not observed in consecutive years. Because these expansions are cumulative and the estimates do not move monotonically, this exercise cannot isolate the contribution of any single restriction. It nevertheless points to sample composition, rather than aggregation, as the main source of the difference. A full reconciliation with the product-level evidence, however, requires further work.\footnote{Consistent with this conclusion, \citet{sui2024incomplete} find incomplete pass-through within continuing firm-level relationships and show that reallocation toward importers with higher pass-through and costlier new supplier relationships raises aggregate pass-through.}

A further caveat is that our analysis is partial equilibrium. Our pricing model holds fixed the trading network, prices in other relationships, wages, and aggregate demand, and studies tariff pass-through within continuing links. Comparing reduced-form and model-implied pass-through therefore tests the model's mechanisms together with its partial-equilibrium approximation: reduced-form pass-through incorporates equilibrium feedback absent from the model.\footnote{This distinction parallels the failure of SUTVA emphasized by \citet{berger2022labor} in oligopsonistic labor markets.} The close agreement between model predictions and data supports the approximation for these price moments, but does not separately identify how big general equilibrium effects are. Equilibrium adjustment can also occur on the extensive margin, as tariffs sever links and redirect sourcing. Our results therefore isolate one component of aggregate incidence. A complete analysis would also allow the network to reorganize, suppliers to reprice across buyers and destinations, and wages and aggregate demand to adjust.\footnote{Recent work embeds bilateral bargaining or two-sided search in equilibrium matching models to study these margins. See, e.g., \cite{GrossmanHelpmanRedding2024, EatonJinkinsTyboutXu2026, KrolikowskiMcCallum2026}.}

Beyond the theoretical and empirical work on pass-through discussed above, our results contribute to two further strands of the literature. The first concerns relationship-specific pricing in firm-to-firm trade. Price dispersion across buyers of the same supplier is pervasive in firm-to-firm data \citep{Fontaine2019,macedoni2024pricing,ignatenko2024market}. Price dispersion can arise from variation in oligopolistic markups even when buyers are price takers \citep[e.g.,][]{huang2024firm}. Buyer power offers another explanation, but evidence linking it to bilateral prices remains limited. \citet{Morlacco2019} estimates buyer power and the resulting misallocation among French importers. Using price and quantity responses to Argentina's discretionary import restrictions, \citet{atkin2024trade} identify which side holds market power and find that buyer power is more common among larger importers. Building on work that treats the buyer--supplier relationship as the unit of analysis \citep{bernard2019production,bernard2022origins}, we develop a model of bilateral oligopoly in which supplier and buyer power jointly determine negotiated prices. The closest antecedent is the bargaining model of \citet{gopinath2011search}, which generates variable wholesale markups and incomplete pass-through. With constant marginal cost, however, its pricing rule depends on supplier share and bargaining power but gives no independent role to buyer share.\footnote{\citet{goldberg2013bargaining} also model bargaining between exporters and importers over prices and invoicing currency in the context of exchange-rate pass-through. Neither model features oligopsony power.} Our model nests markup variation driven by supplier concentration and adds oligopsony markdowns and endogenous marginal-cost adjustments. Our estimates show that these mechanisms jointly account for the decline in both bilateral prices and tariff pass-through with buyer share in U.S. data.

The second literature studies shock propagation through production networks. Existing models typically transmit shocks through quantities under competitive pricing or fixed markups \citep{Acemoglu2012,Grassi2018,barrot2016input,boehm2019input,carvalho2021supply}. \citet{acemoglu2025macroeconomics} introduce bargaining over surplus, but efficient two-part tariffs set unit prices at marginal cost, so prices play no independent role in propagation. Our results show what changes once prices are negotiated rather than taken as given. First, the slope of export supply, not the allocation of bargaining power, determines how a tariff is passed through: when supply is steep, the cost channel dominates regardless of who holds the bargaining power. Second, because buyers are not atomistic, the supply response, and hence pass-through, vary across links; with trade concentrated among a few large firms, this heterogeneity matters for aggregate incidence. Third, bargaining still determines how rents are divided, so a price response that looks competitive can coexist with substantial market power on both sides of the transaction.

\section{Theoretical Framework\label{sec: Theory}}

This section develops our theory of price setting in buyer–supplier relationships. The model yields a structural equation linking the bilateral markup to the supplier and buyer shares, which we later estimate. It also yields predictions about how tariff pass-through varies with these shares, which we later test in the data. Although our application focuses on U.S. importers and their foreign exporters, the framework can be applied to any bilateral market for intermediate inputs.

We maintain three assumptions throughout. First, the model is static and treats each product negotiation separately. Repeated and multi-product relationships are introduced in reduced form later in Section~\ref{sec:Identification-and-estimation}. Second, we take the network of buyer-supplier relationships as given. Third, prices are fully flexible, so we abstract from nominal rigidities.\footnote{This abstraction is unlikely to matter at annual frequency. For example, \citet{amiti2020s} find similar short- and long-run pass-through of the 2018 U.S. tariffs.} Appendix~\ref{subsec: Theory Appendix} provides all proofs and derivations.

\subsection{Environment}
\label{sec:environment}

We focus on the relationship between exporter $i$ and importer $j$ of an intermediate input. We denote by $\mathcal{Z}_{i}$ the set of importers connected to exporter $i$, and by $\mathcal{Z}_{j}$ the set of exporters connected to importer $j$. These sets vary across firms and are taken as given.

\paragraph{Exporters and Supply.}
Exporter $i$ supplies a differentiated input to importers in $\mathcal{Z}_i$. Total exported quantity is $q_i=\sum_{j\in\mathcal{Z}_i} q_{ij}$, where $q_{ij}$ denotes importer $j$'s demand.

The exporter’s marginal cost is
\begin{equation}
MC_i(q_i) = k_i\, q_i^{\frac{1-\theta}{\theta}},
\label{eq:MC_i}
\end{equation}
where $\theta \in (0,1]$ governs returns to scale and $k_i$ captures exogenous cost shifters, such as productivity or foreign wages. When $\theta=1$, marginal cost is constant and equal to $k_i$. When $\theta<1$, the technology exhibits decreasing returns, so marginal cost rises with output.\footnote{The model is static, so $\theta$ describes returns to scale over the horizon at which prices are negotiated, which is a year in our empirical application. We therefore interpret $\theta<1$ as short-run limits to scalability rather than long-run returns to scale. See Section~\ref{sec:Identification-and-estimation} for a discussion.} Integrating the marginal cost yields the exporter's total cost function as $TC_i(q_i) = \theta \, k_i\, q_i^{\frac{1}{\theta}}.$ Average cost is $AC_i(q_i) = \theta\, MC_i(q_i)$, which lies below marginal cost whenever $\theta < 1$.

\paragraph{Importers and Demand.}
Importer $j$ produces a final good $q_j$ by combining a foreign input ($q_j^f$) with a domestic input ($q_j^d$) according to the production function
\begin{equation}
q_j 
= \varphi_j \, (q_j^f)^{\gamma} \, (q_j^d)^{\varrho - \gamma},
\label{nested_ces_1}
\end{equation}
where $\varphi_j$ denotes productivity, $\gamma$ the output elasticity of foreign inputs, and $\varrho$ the returns to scale.\footnote{
Since domestic input purchases are not observed, we impose a unit elasticity of substitution between foreign and domestic inputs, which fixes the cost share of $q_j^d$ and $q_j^f$. 
Under this restriction, $q_j^d$ can be interpreted as a constant-returns aggregator of primary factors such as labor or domestic intermediates.
}

The foreign input is a CES aggregate of the differentiated varieties supplied by exporters in $\mathcal{Z}_j$:
\begin{equation}
q_j^{f}
= 
\left(
\sum_{i \in \mathcal{Z}_j}
\varsigma_{ij} \, q_{ij}^{\,\frac{\rho-1}{\rho}}
\right)^{\!\frac{\rho}{\rho-1}},
\label{nested_ces_2}
\end{equation}
where $\varsigma_{ij}$ is a demand shifter, and $\rho>1$ is the substitution elasticity across input varieties.

Downstream, importer $j$ sells its final good in a monopolistically competitive market and faces CES demand with elasticity $\nu>1$. Revenue is therefore given by
\begin{equation}
R_j(q_j)
= D_j\, q_j^{\frac{\nu-1}{\nu}},
\label{eq:revenue_j}
\end{equation}
where $D_j$ denotes an exogenous demand shifter.

\subsection{Bargaining Protocol and Gains From Trade}
\label{sec:bargaining-and-gft}

We assume that the parties bargain only over the price $p_{ij}$, and that the importer chooses how much to buy at the negotiated price. The bargain therefore takes the importer's demand as given, which we derive first. Given prices and an output target $q_j$, importer $j$ chooses foreign inputs $\{q_{\ell j}\}_{\ell \in \mathcal{Z}_j}$ and domestic input $q_j^d$ to minimize total cost:
\begin{equation}
\min_{\{q_{\ell j}\},\, q_j^d}
\quad 
\sum_{\ell \in \mathcal{Z}_j} p_{\ell j} q_{\ell j} + p_j^d q_j^d
\quad
\text{s.t.} \quad 
\varphi_j (q_j^f)^\gamma (q_j^d)^{\varrho-\gamma} \ge q_j .
\end{equation}

Solving this problem yields the standard CES demand for input varieties:
\begin{equation}
q_{ij} = \varsigma_{ij}^{\rho} \left( \frac{p_{ij}}{p_j^f} \right)^{-\rho} q_j^f,
\label{eq:demand function q_ij}
\end{equation}
where $q_j^f$ is the foreign input aggregator defined in \eqref{nested_ces_2} and $p_j^f$ the associated price index, which satisfies $p_j^f q_j^f = \gamma \, c_j \, q_j$, where $c_j = MC_j(q_j)$ denotes importer $j$'s marginal cost of production.

In the bargaining stage, exporter $i$ and importer $j$ negotiate $p_{ij}$ taking as given that the importer will purchase the quantity in \eqref{eq:demand function q_ij}.\footnote{This protocol is the input-market analogue of right-to-manage bargaining in labor economics, where wages are bargained and firms choose employment \citep{nickell1983unions,manning1987integration}.  It is also closely related to \citet{grennan2013price}, where bargaining determines prices and realized quantities follow downstream demand.} Following the Nash-in-Nash solution concept \citep{Horn1988, collard2019nash}, each pair also takes  prices on all other links, wages, and aggregate demand shifters as given; we suppress these in the notation below.  

Each party's gains from trade are its operating profit when the link is active minus its profit if bargaining fails and the link is severed: $GFT_{ij}^n(p, q) \equiv \pi^n{ij}(p, q) - \tilde{\pi}^n{(-\cdot)}$ for $n \in \{i,j\}$, where $p$ and $q$ denote the price and quantity on link $(i,j)$. When the link is active, the exporter earns revenue from all its buyers net of total cost, $\pi^i_{ij} = \sum_{n \in \mathcal{Z}_i} p_{in} q_{in} - TC_i(q_i)$, and the importer earns downstream revenue net of input costs, $\pi^j_{ij} = R_j(q_j) - \sum_{n \in \mathcal{Z}_j} p_{n j} q_{n j} - p_j^d q_j^d$, with $TC_i$ and $R_j$ as defined in Section~\ref{sec:environment}. If the link is severed, the exporter serves only its other buyers, $\tilde{\pi}_{(-j)}^i = \sum_{n \neq j} p_{in} q_{in} - TC_i(q_i - q_{ij})$, and the importer sources only from its other suppliers at unchanged prices, re-optimizing its input mix and output, which yields $\tilde{\pi}_{(-i)}^j$.\footnote{Appendix \ref{subsec:Generalized-outside-option} discusses an extension with more general nonparametric outside options.}

The negotiated price solves a Nash bargaining problem that divides these gains, with $\phi \in (0,1)$ denoting the importer’s exogenous bargaining weight; the case $\phi=1$ is the limit in which the exporter's gains from trade go to zero:
\begin{equation}
\max_{p,\, q} \;
\left(GFT_{ij}^i(p, q)\right)^{1-\phi}\;
\left(GFT_{ij}^j(p, q)\right)^{\phi}
\quad \text{s.t.} \quad
q
=
q_j^f \varsigma_{ij}^{\rho}
\left( \frac{p}{p_j^f} \right)^{-\rho}.
\label{eq:nash_bargaining1}
\end{equation}

\subsection{Price Characterization}\label{subsec:equilibrium-prices}

The negotiated price can be expressed in terms of two endogenous market shares, each measuring one party's importance to the other. For link $(i,j)$, exporter $i$'s \emph{supplier share} $s_{ij}$ is its share of importer $j$'s foreign input expenditure, and importer $j$'s \emph{buyer share} $x_{ij}$ is its share of exporter $i$'s total shipments:\footnote{The data record only the exporter's shipments to the U.S., so the empirical counterpart of $x_{ij}$ is the importer's share of the exporter's U.S. sales of the product; see Section~\ref{subsec:Measuring-key-variables}.}
\begin{equation}
s_{ij} \equiv \frac{p_{ij} q_{ij}}{\sum_{k \in \mathcal{Z}_{j}} p_{kj} q_{kj}},
\qquad \text{and} \qquad
x_{ij} \equiv \frac{q_{ij}}{\sum_{k \in \mathcal{Z}_{i}} q_{ik}}.
\label{eq:shares}
\end{equation}

Two elasticities govern the response of the importer's gains from trade to the bilateral price. Let $\eta$ denote the elasticity of the importer's foreign input demand with respect to its input price index. This elasticity depends on downstream demand and on the importer's technology:
\begin{equation}\label{eq:eta}
\eta
\equiv
-
\frac{d\ln q_j^f}{d\ln p_j^f}
=
\frac{(\varrho-\gamma)+\nu(1-(\varrho-\gamma))}
     {\varrho+\nu(1-\varrho)}.
\end{equation}
Let $\xi(GFT_{ij}^j, p_{ij})$ denote the elasticity of the importer's gains from trade with respect to the bilateral price, evaluated along the demand schedule:
\begin{equation}
\xi(GFT_{ij}^{j},p_{ij})
\equiv
- \frac{d\ln GFT_{ij}^j}{d\ln p_{ij}}
=
\frac{(\eta-1)s_{ij}}
     {1 - (1 - s_{ij})^{\frac{\eta - 1}{\rho - 1}}}.
\label{eq:xi}
\end{equation}
A higher value of $\xi$ indicates that the importer incurs a larger loss from an increase in the bilateral price and therefore bargains more aggressively. The elasticity depends on the supplier share $s_{ij}$, which measures the fraction of the importer's input expenditure exposed to the price.

Problem \eqref{eq:nash_bargaining1} has a closed-form solution in which the bilateral price-cost ratio depends on the relationship only through the two market shares.
\begin{proposition}\label{prop:markup}
Let $\mu_{ij} \equiv p_{ij}/MC_i$ denote the bilateral markup, or price-cost ratio. The solution to \eqref{eq:nash_bargaining1} satisfies
\begin{equation}
\mu_{ij}
=
(1-\omega_{ij})\,\mu^{\textnormal{oligopoly}}(s_{ij})
+
\omega_{ij}\,\mu^{\textnormal{oligopsony}}(x_{ij}),
\label{eq:bilateral_markup_prop}
\end{equation}
with
\begin{equation}
\omega_{ij}
=
\frac{\phi\,\xi(GFT_{ij}^{j},p_{ij})}
{\phi\,\xi(GFT_{ij}^{j},p_{ij})+(1-\phi)(\varepsilon_{ij}-1)}
\in (0,1).
\label{eq:omega_prop}
\end{equation}
\end{proposition}

\noindent\textbf{Proof.} See Appendix \ref{subsec: Derivations Markup}.

\vspace{0.5em}

We discuss each component of the price-cost ratio in turn. 

\paragraph{Oligopoly Markup.} The term $\mu^{\textnormal{oligopoly}}_{ij}$ in equation \eqref{eq:bilateral_markup_prop} is the standard \emph{oligopoly markup} arising under Bertrand competition with differentiated products:
\begin{equation}
\mu_{ij}^{\text{oligopoly}}
=
\frac{\varepsilon_{ij}}{\varepsilon_{ij}-1}
\ge 1
\qquad \text{where} \qquad 
\varepsilon_{ij}
=
(1-s_{ij})\rho
+
s_{ij}\eta.
\label{eq:mu_oligopoly}
\end{equation}
The residual demand elasticity $\varepsilon_{ij}$ is a weighted average of the within-input elasticity $\rho$ and the outer elasticity $\eta$ in equation \eqref{eq:eta}.\footnote{In standard nested-CES models with Bertrand competition, $\eta$ is a primitive parameter \citep{Atkeson2008}. Here it is derived from the importer's technology ($\gamma,\varrho$) and downstream demand ($\nu$).} When $\rho>\eta$, which we maintain throughout, the markup increases in the supplier share $s_{ij}$, reflecting greater oligopoly power. 

This is the markup in the network oligopoly model of \citet{kikkawa2019imperfect}, which our framework nests as the case $\phi=0$: with no importer bargaining power, $\omega_{ij}=0$, and the negotiated price is $p_{ij} = \mu^{\textnormal{oligopoly}} (s_{ij})\cdot MC_i$. In this case, the buyer share $x_{ij}$ does not affect prices.

\paragraph{Oligopsony Markdown.}  The term $\mu^{\textnormal{oligopsony}}_{ij}$ in equation \eqref{eq:bilateral_markup_prop} is given by
\begin{equation}\label{eq:mu_oligopsony}
\mu_{ij}^{\text{oligopsony}}
=
\theta \cdot
\frac{1-(1-x_{ij})^{1/\theta}}{x_{ij}}
\le 1,
\end{equation}
which corresponds to the price-cost ratio when the importer has full bargaining power, $\phi=1$. In this case, $\omega_{ij}=1$, and the negotiated price is $p_{ij}=\mu^{\textnormal{oligopsony}} (x_{ij})\cdot MC_i$. Because $\mu_{ij}^{\textnormal{oligopsony}}$ is always no greater than one, we refer to it as an \emph{oligopsony markdown}.\footnote{Strictly speaking, $\mu_{ij}^{\textnormal{oligopsony}}$ is a price-cost ratio. Our terminology therefore differs from the conventional use of ``markdown'' to denote the wedge between an input's price and its marginal revenue product.} 

The oligopsony markdown has a cost interpretation. When the importer holds all the bargaining power, the negotiated price leaves the exporter indifferent between serving importer $j$ and not. The price then equals the exporter's average incremental cost of the quantity sold to $j$, $\mathrm{AIC}_{ij} \equiv [TC_i(q_i)-TC_i(q_i-q_{ij})]/q_{ij}$. The markdown is this cost relative to current marginal cost, $\mu^{\text{oligopsony}}_{ij} = \mathrm{AIC}_{ij}/MC_i(q_i)$, and equation~\eqref{eq:mu_oligopsony} follows under the cost function~\eqref{eq:MC_i}.

A markdown below one reflects quasi-rents. When marginal cost is increasing in output, the inframarginal units sold to $j$ are produced at a cost below the marginal unit, so that $\mathrm{AIC}_{ij} < MC_i(q_i)$. The difference $[MC_i(q_i)-\mathrm{AIC}_{ij}]\,q_{ij}$ is the quasi-rent that the exporter earns on the match; under full importer bargaining power, it accrues entirely to the importer.

The markdown deepens with the buyer share. A buyer that accounts for a larger share of the exporter's output purchases units produced farther down the marginal-cost curve, so its average incremental cost lies farther below $MC_i(q_i)$ and the price-cost ratio is lower. This negative price--buyer-share gradient is what we call oligopsony power. As $x_{ij} \to 0$ the buyer is atomistic, $\mathrm{AIC}_{ij} \to MC_i(q_i)$, and $\mu_{ij}^{\textnormal{oligopsony}} \to 1$. As $x_{ij} \to 1$ the buyer accounts for the exporter's entire output, average incremental cost equals average cost, and $\mu_{ij}^{\textnormal{oligopsony}} \to AC_i(q_i)/MC_i(q_i) = \theta$.

Oligopsony power requires both importer bargaining power and decreasing returns. Without bargaining power (i.e., $\phi = 0$), the markdown has no effect on the negotiated price. Without decreasing returns (i.e., $\theta = 1$), average and marginal cost coincide, there are no quasi-rents, and $\mu_{ij}^{\textnormal{oligopsony}} = 1$ for every $x_{ij}$. In this case, importer bargaining can still extract the exporter's oligopoly rents, but the price cannot fall below marginal cost however large the buyer.

\paragraph{Terminology.}
We use \emph{buyer power}, \emph{importer market power}, and \emph{countervailing power} interchangeably for the importer's ability to affect the negotiated price through bargaining, that is, for any $\phi > 0$. Oligopsony power, as defined above, is the narrower case in which that ability operates through the buyer share.

\paragraph{Bargaining Weights.}

Equation~\eqref{eq:bilateral_markup_prop} expresses the price-cost ratio as a weighted average of the oligopoly markup \eqref{eq:mu_oligopoly} and the oligopsony markdown \eqref{eq:mu_oligopsony}. The weight $\omega_{ij}$ governs the relative importance of these two forces. It is increasing in the exogenous bargaining parameter $\phi$, but it is also endogenous because it depends on both the importer’s gains-from-trade elasticity $\xi(GFT_{ij}^{j},p_{ij})$ and the residual demand elasticity $\varepsilon_{ij}$. Thus, even for a given value of $\phi$, effective bargaining power varies systematically across matches.

For $\phi\in(0,1)$, the weight $\omega_{ij}$ is hump-shaped in the supplier share $s_{ij}$. When $s_{ij}$ is small,  the bilateral price has little effect on the importer’s overall costs, so the importer has little to gain from bargaining aggressively. When $s_{ij}$ is very large, the importer is highly exposed to the relationship but also depends heavily on it, which weakens its outside option. Effective bargaining power is therefore strongest at intermediate supplier shares, where cost exposure is significant, but substitution remains feasible.

\paragraph{Equilibrium Allocation.}

\begin{figure}[t]
\caption{Equilibrium Allocations With Different Bargaining Power\label{fig:allocations}}
\begin{centering}
\includegraphics[width=1\textwidth]{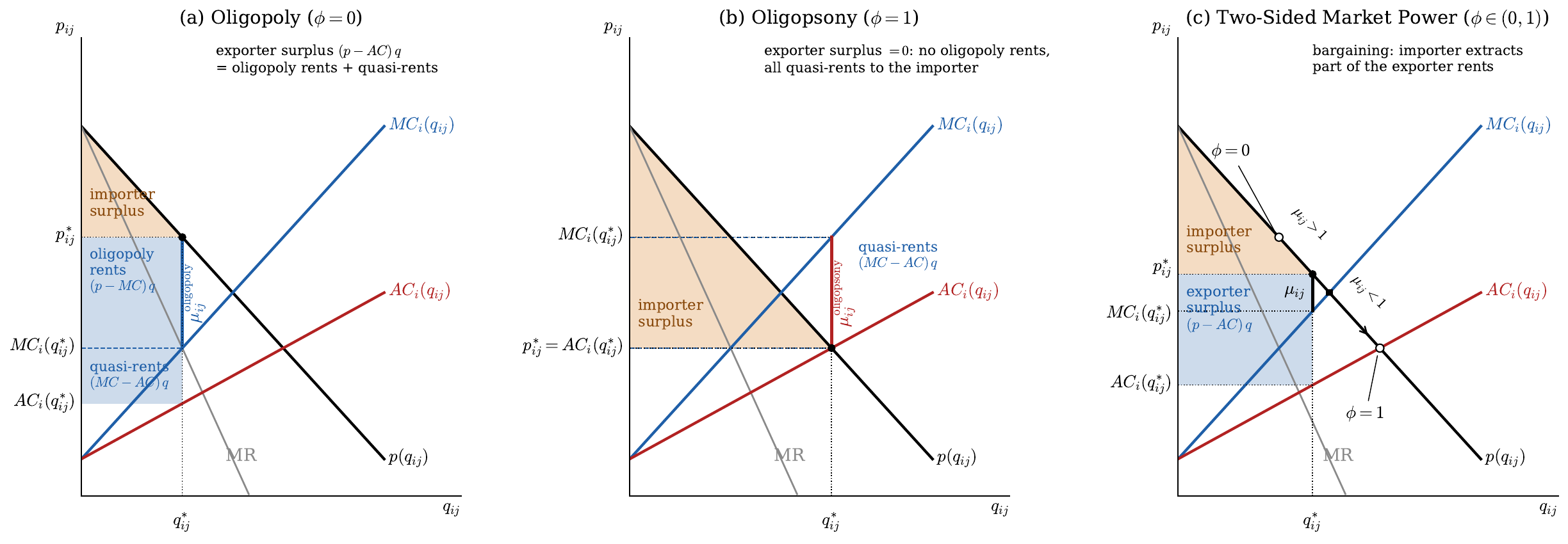}
\par\end{centering}
\footnotesize \textit{Notes}{: Each panel illustrates the negotiated price and quantity under a different value of $\phi$. The exporter has a single buyer, so $q_i=q_{ij}$ and $AC_i=\mathrm{AIC}_{ij}$. The exporter's residual demand function $p(q_{ij})$ is shown in black and its marginal revenue in grey; the exporter's marginal cost $MC_i(q_{ij})$ is shown in blue and its average cost $AC_i(q_{ij})$ in red. Orange areas are the importer's surplus, the area under residual demand above the negotiated price. Blue areas are the exporter's surplus, $(p-AC)\,q$, split in Panel (A) into oligopoly rents $(p-MC)\,q$ in dark blue and quasi-rents $(MC-AC)\,q$ in light blue. The dashed rectangle in Panel (B) is the quasi-rents at the negotiated quantity. In Panel (C), open circles mark the $\phi=0$ and $\phi=1$ allocations and the square marks $p=MC$. All panels assume $\theta < 1$.}
\end{figure}

Figure~\ref{fig:allocations} illustrates the equilibrium allocation for a single $i$-$j$ match under alternative bargaining weights $\phi$.\footnote{For illustration, the exporter serves a single buyer, so average incremental cost equals average cost; with multiple buyers, the corresponding cost object is $\mathrm{AIC}_{ij}$.} The exporter faces a downward-sloping residual demand curve and its associated marginal revenue curve, shown in black and grey, respectively. The importer faces upward-sloping residual marginal and average cost curves, shown in blue and red.

Panel (A) shows the oligopoly benchmark ($\phi=0$). Quantity is pinned down by the intersection of marginal revenue and marginal cost, and price is then determined from residual demand. The exporter's surplus has two components: oligopoly rents, shown in dark blue, equal to $(p-MC)\,q$, and quasi-rents, shown in light blue, equal to $(MC-AC)\,q$. Quasi-rents exist whenever $\theta<1$; under $\phi=0$ the exporter retains all of them.

Panel (B) shows full importer bargaining power ($\phi=1$). The negotiated allocation lies at the intersection of residual demand and average cost curves, leaving the exporter with zero net gains from trade. Relative to the oligopoly benchmark, the negotiated price is lower, the traded quantity is higher, and the importer's surplus is correspondingly larger.

Panel (C) shows the case $\phi \in (0,1)$. The negotiated outcome lies between the two polar allocations, consistent with Proposition~\ref{prop:markup}, and the price may lie above or below marginal cost depending on bargaining power and the size of quasi-rents. The exporter's surplus is the shaded rectangle with height $p-AC$ and width $q$, and the importer's surplus is the area under residual demand above the price. As $\phi$ rises, the negotiated point moves down the demand curve from the $\phi=0$ allocation toward the $\phi=1$ allocation, the exporter's surplus shrinks, and the importer's grows.

\paragraph{Markups and Market Shares.}

Both components of equation \eqref{eq:bilateral_markup_prop} vary systematically with supplier and buyer shares. A higher supplier share $s_{ij}$ raises the oligopoly markup, while a higher buyer share $x_{ij}$ lowers the oligopsony markdown. The co-movements between prices and bilateral shares are thus informative about the presence of the two forces.

\begin{proposition}\label{prop:2}
\emph{The bilateral price-cost ratio $\mu_{ij}$ satisfies:}
\begin{enumerate}
\item \textbf{Supplier Share.} If $\mu_{ij}$ increases in $s_{ij}$, then $\phi < 1$.
\item \textbf{Buyer Share.} If $\mu_{ij}$ decreases in $x_{ij}$, then $\phi > 0$ and $\theta < 1$.
\end{enumerate}
\end{proposition}

\noindent\textbf{Proof.} See Appendix~\ref{appendix:proof-prop2}.

Section~\ref{subsec:Test_prediction} examines whether these cross-sectional predictions hold in the data, and Section~\ref{sec:Identification-and-estimation} uses the same variation in price levels to identify $\phi$ and $\theta$.

\subsection{Pass-Through Characterization \label{subsec: Pass-through (Theory)}}

We now use our model to study the short-run effect of an unanticipated tariff on the bilateral price. Let $T_c=(1+\tau_c)$ denote the gross ad-valorem tariff applied to imports from country $c$. In log terms, the price charged by exporter $i$ from country $c$ to importer $j$ can be written as:
\begin{equation*} \ln p_{ij} = \ln \mu_{ij} + \ln MC_i + \ln T_{c(i)}. %
\end{equation*}
A tariff on country $c$ affects all links involving its exporters. We characterize its effect on $p_{ij}$ holding all other link prices, wages, and aggregate demand shifters fixed. In particular, $dp_{rj} = 0$ for all $r \in \mathcal{Z}_{j} \setminus \{i\}$ and $dp_{i\ell} = 0$ for all $\ell \in \mathcal{Z}_{i} \setminus \{j\}$. The following proposition characterizes this partial elasticity.\footnote{Appendix~\ref{subsec:Generalized-pass-through-elasticity} illustrates selected indirect effects using a truncated network-feedback approximation. For example, a shock to exporter $i$ may alter prices and quantities in its other trading relationships, which may then feed back into $p_{ij}$. Although these adjustment channels are not modeled explicitly in the baseline analysis, the approximation illustrates how spillovers across relationships can generate a wedge between reduced-form and structural pass-through estimates.}
\begin{proposition}\label{prop:3}
\textit{Holding fixed all other link prices and general equilibrium variables, the pass-through elasticity of a tariff into the bilateral import price is
\begin{equation}
\Phi_{ij} \equiv \frac{d \ln p_{ij}}{d \ln T_{c(i)}} = \frac{1}{1 + \Gamma_{ij} + \Lambda_{ij}},
\qquad\text{with} \qquad 
\Gamma_{ij} \equiv -\frac{d \ln \mu_{ij}}{d \ln p_{ij}},
\quad
\Lambda_{ij} \equiv -\frac{d \ln MC_i}{d \ln p_{ij}},
\label{eq:main_pt_formula}
\end{equation}
where $\Gamma_{ij}$ and $\Lambda_{ij}$ are the elasticities of the price--cost ratio and of the exporter's marginal cost with respect to $p_{ij}$.}
\end{proposition}

\noindent\textbf{Proof.} See Appendix~\ref{appendix:proof-prop3}.

Proposition \ref{prop:3} highlights two mechanisms shaping tariff pass-through: a markup channel, capturing endogenous markup responses, and a cost channel, capturing endogenous changes in marginal cost. Following \citet{burstein2014International}, we define both elasticities with a negative sign. Under this convention, positive values correspond to adjustments that dampen pass-through, while negative values correspond to adjustments that amplify it.  We characterize each channel in turn.

\paragraph{The Markup Channel ($\Gamma_{ij}$).}\label{sec:markup_elasticity}

The markup elasticity decomposes as
\begin{equation*}
\Gamma_{ij}
=
\underbrace{
(1-\omega_{ij}^{\Gamma})\Gamma_{ij}^{\text{oligopoly}}
+
\omega_{ij}^{\Gamma}\Gamma_{ij}^{\text{oligopsony}}
\vphantom{\left(
1-\frac{\mu_{ij}^{\text{oligopoly}}}{\mu_{ij}}
\right)}}_{\text{direct price-cost ratio adjustment}}
+
\underbrace{
\left(
1-\frac{\mu_{ij}^{\text{oligopoly}}}{\mu_{ij}}
\right)\Gamma_{ij}^{\omega}
}_{\text{bargaining-weight adjustment}},
\end{equation*}
where
\[
\omega_{ij}^{\Gamma}
\equiv
\omega_{ij}
\frac{\mu_{ij}^{\text{oligopsony}}}{\mu_{ij}}
\in[0,1]
\]
is an adjusted bargaining weight. The terms $\Gamma_{ij}^{\text{oligopoly}}$ and $\Gamma_{ij}^{\text{oligopsony}}$ are the price elasticities of the oligopoly and oligopsony price-cost ratios, respectively; $\Gamma_{ij}^{\omega}$ is defined analogously for the bargaining weight.  Appendix~\ref{subsec: Derivations Pass-through} derives each component.

The first component is the response of the price-cost ratio $\mu_{ij}$ at a given bargaining weight $\omega_{ij}$. It is a weighted average of the oligopoly and oligopsony elasticities, which have opposite signs, with weight $\omega_{ij}^{\Gamma}$ on the latter. The second component arises because the price also moves $\omega_{ij}$ itself. We discuss the three elasticities in turn.

When exporter bargaining power dominates ($\phi \to 0$), $\Gamma_{ij}$ converges to $\Gamma_{ij}^{\text{oligopoly}}$, given by
\[
\Gamma_{ij}^{\text{oligopoly}}
=
-\frac{d \ln \mu_{ij}^{\text{oligopoly}}}{d \ln p_{ij}}
=
\frac{1}{\varepsilon_{ij} - 1}
\cdot
\frac{\rho - \varepsilon_{ij}}{\varepsilon_{ij}}
\cdot
(\rho - 1)(1 - s_{ij})
\ge 0.
\]
In this regime, the markup channel operates through \emph{strategic complementarities}, as is standard in oligopoly models.\footnote{See, e.g., \citealp{amiti2014importers, auer2016market, garetto2016firms, amiti2019international}.}
The markup response is strongest at intermediate supplier shares, implying a hump-shaped relationship between $\Gamma_{ij}^{\text{oligopoly}}$ and $s_{ij}$.\footnote{\cite{amiti2014importers} obtain a pass-through that declines monotonically with the supplier share from a first-order approximation around small shares. The approximation does not apply here, because bilateral supplier shares span the whole unit interval.}

When importer bargaining power dominates ($\phi \to 1$), $\Gamma_{ij}$ converges to $\Gamma_{ij}^{\text{oligopsony}}$, given by
\[
\Gamma_{ij}^{\text{oligopsony}}
=
-\frac{d \ln \mu_{ij}^{\text{oligopsony}}}{d \ln p_{ij}}
=
\left(
\frac{x_{ij}(1-x_{ij})^{\frac{1}{\theta}-1}}
{\theta \left[1-(1-x_{ij})^{\frac{1}{\theta}}\right]}
-1
\right)
(1-x_{ij}) \varepsilon_{ij}
\le 0.
\]
In this case, $\Gamma_{ij}$ is governed by adjustments in the markdown and captures \emph{strategic substitutabilities} among importers. As $p_{ij}$ rises, importer $j$ purchases less, shrinking the quasi-rents and compressing the markdown, which amplifies the price response. This markdown response raises pass-through relative to the cost-only response. The cost channel, however, always more than offsets it (Proposition~\ref{prop:4}).

The elasticity $\Gamma_{ij}^{\text{oligopsony}}$ depends on both the buyer share $x_{ij}$ and the supplier share $s_{ij}$. It is U-shaped in $x_{ij}$: it vanishes as $x_{ij} \to 0$ (atomistic buyer) or $x_{ij} \to 1$ (monopsonist) and is largest in absolute value at intermediate buyer shares. Its absolute size declines in $s_{ij}$ because a higher supplier share reduces the demand elasticity $\varepsilon_{ij}$, limiting variation in $x_{ij}$ and weakening the markdown response.

With two-sided market power, a price change also alters the bargaining weight, with elasticity
\[
\Gamma_{ij}^{\omega}
\equiv -
\frac{d\ln\omega_{ij}}{d\ln p_{ij}},
\]
whose sign and magnitude vary with the supplier share $s_{ij}$.\footnote{Appendix~\ref{subsec: Derivations Pass-through} provides the exact expression for $\Gamma_{ij}^{\omega}$.} For $s_{ij}\to 0$ and $s_{ij}\to 1$, $\omega_{ij}$ converges to the exogenous parameter $\phi$; it is then locally unresponsive to $p_{ij}$, so that $\Gamma_{ij}^{\omega}\to 0$ and only the first component remains.

\paragraph{The Cost Channel ($\Lambda_{ij}$).}
\label{sec:cost_elasticity}

The cost elasticity $\Lambda_{ij}$ captures how the exporter's marginal cost responds to a change in the bilateral price. It is given by
\begin{equation}
\Lambda_{ij}
=
\frac{1-\theta}{\theta}\,
x_{ij}\varepsilon_{ij}
\geq 0.
\label{eq:Lambda}
\end{equation}
Three terms determine this elasticity. The factor $(1-\theta)/\theta$ measures the curvature of the exporter's marginal-cost schedule or, equivalently, the inverse elasticity of export supply at the firm level. The buyer share $x_{ij}$ determines how strongly a contraction in bilateral sales translates into a contraction in the exporter's total output, while the residual demand elasticity $\varepsilon_{ij}$ governs the response of bilateral quantity to the price change.

Equation \eqref{eq:Lambda} thus makes clear the role of supply curvature in the cost channel. An exporter's marginal cost depends on its total output, but a bilateral price change moves only the quantity sold to buyer $j$, so the cost response is the exporter's supply curvature scaled by the buyer's share. The product $\frac{1-\theta}{\theta}\,x_{ij}$ is the \emph{residual} inverse export supply elasticity facing importer $j$. The same scaling clarifies the comparison with the neoclassical benchmark, in which a tariff lowers the aggregate export price \citep{Johnson1953}. Here the adjustment occurs link by link, so pass-through differs across the buyers of a single exporter.

\paragraph{Pass-Through Elasticity and Bilateral Market Shares}
\label{sec:pass_through_bilateral_shares}

\begin{figure}[t]
\caption{Pass-Through Elasticity and Bilateral Market Shares}
\label{fig:hetmap}
\begin{centering}
\includegraphics[width=0.95\columnwidth]{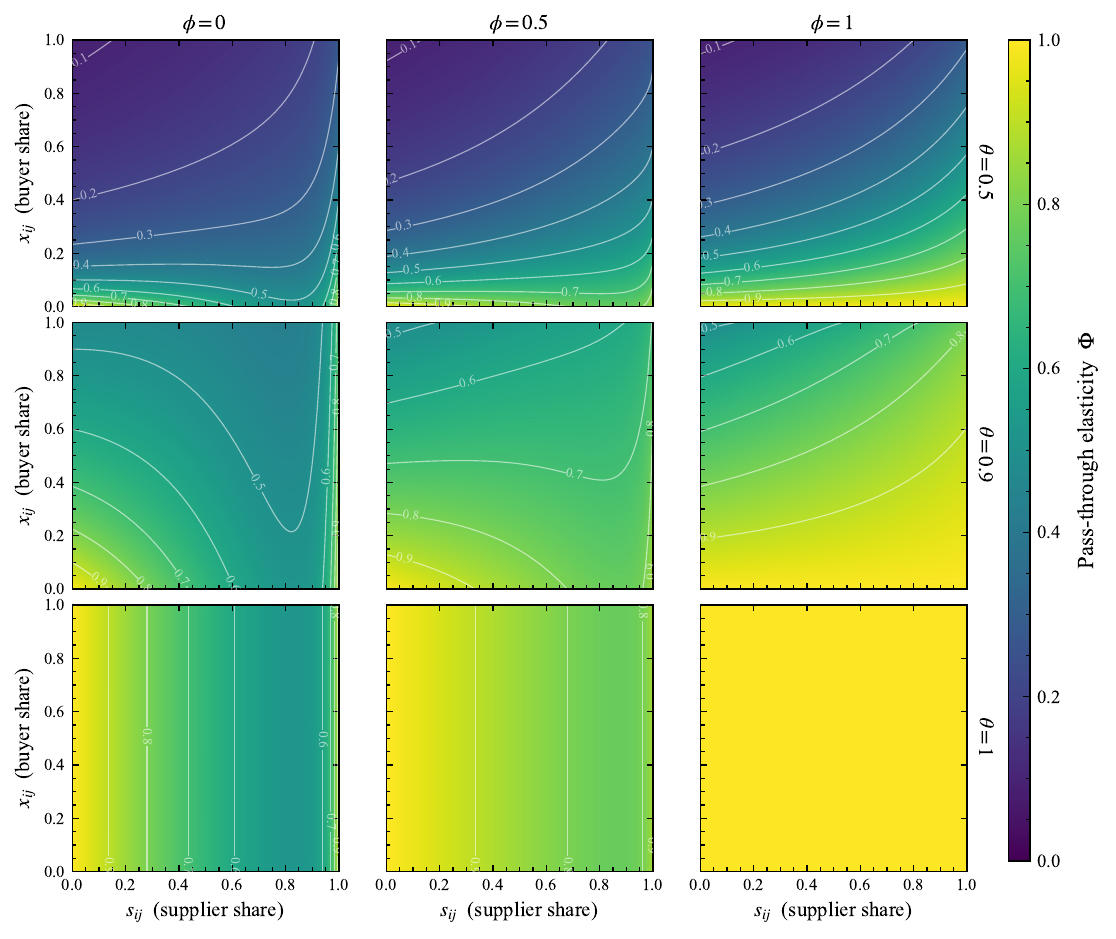}
\par\end{centering}
\footnotesize
\textit{Notes:} The figure presents heatmaps of the pass-through elasticity $\Phi_{ij}$ as a function of the supplier share $s_{ij}$ (horizontal axis) and the buyer share $x_{ij}$ (vertical axis). Columns correspond to $\phi\in\{0,0.5,1\}$, and rows correspond to $\theta\in\{0.5,0.9,1\}$. We use $\theta=0.9$ as an intermediate value of the returns-to-scale parameter, consistent with estimates in the production-function literature, which typically lie between 0.9 and 1; see, for example, \citet{basu1997returns}. The remaining parameters are fixed at $\gamma=0.5$, $\varrho=1$, $\nu=4$, and $\rho=10$.
\end{figure}

Figure~\ref{fig:hetmap} shows how supply curvature and bargaining power jointly shape pass-through. Each panel plots $\Phi_{ij}$ against the supplier share $s_{ij}$ and the buyer share $x_{ij}$. Columns vary the allocation of bargaining power; rows vary $\theta$, with export supply flattening and the cost channel weakening from top to bottom.

Comparing rows shows the first-order role of supply curvature. In the bottom row, $\theta=1$, marginal cost is fixed, and pass-through is governed entirely by the markup channel. Bargaining power then matters: when $\phi=0$, strategic complementarities govern pass-through, making it U-shaped in supplier share; when $\phi=1$, the negotiated price equals marginal cost and pass-through is complete. Moving up the rows, the cost channel strengthens, pass-through becomes more incomplete, and the differences across bargaining regimes shrink. They are smallest in the top row, where $\theta=0.5$.

Under decreasing returns, the cost and markdown responses have opposite effects on the buyer-share gradient of pass-through. The following proposition shows that the markdown response can never outweigh the cost response.

\begin{proposition}\label{prop:4}
\textit{Suppose $\theta<1$. For any bilateral shares $(s_{ij},x_{ij})$ and any $\phi\in[0,1]$:}
\begin{enumerate}
\item \textit{The markdown response is at most half the cost response in absolute value:
\[
-\tfrac{1}{2}\Lambda_{ij}
\le
\Gamma^{\textnormal{oligopsony}}_{ij}
\le 0.
\]}

\item \textit{The sum of the cost response and the weighted markdown response, $\Lambda_{ij}+\omega^{\Gamma}_{ij}\Gamma^{\textnormal{oligopsony}}_{ij}$, is positive and strictly increasing in the buyer share $x_{ij}$.}
\end{enumerate}
\end{proposition}

\noindent\textbf{Proof.} See Appendix~\ref{appendix:proof-prop4}.

\vspace{0.5em}

By showing that the cost channel dominates the markdown response regardless of the allocation of bargaining power, Proposition~\ref{prop:4} establishes two results. First, the combined cost and weighted markdown contribution to the denominator of pass-through is positive. Second, this contribution rises with the buyer share.
Under constant returns, by contrast, the buyer share drops out of the pricing equation and pass-through does not vary with it. A decline of pass-through with the buyer share therefore identifies decreasing returns, and it is the prediction we take to the data.

There is no comparable sign restriction for the supplier-share gradient. Strategic complementarity tends to make pass-through U-shaped in $s_{ij}$, while strategic substitutability and the cost channel tend to make it increasing. Which force dominates is an empirical matter.

Proposition~\ref{prop:4} compares the cost channel only with the opposing markdown response. The markup channel also contains the oligopoly markup response and changes in the effective bargaining weight. The next result establishes when the cost channel dominates this full markup channel.

\begin{proposition}\label{prop:phi_theta}
\textit{Fix a relationship with $(s_{ij},x_{ij})\in(0,1)^2$. There is a finite constant $B_{ij}$ such that $|\Gamma_{ij}|\le B_{ij}$ for every $\phi\in[0,1]$ and every $\theta\in(0,1)$, whereas $\Lambda_{ij}=\frac{1-\theta}{\theta}x_{ij}\varepsilon_{ij}$ grows without bound as $\theta\to 0$. Hence, as $\theta\to 0$,
\[
\frac{|\Gamma_{ij}|}{\Lambda_{ij}}\to 0
\qquad\text{and}\qquad
\Phi_{ij}\,(1+\Lambda_{ij})\to 1,
\]
uniformly in $\phi$ and at rate $\theta$.}
\end{proposition}

\noindent\textbf{Proof.} See Appendix~\ref{appendix:proof-prop-phi-theta}.

\vspace{0.5em}

Proposition~\ref{prop:phi_theta} formalizes the convergence across columns in Figure~\ref{fig:hetmap}. As export supply steepens, the markup channel stays bounded while the cost channel grows, so pass-through is governed by the cost channel regardless of how bargaining power is allocated. The result also has a cross-sectional implication. Pass-through is governed by the total elasticity $\Lambda_{ij}+\Gamma_{ij}$, and its variation across relationships can be attributed to the two channels through their covariances with the total.

\begin{corollary}\label{cor:cov}
\textit{Consider a finite cross-section of relationships with $(s_{ij},x_{ij})\in(0,1)^2$ in which $x_{ij}\varepsilon_{ij}$ varies, and let $D_{ij}\equiv\Lambda_{ij}+\Gamma_{ij}$. As $\theta\to 0$,
\[
\frac{\operatorname{Cov}(\Lambda_{ij}, D_{ij})}{\operatorname{Var}(D_{ij})}\to 1,
\]
uniformly in $\phi$.}
\end{corollary}

\noindent\textbf{Proof.} See Appendix~\ref{appendix:proof-prop-phi-theta}.

\vspace{0.5em}

The ratio in Corollary~\ref{cor:cov} is the share of the cross-sectional variance of the total elasticity accounted for by the cost channel; the corresponding share for the markup channel is one minus this ratio and therefore vanishes in the limit. Overall, as $\theta\to 0$, both the level of pass-through and its variation across relationships reflect the cost channel alone, regardless of the allocation of bargaining power.

These are limiting results, and the approach to the limit need not be monotonic. Whether export supply is steep enough for the cost channel to dominate is therefore an empirical question. We estimate the model's parameters and return to this discussion in Sections~\ref{sec:Identification-and-estimation} and \ref{sec:Implications}.

\subsection{Discussion}
\label{subsec:Discussion-and-Extensions.}

We briefly review our key modeling choices before turning to the empirical analysis.

\paragraph{Bargaining Protocol and Quantities}

Our baseline model assumes \emph{demand-driven quantities}, whereby the importer chooses quantities given negotiated prices. This ensures allocative pricing and nests the standard monopolistic-competition benchmark as a limiting case. This assumption is standard in the literature and is consistent with firm-level evidence \citep{gopinath2011search}.

We derive two alternative protocols in the Appendix. Appendix~\ref{sec:APP_efficient-bargaining} discusses \emph{efficient bargaining}, which maximizes joint surplus but makes the negotiated price non-allocative. Appendix~\ref{sec:APP_supply-bargaining} discusses the case of \emph{supply-driven bargaining}, in which the parties bargain over quantity along the exporter's supply schedule. This protocol yields a simple price equation, but its bargaining wedge is not pinned down by observable bilateral shares without further restrictions.

One caveat of our framework is that when $\phi$ is sufficiently high, the negotiated price falls below the exporter's marginal cost ($\mu_{ij}<1$). Because the importer chooses the quantity, it then buys more than the efficient quantity, at which price equals marginal cost (the square in Panel~(C) of Figure~\ref{fig:allocations}). Recent work on bilateral oligopoly rules out this region \citep[e.g.,][]{demirer2025welfare, avignon2024markups}, letting quantities be determined endogenously on either the demand or the supply side. Adopting a similar restriction here would make the allocation piecewise and would require taking a stance on which relationships are constrained. We instead keep the demand-driven protocol as a transparent identifying assumption, because its allocative pricing equation maps bilateral shares into markups we can take to the data. 

\paragraph{General Equilibrium and Network Spillovers}
Our theoretical results are derived in partial equilibrium, holding wages, aggregate demand shifters, and prices on all other links fixed. The pass-through elasticity derived in Section~\ref{subsec: Pass-through (Theory)} therefore isolates the direct, intensive-margin response of markups and marginal costs to a tariff shock.

Embedding this framework in a general equilibrium environment introduces two main complexities. First, under decreasing returns, quantity adjustments along one link affect the exporter’s marginal cost across all links. Second, aggregate shocks or network spillovers may differentially affect outside options across matches of the same exporter. Fully accommodating these channels requires committing to a specific macroeconomic closure for wages, final demand, and network formation, which is beyond the scope of this paper.\footnote{A complementary literature embeds bilateral bargaining between exporters and importers in equilibrium models of search and matching \citep[e.g.,][]{KrolikowskiMcCallum2021, KrolikowskiMcCallum2026, EatonJinkinsTyboutXu2026, GrossmanHelpmanRedding2024}. Its focus is the formation and dissolution of relationships, and its pricing blocks are correspondingly more stylized.} As we show below, the model matches tariff-induced price changes within continuing relationships, lending support to the partial-equilibrium approach for the question at hand. This does not establish that network feedback is small.

\section{Data and Stylized Facts \label{sec:data-stylized-facts}}
This section introduces the data and the empirical objects used in the analysis. We first describe the data sources, adapt the model’s shares to the customs data, and construct the sample. We then present reduced-form evidence on bilateral prices and tariff pass-through.

\subsection{Data Sources\label{subsec:Data-sources}}
Our main dataset is the U.S. Census Bureau's Linked/Longitudinal Firm Trade Transaction Database (LFTTD), which covers the universe of U.S. import transactions from 2001 to 2018. Each observation corresponds to a shipment from a foreign exporter to a U.S. importer and includes the transaction date, product classification at the 10-digit Harmonized System (HS10) level, Free on Board (FOB) import value in U.S. dollars, physical quantity, transportation mode, and country of origin. Exporters are identified using a manufacturer ID (MID) constructed by the Census Bureau from the exporter's name, street address, city, and country.\footnote{The MID combines country code, elements of the firm name, city, and address \citep{Kamal2018}. Because the algorithm is not standardized, identifiers may be inconsistent: one firm may have multiple MIDs or multiple firms may share one. Following \citet{Kamal2018}, we construct a robustness measure that truncates location fields to improve consistency. Results are robust.}

We identify related-party transactions using two sources. The LFTTD includes a related-party indicator based on a mandatory field in U.S. Customs forms, which flags relationships with ownership stakes of at least five percent. Because this measure relies on self-reporting and a low ownership threshold, it may misclassify firms \citep{kimintrafirmAERPP}. We therefore complement it with an indicator constructed from ORBIS, which records firm-level cross-border ownership links; Appendix~\ref{subsec:Merging-foreign-exporter-id-orbis} describes the merge. 

We then supplement the transaction-level data with information on statutory U.S. import tariffs introduced during the 2018 trade war. We use the dataset from \citet{fajgelbaum2020return}, which records the timing, product coverage, and country-specific scope of these measures at the HS8-month level. The tariffs averaged 25 percentage points and were imposed on top of existing rates, targeting selected goods and countries. Tariff changes are annualized based on the number of months each measure was in effect.

\subsection{Measuring Key Variables of the Model\label{subsec:Measuring-key-variables}}
The model treats each exporter as supplying a single input, whereas in the data an exporter sells one or more HS10 products, and prices and shares are observed at the product level. To accommodate this, we extend the model to multiple foreign inputs, indexed by $h$, each corresponding to an HS10 product, and treat the foreign input bundle as a Cobb-Douglas composite of individual product quantities:
\begin{align*}
q_{j}^{f} & =\prod_{h\in\mathcal{H}_{j}}\left(q_{jh}^{f}\right)^{\alpha_{jh}},\ \ \ \text{where}\ \ \ q_{jh}^{f}=\Big(\sum_{i\in\mathcal{Z}_{j}^h}\varsigma_{ijh}\cdot\left(q_{ijh}\right)^{\frac{\rho-1}{\rho}}\Big)^{\frac{\rho}{\rho-1}},
\end{align*}
and $\alpha_{jh} \in (0,1)$ denotes the (observed) Cobb-Douglas share of input $h$ in firm $j$’s total imports of foreign intermediates. 
At the baseline $\varrho=1$ (constant returns to scale for importers), the elasticity of the importer's marginal cost with respect to the price of foreign input $h$ is $\frac{d\ln c_j}{d\ln p_{jh}^{f}}=\alpha_{jh}\gamma$.\footnote{More generally, allowing downstream output to adjust as in Appendix~\ref{subsec: Demand Elasticity}, it is $\alpha_{jh}\gamma/[\varrho+\nu(1-\varrho)]$.}
We also note that allowing this elasticity to vary across importer-product pairs in turn makes $\eta$ in equation \eqref{eq:eta} vary across the same dimension, $\eta_{jh}$.

We construct the exporter's supplier share as
$s_{ijh}\equiv\frac{p_{ijh}q_{ijh}}{\sum_{k\in\mathcal{Z}_{j}^h}p_{kjh}q_{kjh}}$,
where $\mathcal{Z}_{j}^h$ denotes the set of firm $j$'s foreign suppliers of input $h$. The numerator captures the total value of imports of product $h$ from exporter $i$ (a MID in our data) to firm $j$ in a given year. The denominator aggregates imports of product $h$ from all foreign suppliers to $j$. 

In contrast, the importer’s buyer share is constructed as $x_{ijh}\equiv\frac{q_{ijh}}{\sum_{k\in\mathcal{Z}_{i}^{h}}q_{ikh}}$, where $\mathcal{Z}_{i}^{h}$ is the set of all U.S. importers buying product $h$ from exporter $i$. Since our dataset only includes U.S. importers, we assume that exporter $i$ operates product- and destination-specific production lines. Under this assumption, the denominator of $x_{ijh}$, which captures the total quantity of product $h$ from exporter $i$, includes only those sold to U.S. buyers. This restriction reflects a data limitation, as we do not observe importer destinations beyond the U.S. and thus cannot account for the full set of an exporter’s buyers.\footnote{To address potential overstatement of buyer share $x_{ij}$ due to unobserved exports to other destinations, we replicate the analysis for Canada and Mexico, which send most exports to the U.S. (71\% and 73\% in 2019). Estimates are similar in this subsample. Full results available upon request.}

\subsection{Sample Construction and Summary Statistics}
\label{subsec:Selection}

We apply a sequence of restrictions to the LFTTD to construct the estimation sample. Some are required for identification and counterfactuals; others align the baseline sample with the model's environment. Appendix~\ref{subsec:Sample-Selection} provides details, and Appendix Table~\ref{tab:sample_stats_bec1} traces the sample under each restriction.

\paragraph{Essential restrictions.}
First, we require importer\allowbreak-exporter\allowbreak-product triplets to be observed in two consecutive years, which is necessary to compute bilateral price changes. Second, because identification of the bargaining and returns-to-scale parameters relies on cross-sectional price dispersion across importers sourcing the same exporter-product-year, we restrict the sample to exporter-product pairs that transact with at least two U.S.\ buyers in consecutive years.

\paragraph{Other restrictions.}
We impose three additional restrictions to align the data with the model. First, to limit measurement error and ensure that unit values reflect prices rather than compositional changes within HS10 products, we follow standard practice and (i) exclude energy goods, (ii) drop unit values outside the 1st--99th percentile within product, and (iii) remove observations with absolute log price changes exceeding four.

Second, we exclude related-party transactions, as prices in such relationships may reflect forces outside our framework, such as transfer pricing \citep{bernard2006transfer}. We classify a pair as related if ORBIS identifies a common corporate parent. The baseline excludes all such pairs. Alternative definitions based solely on the LFTTD indicator or on multinational status yield similar results.

Third, the baseline covers capital and intermediate goods and excludes consumption goods under the Broad Economic Categories (BEC) classification. This restriction matches the model's interpretation of supply curvature as arising in production relationships, where relationship-specific investments may limit short-run scalability \citep{Nunn2007}. The bilateral oligopoly mechanism may also apply to foreign sourcing by retailers and wholesalers. However, because the estimated bargaining weight, supply curvature, and pass-through are likely to differ for consumption goods, we report those estimates separately.

After all restrictions, the baseline sample covers approximately \$160 billion in import value and 250,000 buyer-supplier-product-year observations over the 2017--2018 period. This corresponds to about 8\% of total U.S.\ imports by value and 5\% of all buyer--supplier--product relationships observed in the LFTTD (Appendix Table~\ref{tab:sample_stats_bec1}).

\vspace{1ex}
\begin{table}[t]
\caption{Summary Statistics for Main Estimation Sample (2001-2018)}
\label{tab:summary_stats}
\centering
{\small
\renewcommand{\arraystretch}{1.2}
\arrayrulecolor{gray!50}
\begin{tabular}{
  >{\raggedright\arraybackslash}p{6cm} 
  >{\centering\arraybackslash}p{1.6cm} 
  >{\centering\arraybackslash}p{1.6cm} 
  >{\centering\arraybackslash}p{1.6cm}
}
\toprule
\text{Variable} & \text{Mean} & \text{Median} & \text{P25--P75} \\
\midrule
\addlinespace
\multicolumn{4}{c}{\textit{Panel A}: Trade Relationships } \\
\midrule
\addlinespace
$s_{ijh}$: Supplier share & 0.32 & 0.15 & 0.03--0.57 \\
$x_{ijh}$: Buyer share & 0.25 & 0.10 & 0.02--0.40 \\
Relationship length & 4.0 & 3.5 & 2.5--5.5 \\
Relationship length (\emph{all products}) & 4.8 & 4.5 & 2.5--6.5 \\
\# Transactions  & 120 & 16 & 6.5--50 \\
\# Products per pair & 3.8 & 2.5 & 1.5--4.5 \\
Buyer tenure (product $h$)    & 6.9 & 6.5 & 3.5--10.0 \\
Supplier tenure (product $h$)  & 6.4 & 6.5 & 3.5--9.5 \\
\midrule
\addlinespace
\multicolumn{4}{c}{\textit{Panel B}: Prices (log unit values)} \\
\midrule
\addlinespace
$\log p$ (pre-duty, excl.\ charges) & 3.40 & 3.00 & 1.30--5.40 \\
$\log p$ (pre-duty) & 3.50 & 3.10 & 1.40--5.40 \\
$\log p^{\text{duty}}$ (post-duty) & 3.50 & 3.10 & 1.40--5.40 \\
\bottomrule
\end{tabular}
}
\begin{minipage}{14cm}\vspace{0.5em}
\scriptsize \textit{Notes}: This table reports summary statistics for the main estimation sample at the importer-supplier-product level. Panel~A reports trade relationship characteristics. $s_{ijh}$ is exporter $i$'s share in buyer $j$'s imports of product $h$; $x_{ijh}$ is buyer $j$'s share in exporter $i$'s U.S. exports of product $h$. Relationship length is the number of years between the supplier-buyer pair's first transaction in product $h$ and year~$t$; Relationship length (all products) is defined analogously, using the pair's first transaction in any product. \# Transactions is the cumulative number of transactions between the pair in product $h$, from their first transaction through year~$t$. \# Products per pair is the number of distinct HS10 products traded by the pair in year~$t$. Buyer tenure (Supplier tenure) is the number of years between the buyer's (supplier's) first recorded transaction in product $h$ with any partner and year~$t$; Appendix Table~\ref{tab:summary_stats_main} also reports tenure pooled across products. Panel~B reports log unit values (value divided by quantity). $\log p$ (pre-duty, excl.\ charges) corresponds to the FOB (Free on Board) unit value, excluding shipping and insurance charges; $\log p$ (pre-duty) includes these charges to obtain the CIF (Cost, Insurance, and Freight) unit value; and $\log p^{\text{duty}}$ (post-duty) further incorporates import duties. Statistics are based on confidential LFTTD data and rounded according to US Census Bureau disclosure guidelines. \textit{Source}: FSRDC Project Number 2109 (CBDRB-FY25-P2109-R12520).
\end{minipage}
\end{table}

\paragraph{Summary Statistics} Table~\ref{tab:summary_stats} reports summary statistics for our final sample.\footnote{Appendix~\ref{subsec:Sample-Selection} details the sample construction, Appendix Table~\ref{tab:summary_stats_main} reports additional statistics, and Appendix Tables~\ref{tab:sample_stats_bec0} and~\ref{tab:summary_stats_bec0} report sample composition and summary statistics for the pooled sample including all BEC categories.} Panel~A reports trade relationship characteristics. It documents substantial concentration on both sides of the market. On average, an exporter supplies 32\% of an importer’s total imports of a given HS10 product, with a median share of 15\%. The average buyer share is lower, at 25\%, with a median of 10\%, likely reflecting the exclusion of cases with buyer share equal to one. The two shares are highly dispersed and essentially uncorrelated (correlation 0.04).

Trading relationships are also persistent: the typical importer-exporter pair has been transacting in a given HS10 product for about four years, and in any product for nearly five. Activity within the relationship is intensive: at the median, a pair has accumulated 16 transactions in a product to date and trades 2.5 distinct HS10 products in a given year. Buyer and supplier tenure in a given product, counted across all partners, average about seven and six years, respectively.

Panel B reports descriptive statistics for three bilateral (log) price measures, each defined as transaction value divided by quantity: an FOB (Free on Board) pre-duty price that excludes shipping and insurance charges; a CIF (Cost, Insurance, and Freight) pre-duty price that adds those charges; and a post-duty price that further incorporates import duties. All three are highly dispersed across importer-exporter-product-year observations, with interquartile ranges of about four log points.

To assess the sources of this variation, we conduct a variance decomposition following  \cite{Fontaine2019}, described in Appendix~\ref{sec:price_dispersion}. Product-year fixed effects account for about 50\% of the total variance, while match-specific residuals account for 4\%. Focusing on within supplier--product--year variation, 77\% of the remaining dispersion is explained by match-specific factors. This pattern holds across price definitions and underscores the importance of relationship-specific forces in pricing.

\subsection{Reduced-Form Evidence on Prices and Pass-Through\label{subsec:Test_prediction}}
Before estimating the model, we examine how bilateral prices and pass-through of the 2018 tariffs vary with buyer and supplier shares. These regressions do not recover structural parameters: shares are endogenous and enter prices nonlinearly, while tariff responses may reflect general-equilibrium adjustments outside the model. The price regressions assess whether the data support the model's mechanisms and preview the variation identifying $\theta$ and $\phi$ in Section~\ref{subsec:Estimation-of-parameters-theta-phi}. The pass-through regressions instead provide untargeted moments for the out-of-sample evaluation in Section~\ref{subsec:ModelTest}, since 2017--2018 is excluded from estimation.

Because bilateral price changes require links observed in consecutive years, our pass-through estimates pertain to continuing relationships. Section~\ref{sec:comparison_product_level} distinguishes this relationship-level object from broader product-level estimates in the literature.

\paragraph{Markups and Bilateral Market Shares\label{subsubsec:Test_P1}}
Proposition~\ref{prop:2} provides sufficient conditions linking bilateral price-cost margins to the two market shares. Conditional on marginal cost, a positive comovement between price and the supplier share, \(s_{ijht}\), implies \(\phi<1\), while a negative comovement between price and the buyer share, \(x_{ijht}\), implies \(\phi>0\) and \(\theta<1\). We examine whether the data exhibit these signs using
\begin{align}
\ln p_{ijht} & = \beta_{s}\, s_{ijht} + \beta_{x}\, x_{ijht} + \mathbf{X}_{ijht} \boldsymbol{\gamma} + \mathbf{FE} + \upsilon_{ijht}. \label{eq:fact_3}
\end{align}
The coefficients of interest are \(\alpha_s\) and \(\alpha_x\). Because price-cost margins are unobserved, the dependent variable is the log bilateral price. Our most informative specifications include supplier--product-year fixed effects, which absorb marginal cost and other price determinants common across the supplier's buyers. The remaining variation is across buyers within supplier--product-year cells and, through the model, maps into differences in bilateral price-cost margins. The structural estimator in Section~\ref{subsec:Estimation-of-parameters-theta-phi} exploits the same cross-sectional variation.
\vspace{1ex}
\begin{table}[t]
\caption{Prices and Bilateral Concentration}
\label{Table 2 - Prices}
\centering
{\small
\renewcommand{\arraystretch}{1.2}
\arrayrulecolor{gray!50}
\begin{tabular}{
  >{\raggedright\arraybackslash}p{4cm} 
  >{\centering\arraybackslash}p{1.6cm} 
  >{\centering\arraybackslash}p{1.6cm} 
  >{\centering\arraybackslash}p{1.6cm} 
  >{\centering\arraybackslash}p{1.6cm} 
  >{\centering\arraybackslash}p{1.6cm} 
  >{\centering\arraybackslash}p{1.6cm}
}
\toprule
\small{Dependent Variable:} & \multicolumn{6}{c}{$\ln p_{ijht}$} \\
\cmidrule(lr){2-7}
& (1) & (2) & (3) & (4) & (5) & (6) \\
& OLS & IV & OLS & IV & OLS & IV \\
\midrule
$s_{ijht}$ & 0.166 & 0.170 & 0.237 & 0.169 & 0.250 & 0.642 \\
& (0.026) & (0.047) & (0.040) & (0.039) & (0.058) & (0.074) \\
$x_{ijht}$ & -0.559 & -0.093 & -0.617 & -0.237 & -0.549 & -0.783 \\
& (0.036) & (0.024) & (0.037) & (0.036) & (0.047) & (0.042) \\
\midrule
$\text{FE}_i + \text{FE}_j + \text{FE}_{ht}$     & Yes & Yes & No  & No  & No  & No \\
$\text{FE}_{iht} + \text{FE}_j$           & No  & No  & Yes & Yes & No  & No \\
$\text{FE}_{iht} + \text{FE}_{jht}$       & No  & No  & No  & No  & Yes & Yes \\
\midrule
Observations & 1,200,000 & 784,000 & 1,200,000 & 784,000 & 1,200,000 & 784,000 \\
R-squared & 0.958 & --- & 0.976 & --- & 0.991 & --- \\
First-stage F stat. & --- & 637 & --- & 781 & --- & 169 \\
\makecell[l]{SW F stat  ($s_{ijht}$)} & --- & 2,227 & --- & 1,519 & --- & 4,595 \\
\makecell[l]{SW F stat  ($x_{ijht}$)} & --- & 999 & --- & 1,696 & --- & 7,768 \\
\bottomrule
\end{tabular}
}
\begin{minipage}{17cm}\vspace{0.5em}
\footnotesize \textit{Notes:} This table reports OLS and IV estimates of equation~(\ref{eq:fact_3}), where the dependent variable is the log CIF duty-exclusive unit value of product $h$ imported by buyer $j$ from supplier $i$ in year $t$. Columns alternate between OLS and IV specifications. All regressions control for log relationship length (in years) within HS10 products. Columns~(1)--(2) include buyer ($\text{FE}_j$), supplier ($\text{FE}_i$), and product-year ($\text{FE}_{ht}$) fixed effects. Columns~(3)--(4) use supplier--product-year ($\text{FE}_{iht}$) and buyer ($\text{FE}_j$) fixed effects. Columns~(5)--(6) include fully interacted buyer--product-year and supplier--product-year fixed effects ($\text{FE}_{jht}$, $\text{FE}_{iht}$). IV estimates (even-numbered columns) use leave-one-out instruments: $s_{ijht}$ is instrumented with exporter $i$'s average supplier share across its other buyers (excluding $j$), and $x_{ijht}$ with importer $j$'s average buyer share across its other suppliers (excluding $i$), within product $h$ and year $t$. Because the model includes multiple endogenous regressors, we report both first-stage and conditional F-statistics from \citet{SANDERSON2016212}. Standard errors are clustered at the HS8 product and exporter-country level. The number of observations is rounded to four significant digits in accordance with US Census Bureau disclosure guidelines. \textit{Source}: FSRDC Project Number 2109 (CBDRB-FY25-P2109-R12520).
\end{minipage}
\end{table}

The expenditure-based supplier share mechanically contains price, and both shares may respond to unobserved match characteristics. We therefore report OLS and IV estimates based on leave-one-out measures of market structure. We instrument $s_{ijht}$ with the exporter's average supplier share across its other buyers and $x_{ijht}$ with the importer's average buyer share across its other suppliers, excluding the focal match from both constructions.

Table~\ref{Table 2 - Prices} reports the results. Columns~(1)--(2) include exporter, importer, and product-year fixed effects. Columns~(3)--(4) replace exporter fixed effects with exporter--product-year fixed effects, while Columns~(5)--(6) also absorb importer--product-year shocks. All specifications control for relationship duration; odd-numbered columns report OLS estimates and even-numbered columns report IV estimates.

Across specifications, we find \(\hat{\beta}_s>0\) and \(\hat{\beta}_x<0\). In the most saturated IV specification, a 10 percentage point increase in supplier share is associated with an approximately 6.4 percent higher price, while the same increase in buyer share is associated with an approximately 7.8 percent lower price. Through the sufficient conditions in Proposition~\ref{prop:2}, these signs are consistent with \(\phi\in(0,1)\) and \(\theta<1\). 

\paragraph{Tariff Pass-Through and Decreasing Returns}
\label{subsubsec:Test_P3}
As shown in Section~\ref{subsec: Pass-through (Theory)}, pass-through can decline with the buyer share \(x_{ijht}\) only under decreasing returns. We examine this prediction during the held-out 2017--2018 tariff episode using
\begin{align}
\Delta \ln p^*_{ijht} =\ & \alpha_0 + \alpha_1 \, \Delta \ln(1 + \tau_{cht}) 
+ \alpha_s \, \Delta \ln(1 + \tau_{cht}) \cdot s_{ijh,t-1} 
+ \alpha_x \, \Delta \ln(1 + \tau_{cht}) \cdot x_{ijh,t-1} \notag \\
& + \alpha_{2} \, s_{ijh,t-1} + \alpha_{3} \, x_{ijh,t-1} 
+ \mathbf{X}_{ijht}\boldsymbol{\gamma} + \mathbf{FE} + \epsilon_{ijht},
\label{eq:pass_through}
\end{align}
where $c$ denotes supplier $i$'s country. The dependent variable is the change in the duty-exclusive price, $p^*_{ijht}$, and the tariff shock is the change in the statutory tariff. Focusing on duty-exclusive prices isolates the response of the exporter's received price.\footnote{In the model, the buyer's effective price is $p_{ijh}=p^*_{ijh}(1+\tau)$, where $p^*_{ijh}$ is the exporter's received price and $\tau$ the statutory tariff rate.}

The interaction terms measure how pass-through varies with the supplier and buyer shares, $s_{ijh,t-1}$ and $x_{ijh,t-1}$, both measured before the tariff change. The vector $\mathbf{X}_{ijht}$ includes changes in exporter $i$'s sales to its other U.S. buyers and the average price change charged to importer $j$ by its other suppliers. These controls absorb observable spillovers through the parties' other relationships and bring the reduced-form estimate closer to the bilateral response the model characterizes, though they capture only part of the equilibrium adjustment that may enter the reduced-form coefficient.
\begin{table}[t]
\caption{Pass-Through and Relationship Heterogeneity}
\label{Table_PT_FEhtcs}
\label{tab:pass_through_main}
\centering
{\small
\renewcommand{\arraystretch}{1.2}
\arrayrulecolor{gray!50}
\begin{tabular}{
  >{\raggedright\arraybackslash}p{4.7cm} 
  >{\centering\arraybackslash}p{1.45cm} 
  >{\centering\arraybackslash}p{1.45cm} 
  >{\centering\arraybackslash}p{1.45cm} 
  >{\centering\arraybackslash}p{1.45cm} 
  >{\centering\arraybackslash}p{1.45cm} 
  >{\centering\arraybackslash}p{1.45cm}
}
\toprule
Dependent variable: & \multicolumn{6}{c}{$\Delta\ln p^*_{ijht}$} \\
\cmidrule(lr){2-7}
\cmidrule(lr){2-4}\cmidrule(lr){5-7}
& (1) & (2) & (3) & (4) & (5) & (6) \\
\midrule
$\Delta\ln(1+\tau_{cht})$ & -0.126 & -0.019 & -0.032 & -0.184 & -0.113 & -0.114 \\
                          & (0.091) & (0.097) & (0.091) & (0.107) & (0.126) & (0.104) \\
$\Delta\ln(1+\tau_{cht}) \cdot s_{ijh,t-1}$ 
                          &        &         & 0.033   &         &         & 0.004   \\
                          &        &         & (0.080) &         &         & (0.165) \\
$\Delta\ln(1+\tau_{cht}) \cdot x_{ijh,t-1}$ 
                          &        & -0.407 & -0.407 &         & -0.282 & -0.280 \\
                          &        & (0.112) & (0.112) &         & (0.125) & (0.134) \\
\midrule
$\text{FE}_{ht} + \text{FE}_{cs}$       & Yes & Yes & Yes & No & No & No \\
$\text{FE}_{ht} + \text{FE}_{ct} + \text{FE}_{jt}$ & No & No & No & Yes & Yes & Yes \\
\midrule
Observations              & 249,000& 249,000 & 249,000 & 249,000 & 249,000 & 249,000 \\
R-squared                 & 0.04 & 0.04 & 0.04 & 0.31 & 0.31 & 0.31 \\
\bottomrule
\end{tabular}
}
\begin{minipage}{16cm}\vspace{0.5em}
\footnotesize \textit{Notes:} This table reports estimates of equation~\eqref{eq:pass_through}, relating statutory tariff changes, $\Delta \ln (1 + \tau_{cht})$, to changes in CIF duty-exclusive unit values at the exporter-importer-HS10 product-year level, $\Delta \ln p^*_{ijht}$. Columns~(1)--(3) use the baseline product-year and exporter country-sector fixed effects ($\text{FE}_{ht}+\text{FE}_{cs}$). Columns~(4)--(6) use the more stringent product-year, exporter country-year, and importer-year fixed effects ($\text{FE}_{ht}+\text{FE}_{ct}+\text{FE}_{jt}$); Column~(4) is the preferred headline specification for average pass-through. Columns~(2) and~(5) interact tariffs with the lagged buyer share, $x_{ijh,t-1}$, defined as buyer $j$'s quantity share in supplier $i$'s U.S. exports of product $h$. Columns~(3) and~(6) also interact tariffs with the lagged supplier share, $s_{ijh,t-1}$, defined as supplier $i$'s expenditure share in buyer $j$'s imports of product $h$. Every regression controls for log relationship longevity, the change in supplier $i$'s sales of $h$ to other U.S. buyers, and the lag-share-weighted average price change charged to buyer $j$ by other suppliers of $h$. The sample is the ``+ Supplier Multi-Buyer'' sample in Table~\ref{tab:sample_stats_bec1}: continuing, arm's-length relationships in intermediate and capital goods during 2017--2018. Standard errors, in parentheses, are clustered at the HS8 product and exporter-country level. Observation counts are rounded to four significant digits under US Census Bureau disclosure guidelines. \textit{Source}: FSRDC Project Number 2109 (CBDRB-FY25-P2109-R12520).
\end{minipage}
\end{table}

Table~\ref{tab:pass_through_main} reports the results, with the baseline fixed effects in Columns~(1)--(3) and the more stringent set in Columns~(4)--(6). The main effects in Columns~(1) and~(4) imply duty-inclusive pass-through of about 87\% and 82\%, respectively. Thus, foreign suppliers absorb 13–18\% of the tariff within continuing relationships. Point estimates remain below one across fixed-effects structures, both here and in every specification of Appendix Table~\ref{Table_PT_strictFE}. Although imprecisely estimated, complete pass-through can be rejected at the 10\% level under the stringent fixed effects in Column~(4).

The buyer-share gradient is more precisely estimated. Under both sets of fixed effects, Columns~(2) and~(5) show that pass-through declines significantly with the importer's buyer share. Columns~(3) and~(6) add the interaction with the supplier share; this interaction is small and statistically insignificant under both fixed-effect structures, while the buyer share gradient remains negative. The results therefore identify buyer share as the principal dimension of pass-through heterogeneity across buyer-supplier relationships.

The patterns are robust to alternative specifications. Appendix Table~\ref{Table_PT_strictFE} adds interactions with relationship longevity, while Appendix Table~\ref{tab:robustness_pass_through} varies the price measure and the controls for observable equilibrium spillovers. Figure~\ref{fig:PT_nonlinearities} allows for nonlinear gradients by interacting tariff changes with quartiles of each share. It shows no systematic pattern across supplier-share quartiles but a monotonic decline across buyer-share quartiles under both fixed-effect structures. 

Together, Tables~\ref{Table 2 - Prices} and~\ref{tab:pass_through_main} provide reduced-form evidence on the model's two key ingredients, bargaining and decreasing returns to scale, respectively. Table~\ref{tab:pass_through_main} also suggests that average pass-through in the baseline sample is incomplete, which contrasts with the near-complete pass-through documented at the product level in the literature. Before turning to the structural estimation, Section~\ref{sec:comparison_product_level} clarifies how the focus on continuing relationships accounts for this difference.

\subsection{From Relationship-Level to Product-Level Pass-Through}
\label{sec:comparison_product_level}

Our reduced-form estimates point to incomplete tariff pass-through within continuing buyer--supplier--product relationships, although the average response is not always precisely estimated. Much of the literature on the 2018 tariffs finds near-complete pass-through \citep{amiti2019impact, amiti2020s, fajgelbaum2020return, cavallo2020tariff}. These studies differ from ours in both price measurement and sample composition: they typically use monthly product--source-country prices rather than annual relationship-level prices and cover a broader set of transactions. Table~\ref{tab:reconciliation} separates the role of price measurement from that of sample composition.\footnote{We use the baseline fixed effects, which are closer to the specifications in the literature. Appendix Table~\ref{tab:price_changes_sample_CTCS} reports the results under the more stringent fixed effects.} Panel~A uses annual relationship-level prices, while Panel~B uses monthly product--source-country prices. Column~(1) contains the baseline sample, and subsequent columns cumulatively restore transactions excluded from that sample. Within each column for which both measures are available, Panel~B is built from the same set of continuing relationships as Panel~A, aggregated to the product--source-country--month level.

Panel~A shows how pass-through changes as the sample broadens while price measurement remains fixed. Baseline pass-through is 87\%; the point estimate is below one, although complete pass-through cannot be rejected at conventional levels under the standard fixed effects. Adding consumption goods and then suppliers with a single U.S. buyer leaves the estimate largely unchanged, at 85\% and 86\%, respectively. Precision improves substantially once single-buyer suppliers are included, allowing us to reject complete pass-through in Column~(3). Restoring energy goods, related-party trade, and outliers raises estimated pass-through to 94\%, which is no longer statistically distinguishable from one.

\begin{table}[t]
\caption{Tariff Pass-Through: Aggregation vs. Sample Composition}
\label{tab:reconciliation}
\centering
{\small
\renewcommand{\arraystretch}{1.2}
\arrayrulecolor{gray!50}
\begin{tabular}{
  >{\raggedright\arraybackslash}p{3.3cm}
  >{\centering\arraybackslash}p{1.9cm}
  >{\centering\arraybackslash}p{2.7cm}
  >{\centering\arraybackslash}p{2.3cm}
  >{\centering\arraybackslash}p{1.9cm}
  >{\centering\arraybackslash}p{2.3cm}
}
\toprule
& (1) & (2) & (3) & (4) & (5) \\
Sample: & Baseline & + Consumption Goods & + Suppliers w/ $<$2 Buyers & + Energy/ RPT/ Outliers & + Non-Consecutive Relationships \\
\midrule
\multicolumn{6}{l}{\textit{\textbf{Panel A:} Relationship level (annual)}, $\Delta\ln p_{ijht}$} \\[0.3em]
$\Delta\ln(1+\tau_{cht})$ & -0.126  & -0.151  & -0.138  & -0.060  & ---     \\
                          & (0.091) & (0.089) & (0.034) & (0.044) &         \\[0.2em]
Observations              & 249,000 & 473,000 & 1,490,000 & 1,768,000 & --- \\
\midrule
\multicolumn{6}{l}{\textit{\textbf{Panel B:} Product level (monthly)}, $\Delta\ln p_{cht}$} \\[0.3em]
$\Delta\ln(1+\tau_{cht})$ & -0.654  & -0.353  & -0.063  & -0.054  & -0.017  \\
                          & (0.245) & (0.156) & (0.073) & (0.068) & (0.068) \\[0.2em]
Observations              & 124,000 & 213,000 & 665,000 & 805,000 & 1,071,000 \\
\bottomrule
\end{tabular}
}
\begin{minipage}{17cm}\vspace{0.5em}
\footnotesize \textit{Notes:} Each cell reports the coefficient on the statutory tariff change, $\Delta\ln(1+\tau_{cht})$, from a separate regression of the change in the duty-exclusive CIF price; columns add the indicated transactions to the baseline sample cumulatively. Panel~A is estimated at the buyer--supplier--product level with annual data; the specification includes product-year and exporter country-sector fixed effects ($\text{FE}_{ht} + \text{FE}_{cs}$) and the controls of Table~\ref{tab:pass_through_main}. Panel~B aggregates transactions from the same set of continuing relationships to the product--source-country level at monthly frequency and includes the same product-year and exporter country-sector fixed effects ($\text{FE}_{ht} + \text{FE}_{cs}$); Appendix~\ref{sec:additional_PT} details the construction. The dash in Column~(5) of Panel~A reflects that no within-relationship price change is defined for transactions not observed in consecutive years. RPT denotes related-party trade. Standard errors, in parentheses, are clustered at the HS8 product and exporter-country level. Observation counts are rounded to four significant digits per US Census Bureau disclosure guidelines. Appendix Tables~\ref{tab:price_changes_sample_CTCS} and~\ref{tab:price_changes_sample_product_level} report additional fixed-effects specifications and the model-implied counterpart of Panel~A. \textit{Source}:
FSRDC Project Number 2109 (CBDRB-FY25-P2109-R12520).
\end{minipage}
\end{table}

Panel~B shows how the estimates change when we adopt the price measurement used in the literature. If aggregation and monthly measurement accounted for the gap, applying them to the baseline sample should bring estimated pass-through closer to one. Instead, pass-through in Column~(1) falls to 35 percent, well below the relationship-level estimate and significantly below one. Pass-through approaches one as the sample broadens: it rises to 65 percent with the inclusion of consumption goods, to 94 percent with single-buyer suppliers, and ultimately to 98 percent when relationships not observed in consecutive years are included. From Column~(3) onward, complete pass-through cannot be rejected.\footnote{Appendix~\ref{sec:additional_PT} describes the construction of the product-level prices, and Appendix Table~\ref{tab:price_changes_sample_product_level} reports the underlying specifications.}

In Column~(4), where both panels use the broadest common sample, the two price measures yield nearly identical pass-through estimates of 94 and 95\%. These comparisons show that the gap between our incomplete baseline estimates and the near-complete pass-through documented in broader samples reflects sample composition rather than aggregation. The sources of the gap are difficult to isolate, however, because the sample expansions are cumulative and, as Panel~A shows, do not uniformly increase estimated pass-through. Our analysis focuses on repricing within continuing relationships in the baseline sample. A systematic examination of how trade composition shapes pass-through across samples lies beyond the scope of this paper but remains a promising avenue for future research.

\section{Structural Evidence\label{sec:Identification-and-estimation}}
The reduced-form patterns in Section~\ref{subsec:Test_prediction} are consistent with two-sided market power and decreasing returns, but they do not identify the structural parameters. We therefore adopt a structural approach. Estimation focuses on $\phi$ and $\theta$, which govern bargaining power and returns to scale; the remaining parameters are calibrated from the literature.

We set the elasticity of substitution across foreign varieties to $\rho = 10$, consistent with \citet{Anderson2004} and \citet{edmond2023costly}, who use similar values to match observed U.S. markups, and the downstream demand elasticity to $\nu = 4$, based on variety-level import demand estimates in \citet{Broda2006}.\footnote{The \citet{Broda2006} import demand elasticities have a mean of 3.85 and a median of 2.8 at the HS10 level. Because $\nu$ enters the model only through $\eta$, together with $\gamma$ and $\varrho$, the $\varrho=0.5$ specification in Table~\ref{tab:robust_model_estimates_d} also covers alternative values of $\nu$.}  With multiple foreign inputs (Section~\ref{subsec:Measuring-key-variables}), the output elasticity of input $h$ is $\alpha_{jh}\gamma$, where $\alpha_{jh}$ is the observed expenditure share of input $h$ in firm $j$'s imported intermediates and $\gamma = 0.5$ is the share of imported inputs in total material costs for U.S. manufacturers \citep{Eldridge2018}.

Finally, the importer’s returns-to-scale parameter $\varrho$ has no direct counterpart in the literature. We therefore adopt $\varrho = 1$ as a parsimonious benchmark and assess robustness to alternative values, including $\varrho = 0.5$. Because equilibrium outcomes depend on importer technology $(\nu,\gamma,\varrho)$ only through the composite elasticity $\eta_{jh}$, these exercises are equivalent to varying $\eta_{jh}$ directly.

\subsection{Identification and Estimation of the Parameters $\theta$ and $\phi$}
\label{subsec:Estimation-of-parameters-theta-phi}

Let $\Omega_{ijht}$ denote the observed state of relationship $(i,j,h)$ in year $t$, including the bilateral market shares and calibrated parameters that enter the markup formula. Equation~\eqref{eq:bilateral_markup_prop} then implies a bilateral price-cost ratio $\mu_{ijht}=\mu(\phi,\theta\mid\Omega_{ijht})$.

We write the bilateral log price as
\[
\ln p_{ijht}=\ln \mu\!\left(\phi,\theta \mid \Omega_{ijht}\right)
+\frac{1-\theta}{\theta}\ln q_{iht}+\ln k_{ijht}.
\]
In the baseline theory, marginal cost is common across buyers of an exporter-product-year. For estimation, $\ln k_{ijht}$ combines the common cost shifter with a match-specific pricing residual that could reflect unobserved quality, bundled services, or other relationship characteristics. We assume these departures enter log prices additively, and that the scale term $\frac{1-\theta}{\theta}\ln q_{iht}$ remains common across the exporter's buyers; the baseline markup function is unchanged.

Consider exporter $i$ serving two buyers, $j$ and $\ell$. Differencing their log prices removes the common exporter-product-year cost component but leaves the match-specific difference $\ln k_{ijht}-\ln k_{i\ell ht}$. Let $\mathbf Z_{ij\ell ht}$ denote the instrument vector for this pairwise comparison. We allow the match-specific pricing residual to be correlated with the bilateral shares themselves and instead impose the IV orthogonality condition
\begin{equation}
\mathbb{E}\!\left[\mathbf Z_{ij\ell ht}\left(\ln k_{ijht}-\ln k_{i\ell ht}\right)\right]=\mathbf 0,
\label{mean_indep}
\end{equation}
after partialling out the included controls and fixed effects. The resulting estimating moment is
\begin{equation}
\mathbb{E}\!\left[
\mathbf Z_{ij\ell ht}
\left\{
\ln p_{ijht}-\ln p_{i\ell ht}
-\left[
\ln\mu(\phi,\theta\mid\Omega_{ijht})
-\ln\mu(\phi,\theta\mid\Omega_{i\ell ht})
\right]
\right\}
\right]=\mathbf 0.
\label{eq:moment-cond}
\end{equation}

\paragraph{Identifying variation}
The parameters are estimated from cross-sectional variation in price levels across buyers served by the same exporter in the same product and year. This variation is distinct from the 2017--2018 tariff-induced price changes, which are excluded from estimation and used for out-of-sample evaluation in Section~\ref{subsec:ModelTest}.

When $\phi=0$, the buyer share drops out of the price–cost ratio, whatever the value of $\theta$. 
The negative buyer-share gradient in Table~\ref{Table 2 - Prices} therefore indicates $\phi>0$ and $\theta<1$ (Proposition~\ref{prop:2}). 
The variation in the supplier share identifies $\phi$ through the bargaining weight $\omega_{ij}$, and the variation in the buyer share identifies $\theta$ through the markdown, given $\phi$. 
How fast prices fall with the buyer share cannot by itself separately identify the two parameters, because they fall faster both when $\phi$ is higher and when $\theta$ is lower.

\paragraph{Threats to identification}
Endogenous match formation and omitted relationship characteristics can violate equation~\eqref{mean_indep} if the instruments are correlated with match-specific pricing residuals. Pairwise differencing removes shocks common to all buyers of an exporter-product-year, but it does not remove these match-specific components. Our IV strategy addresses this concern under the maintained exclusion restriction that, conditional on controls and fixed effects, the instruments are orthogonal to the differenced $k_{ijht}$ terms. Appendix~\ref{endogenousrelationships} discusses how selection into observed matches can threaten this restriction.

The estimation does not require the tariff changes used in the out-of-sample exercise to be exogenous, because those observations do not enter the structural estimation. It does require that aggregate or equilibrium forces affecting pre-2017 price levels other than through the two shares be either removed by pairwise differencing and any included fixed effects, or orthogonal to the instruments.

\paragraph{Estimation}
For each pairwise comparison, define the residual inside braces in equation~\eqref{eq:moment-cond} as $u_{ij\ell ht}(\phi,\theta)$ and the corresponding sample moment as
\[
\overline{\mathbf m}(\phi,\theta)=\frac{1}{N}\sum_{i,j,\ell,h,t}\mathbf Z_{ij\ell ht}\,u_{ij\ell ht}(\phi,\theta).
\]
We estimate the parameters by GMM:
\begin{equation}
(\hat\phi,\hat\theta)=\arg\min_{(\phi,\theta)}
\overline{\mathbf m}(\phi,\theta)^{\prime}\mathbf W\overline{\mathbf m}(\phi,\theta),
\label{eq:GMM_IV-1}
\end{equation}
where $\mathbf W$ is the optimal weighting matrix. The instrument vector contains the numbers of exporters and importers in each HS10 product-year and lagged leave-pair-out means and medians of the supplier- and buyer-share distributions. Leave-pair-out means that the focal buyer--supplier pair is excluded, so these means and medians take a different value for each of an exporter's buyers. The counts proxy for the pool of potential foreign suppliers and U.S. buyers in a variety. Specifications with fixed effects additionally partial out HS10-product, year, and buyer components, as reported in Table~\ref{tab:model_primitives}.

Because the pairwise construction reuses individual buyer relationships within exporter-product cells, the resulting moments are correlated within those cells. We therefore cluster standard errors at the exporter-product level and use the delta method for functions of the estimated parameters, including the implied bargaining weight $\phi$.

\paragraph{Extension: pair-specific bargaining weights}
The baseline imposes a common bargaining weight $\phi$. To allow bargaining power to vary across buyer-supplier pairs, we parameterize
\begin{equation}
\phi_{ijt}=\frac{\exp\left({\bf X}_{ijht}\,\boldsymbol{\delta}\right)}{1+\exp\left({\bf X}_{ijht}\,\boldsymbol{\delta}\right)}\in[0,1],
\label{phi_ijt}
\end{equation}
where $\boldsymbol{\delta}$ is estimated jointly with $\theta$. The vector ${\bf X}_{ijht}$ contains observable relationship characteristics excluded from the baseline gains-from-trade formulas: (i) relationship longevity, (ii) the annual number of transactions, (iii) the lagged ratio of exporter $i$'s sales to buyers other than $j$ to importer $j$'s purchases from suppliers other than $i$, and (iv) an indicator for whether the buyer and supplier transact in multiple HS10 products. This parsimonious parameterization permits observed heterogeneity in bargaining power without estimating a separate unrestricted weight for every pair or allowing bargaining weights to vary freely over time.

\subsection{Estimation Results\label{subsec:Estimation-results}}

We estimate \eqref{eq:GMM_IV-1} using data from 2001 to 2016. Because $\phi$ lies in the unit interval, we estimate the unbounded transformation $\bar{\phi} \equiv \ln \frac{\phi}{1 - \phi}$, which avoids convergence problems at the boundary.
\begin{table}[p]
\caption{Estimated Model Primitives}
\label{tab:model_primitives}
\centering
{\small
\renewcommand{\arraystretch}{1.2}
\arrayrulecolor{gray!50}
\begin{adjustbox}{max width=\textwidth,center}
\begin{tabular}{
  >{\raggedright\arraybackslash}p{6cm}
  >{\centering\arraybackslash}p{2.2cm}
  >{\centering\arraybackslash}p{2.2cm}
  >{\centering\arraybackslash}p{2.2cm}
  >{\centering\arraybackslash}p{2.2cm}
}
\toprule
\multicolumn{5}{c}{\textit{Panel A}: Calibrated Parameters} \\
\midrule
\multicolumn{1}{c}{$\hat{\nu}$} & \multicolumn{2}{c}{$\hat{\gamma}$} & \multicolumn{2}{c}{$\hat{\rho}$} \\
\multicolumn{1}{c}{4} & \multicolumn{2}{c}{0.5} & \multicolumn{2}{c}{10} \\
\midrule
\multicolumn{5}{c}{\textit{Panel B}: Estimated Parameters (GMM)} \\
\midrule
& (1) & (2) & (3) & (4) \\
\midrule
Rel. bargaining power: $\ln \widehat{\frac{\phi}{1-\phi}}$  & 1.563 &  & 1.168 &  \\
 & (0.222) &  & (0.179) &  \\
\addlinespace
Returns to scale ($\hat{\theta}$) & 0.455 & 0.505 & 0.417 & 0.508 \\
& (0.016) & (0.016) & (0.019) & (0.018) \\
Constant &   & 4.327 &   & 2.099 \\
&  & (1.153) &  & (0.761) \\
Longevity &  & -0.216 &  & 0.476 \\
&  & (0.200) &  & (0.272) \\
Number of HS10 transactions &  & -0.342 &  & -0.115 \\
&  & (0.110) &  & (0.135) \\
Multiple HS10 dummy &  & 0.004 &  & 0.113 \\
&  & (0.151) &  & (0.134) \\
Lagged outside option &  & -0.285 &  & -0.242 \\
&  & (0.100) &  & (0.106) \\
\midrule
None & \multicolumn{1}{c}{Yes} & \multicolumn{1}{c}{Yes} & \multicolumn{1}{c}{No} & \multicolumn{1}{c}{No} \\
$\text{FE}_{h} + \text{FE}_{t} + \text{FE}_{j}$ & \multicolumn{1}{c}{No} & \multicolumn{1}{c}{No} & \multicolumn{1}{c}{Yes} & \multicolumn{1}{c}{Yes} \\
\midrule
Observations & \multicolumn{4}{c}{3,120,000} \\
\midrule
\multicolumn{5}{c}{\textit{Panel C}: Implied Bargaining Powers ($\hat{\phi}$)} \\
\midrule
Mean & 0.827 & 0.931 & 0.763 & 0.904 \\
& (0.032) & (0.081) & (0.032) & (0.079) \\
Median & -- & 0.957 & --  & 0.927 \\
& -- & (0.081) & -- & (0.053) \\
\bottomrule
\end{tabular}
\end{adjustbox}
}
\begin{minipage}{16.5cm}\vspace{0.25em}
\footnotesize
\justifying
\textit{Notes}: This table presents model estimates using our main estimation sample of U.S. imports of intermediate inputs and capital goods between 2001 and 2016. Panel A reports calibrated parameters: demand elasticity ($\nu$), cost elasticity to foreign input prices ($\gamma$), and the elasticity of substitution across foreign varieties ($\rho$). The importer's returns to scale, $\varrho$, is set to 1. Panel B reports GMM estimates. Columns (1) and (3) impose a constant $\phi$, while Columns (2) and (4) allow heterogeneous bargaining power via $\boldsymbol{\delta}$. Specifications differ in fixed effects. Controls include relationship longevity, transaction intensity, the relative outside option (lagged), and a multi-product indicator. Panel C reports the mean and median implied bargaining power. Standard errors are clustered at the exporter-product level; Panel C uses the delta method. Instruments are the numbers of importers and exporters in each HS10 product-year, and lagged leave-pair-out means and medians of the supplier- and buyer-share distributions. All estimations use CIF duty-exclusive unit values. The number of observations is rounded to four significant digits in accordance with US Census Bureau disclosure guidelines. \textit{Source}: FSRDC Project Number 2109 (CBDRB-FY25-P2109-R12520).
\end{minipage}
\end{table}

Table \ref{tab:model_primitives} presents the estimation results. Panel B reports the GMM estimates. Columns (1) and (3) assume a constant $\phi$, while Columns (2) and (4) allow $\phi_{ij}$ to vary by trade pair as specified in equation (\ref{phi_ijt}). The specifications in Columns (1) and (2) are estimated without fixed effects; those in Columns (3) and (4) include year, product, and importer fixed effects. Panel C shows the implied values of $\phi_{ijt}$.  %

The parameters are precisely estimated. Across specifications, U.S. importers appear to wield substantial bargaining power, with estimated values of $\phi$ ranging from 0.7 to 0.9. Our preferred estimate, reported in Column~(1), is $\hat\phi = 0.83$, implying that U.S. importers have, on average, nearly five times the bargaining power of their foreign suppliers.

The returns to scale parameter \(\hat{\theta}\) is consistently estimated below one, ranging from 0.42 to 0.51 across specifications, with a preferred estimate of 0.46.\footnote{Our estimate of $\theta$ lies below standard production-function estimates, which are typically closer to one. Those estimates rely on firm-level variation in inputs and outputs and recover a longer-run parameter than the short-run object we identify here.} This implies a residual inverse export supply elasticity of about 0.24--0.35 for the average importer.\footnote{Computed as $\frac{1-\theta}{\theta}\,x_{ijh}$, the elasticity of the exporter's marginal cost with respect to its sales to importer $j$ (Section~\ref{sec:cost_elasticity}), at the average buyer share of 0.25 from Table~\ref{tab:summary_stats}.} These values are consistent with evidence from U.S. manufacturing under short-run constraints: \citet{boehm2022convex} report median inverse elasticities around 0.3 at typical capacity levels, and \citet{broda2008optimal} document similarly low elasticities across many traded goods.

Moving to the estimates of the vector $\hat{\boldsymbol{\delta}}$, the relationship-level controls are imprecisely estimated. Relationship longevity changes sign across the two specifications and is only significant at the 10 percent level; transaction frequency is negative and significant without fixed effects but not with them; and the multiple-product indicator is small and insignificant in both. The one stable and precisely estimated coefficient is the relative outside option: importers hold less bargaining power when their supplier has a stronger outside option, that is, when the supplier's past sales to other buyers are large relative to the importer's purchases from other suppliers.

\paragraph{Implied Markups} 
Using equation~\eqref{eq:bilateral_markup_prop}, we compute markups for all buyer-supplier-product matches given the estimated parameters and the observed market shares.

The resulting markup distribution is clustered near the competitive benchmark, with a median of 1.01. This contrasts with the oligopoly benchmark, whose median is 1.21 and whose interquartile range spans 1.11 to 2.45. In the model, implied markups are a convex combination of oligopoly and oligopsony markups, where the median oligopsony markup is 0.95. The large estimated bargaining power of buyers ($\hat\phi \approx 0.8$) shifts weight toward oligopsony, allowing importers to capture a substantial share of the surplus from exporters.

\subsection{Robustness and the Consumption-Goods Sample}
\label{subsec:robustness}

Table~\ref{tab:robust_model_estimates_d} reports structural estimates under alternative sample definitions and calibrations. For consumption goods, we also repeat the baseline reduced-form analysis, providing a direct comparison of pricing and pass-through across the two samples. This comparison allows us to assess how differences in bargaining power and supply curvature account for the patterns observed beyond the baseline sample.

\begin{table}[th]
\caption{Robustness of Model Estimates}
\label{tab:robust_model_estimates_d}
\centering
\renewcommand{\arraystretch}{1.3}
\arrayrulecolor{gray!50}

\begin{adjustbox}{max width=\textwidth}
\begin{tabular}{
  >{\raggedright\arraybackslash}p{4.5cm} 
  >{\centering\arraybackslash}p{2cm} 
  >{\centering\arraybackslash}p{2cm} 
  >{\centering\arraybackslash}p{2cm}
  >{\centering\arraybackslash}p{2cm}
  >{\centering\arraybackslash}p{2cm}
}
\toprule
& Baseline & Only Consumption & All Products & $\rho=5$ & $\varrho=0.5$  \\
\midrule
& (1) & (2) & (3) & (4) & (5) \\
\midrule

Bargaining power ($\phi$) 
    & 0.827 & 0.718 & 0.765 & 0.877 & 0.856 \\
    & (0.032) & (0.017) & (0.017) & (0.015) & (0.031) \\

Returns to scale ($\theta$)
    & 0.455 & 0.558 & 0.508 & 0.465 & 0.470  \\
    & (0.016) & (0.009) & (0.009) & (0.009) & (0.015)  \\

\midrule

Observations 
    & \multicolumn{1}{c}{3,120,000} 
    & \multicolumn{1}{c}{3,024,000} 
    & \multicolumn{1}{c}{6,143,000} 
    & \multicolumn{1}{c}{3,120,000} 
    & \multicolumn{1}{c}{3,120,000} 
  \\
\bottomrule
\end{tabular}
\end{adjustbox}
\begin{minipage}{\textwidth}\vspace{0.5em}
\footnotesize \textit{Notes}: This table reports robustness checks for the structural estimates. Columns vary either the sample definition or the calibrated parameters. “Only Consumption” restricts the sample to consumption goods. “All Products” includes all product categories. Columns $\rho=5$ and $\varrho=0.5$ modify the calibrated elasticities. All columns report GMM estimates without fixed effects. Standard errors are clustered at the exporter-product level. All estimations use CIF duty-exclusive unit values. Observation counts are rounded in accordance with US Census Bureau disclosure guidelines. \textit{Source}: FSRDC Project Number 2109 (CBDRB-FY25-P2109-R12520).
\end{minipage}
\end{table}

\paragraph{Consumption goods.}
Restricting the sample to consumption goods (Column~(2)) lowers $\hat\phi$ from 0.83 to 0.72 and raises $\hat\theta$ from 0.46 to 0.56. Both differences are statistically significant. The higher $\hat\theta$ implies less pronounced decreasing returns, consistent with our interpretation of supply curvature as reflecting short-run capacity constraints that are more severe for intermediate inputs. The lower $\hat\phi$ indicates less bargaining power for consumption-goods importers, who are largely retailers and wholesalers. The estimates for all products (Column~(3)) lie between those for the baseline and consumption-goods samples. Full estimates for the all-products and consumption-goods samples, including the fixed-effects and pair-specific-weight specifications, are reported in Appendix Tables~\ref{tab:model_primitives_robust_bec0} and~\ref{tab:model_primitives_robust_bec2}.

Appendix Tables~\ref{tab:prices_consumption} and~\ref{tab:PT_consumption} reproduce the price-level and pass-through regressions of Tables~\ref{Table 2 - Prices} and~\ref{tab:pass_through_main} for consumption goods. In the price-level regressions, the buyer-share coefficient moves from $-0.56$ to $-0.31$, while the supplier-share coefficient changes only marginally. In the pass-through regression, the buyer-share interaction moves from $-0.407$ to $0.063$ (s.e.~$0.112$). The attenuation is directionally consistent with a weaker cost channel, though the predicted gradient remains negative at $\hat\theta=0.56$ (Appendix Figure~\ref{fig:hetmap_consumption}). The supplier-share interaction becomes larger, but its theoretical interpretation is ambiguous (Section~\ref{sec:pass_through_bilateral_shares}). The average tariff coefficient is $-0.48$ (s.e.~$0.32$). Although the point estimate suggests substantially incomplete pass-through, the confidence interval includes both the baseline estimate and complete pass-through.

The pass-through comparisons are limited by the precision of the estimates: despite similar sample sizes, the standard error of the average tariff coefficient is roughly three and a half times the value in the baseline sample of intermediate and capital goods. This difference may partly reflect the timing of tariff exposure, as consumption goods were affected mainly by the September 2018 wave, leaving only a few months of post-tariff observations within the year \citep{fajgelbaum2020return,amiti2019impact}. Although the pass-through evidence is less conclusive, these results support applying the same framework to foreign sourcing by retailers and wholesalers, with bargaining power and supply curvature that vary across product categories.

\paragraph{Calibrated parameters.}
We next vary the two calibrated elasticities. Column~(4) lowers the elasticity of substitution across foreign varieties to $\rho=5$, at the lower end of the range in \citet{Anderson2004}. Column~(5) introduces decreasing returns downstream by setting $\varrho=0.5$, close to the supplier-side estimate. Both parameters affect the relationship between supplier shares and cross-sectional price dispersion. The structural estimates remain very close to the baseline: $\hat\phi$ rises from 0.83 to 0.88 and 0.86, respectively, while $\hat\theta$ increases from 0.46 to approximately 0.47 in both specifications.\footnote{Results are also robust to alternative related-party classifications; see Appendix~\ref{sec:model_estimation_robustness} and Table~\ref{tab:model_primitives_robust_RPTcensus}.}  Our results therefore do not hinge on these calibration choices.

\subsection{Out-of-Sample Evaluation}
\label{subsec:ModelTest}

Having estimated the model, we next evaluate how well it captures pass-through within buyer--supplier relationships. Using 2017 shares and parameters estimated from pre-2017 price levels, we ask whether the model predicts the level and heterogeneity of price responses to the 2017--2018 tariff increases. The model-implied change in duty-exclusive prices is
\begin{equation}
\widehat{\Delta \ln p^*_{ijht}}
=
\Phi^*_{ijht}\!\left(s_{ijh,t-1}, x_{ijh,t-1} \mid \boldsymbol{\hat{\Theta}}\right)
\cdot
\Delta \ln \left(1+\tau_{c(i)ht}\right),\label{eq:implied}
\end{equation}
where $\tau_{cht}$ is the statutory tariff rate and $\Phi^*_{ijht} = \Phi_{ijht} - 1$ is the model-implied pass-through elasticity into duty-exclusive prices. For this exercise, the elasticities are evaluated at lagged (2017) bilateral shares.

Table~\ref{tab:table5_new} compares empirical and model-implied pass-through elasticities under baseline fixed effects. Appendix Table~\ref{tab:table5_new_stringentFE} uses the more stringent fixed effects, yielding similar results. To isolate the roles of importer bargaining power and decreasing returns, we compare the baseline model with three restricted variants that remove these mechanisms separately and jointly.\footnote{We evaluate equation~\eqref{eq:implied} using the baseline estimates from Column~(1) of Table~\ref{tab:model_primitives}, imposing the relevant restrictions on $\phi$ and $\theta$ without re-estimating the remaining parameters.} Columns~(1) and~(2) report tariff pass-through in the data and in the baseline model, respectively. Column~(3) imposes $\phi=0$ and $\theta=1$, yielding a standard Nash--Bertrand model with constant marginal cost, as in \citet{kikkawa2019imperfect}. Column~(4) retains bargaining but imposes constant returns ($\theta=1$), as in \citet{gopinath2011search}. Column~(5) retains decreasing returns but removes importer bargaining power ($\phi=0$).

\begin{table}[htbp]
\caption{{Price Responses and Relationship Heterogeneity: Data vs. Model}}
\label{tab:table5_new}
\centering
{\small
\renewcommand{\arraystretch}{1.2}
\arrayrulecolor{gray!50}
\begin{tabular}{
  >{\raggedright\arraybackslash}p{4.6cm}
  >{\centering\arraybackslash}p{1.8cm}
  >{\centering\arraybackslash}p{1.8cm}
  >{\centering\arraybackslash}p{1.8cm}
  >{\centering\arraybackslash}p{1.8cm}
  >{\centering\arraybackslash}p{1.8cm}
}
\toprule
\multicolumn{6}{c}{{Panel A: Main Effect}} \\
\midrule
& Data & Baseline & $\phi=0$, $\theta=1$ & $\theta=1$ & $\phi=0$ \\
& (1) & (2) & (3) & (4) & (5) \\
\midrule
$\Delta \ln(1+\tau_{cht})$ 
    & -0.126  &	-0.222  &	-0.094	& -0.013	& -0.279 \\
    & (0.091) &	(0.011) & (0.004) &	(0.003)	& (0.009) \\
\midrule
R-squared    & 0.04	& 0.31 &	0.22 &	0.21	& 0.34 \\
\midrule

\addlinespace[1.5em]
\multicolumn{6}{c}{{Panel B: Including Interactions}} \\
\midrule
& Data & Baseline & $\phi=0$, $\theta=1$ & $\theta=1$ & $\phi=0$ \\
& (1) & (2) & (3) & (4) & (5) \\
\midrule
$\Delta \ln(1+\tau_{cht})$ 
    & -0.032 & -0.145 & -0.049 & -0.005 & -0.219 \\
    & (0.091) & (0.011) & (0.009) & (0.001) & (0.010) \\

$\Delta \ln(1+\tau_{cht}) \cdot s_{ijht-1}$ 
    & 0.033 & 0.125 & -0.079 & -0.011 & 0.122 \\
    & (0.080) & (0.013) & (0.014) & (0.004) & (0.011) \\

$\Delta \ln(1+\tau_{cht}) \cdot x_{ijht-1}$ 
    & -0.407 & -0.421 & -0.046 & -0.013 & -0.366 \\
    & (0.112) & (0.023) & (0.023) & (0.004) & (0.022) \\
\midrule
R-squared 
    & 0.04 & 0.44 & 0.23 & 0.23 & 0.41 \\
  \midrule \addlinespace[1em]

$FE_{ht} + FE_{cs}$ & Yes & Yes & Yes & Yes & Yes \\
  \midrule
    Observations & \multicolumn{5}{c}{249,000} \\
\bottomrule
\end{tabular}
}
\begin{minipage}{\textwidth}\vspace{0.5em}
\footnotesize \textit{Notes}: This table reports the response of duty-exclusive CIF prices to tariff changes at the exporter-importer-product level. Column (1) presents reduced-form estimates using observed price changes in the data. Columns (2)--(5) report the corresponding responses implied by model-predicted price changes under each specification. Panel A reports the main effect of tariff changes, while Panel B allows the response to vary with lagged supplier shares ($s_{ijht-1}$) and buyer shares ($x_{ijht-1}$). All specifications include product--time and country-sector fixed effects ($FE_{ht} + FE_{cs}$). Standard errors in parentheses combine the sampling variance of the coefficients, clustered by HS8 product and exporter country, with the variance contributed by the estimation of $(\hat{\phi},\hat{\theta})$, computed by a parametric bootstrap with 1,000 replications. In column (1), standard errors are clustered at the HS8 product and exporter-country level only. Observation counts are rounded to four significant digits per US Census Bureau disclosure guidelines. \textit{Source}: FSRDC Project Number 2109 (CBDRB-FY25-P2109-R12520).
\end{minipage}
\end{table}

Panel~A of Table~\ref{tab:table5_new} reports average pass-through of about 87\% under the baseline fixed effects. The baseline model predicts 78\%, within the empirical confidence interval. The restricted variants predict pass-through between 72 and 99\%; because the empirical average is imprecisely estimated, none can be rejected based on the average alone at the 5\% level. Figure~\ref{fig:hetmap} explains why these variants nonetheless differ: incomplete pass-through can arise from strategic complementarities in markups or from the cost channel, and the variants combine these mechanisms differently. Average pass-through therefore cannot distinguish among them.

Panel~B shows how tariff pass-through varies with bilateral shares. In the data, the buyer-share gradient is negative and statistically significant, at $-0.41$, while the supplier-share gradient is near zero and imprecisely estimated. The baseline model closely matches the buyer-share gradient, predicting $-0.42$. The constant-returns variants in Columns~(3) and~(4) substantially attenuate this gradient and are rejected. The decreasing-returns variant without importer bargaining power in Column~(5) instead cannot be rejected.

Price-level moments distinguish this variant from the baseline. Appendix Table~\ref{Table_simulated_markups} regresses model-implied markups on the two shares and shows that the buyer-share coefficient becomes economically negligible when $\phi=0$, although it need not equal zero in a linear projection of the nonlinear markup function. Among the variants considered, only the baseline captures the negative buyer-share patterns in both price levels and pass-through. The supplier-share coefficient is positive under both specifications, as in the data. The magnitude is hard to interpret, however, because the price-cost ratio is strongly non-monotonic in the supplier share. The informative comparison is the buyer share, in which the price--cost ratio is monotonic in the buyer share.

\subsection{Additional Exercises}
\label{subsec:additional_exercises}

We complement the above evidence with two additional exercises. We first run a formal test of the model's fit to average pass-through, and then we extend the analysis to quantities and expenditure.

\paragraph{A formal test.}
The model-implied and reduced-form estimates in Table~\ref{tab:table5_new} both measure the bilateral price response to tariff changes; yet the table compares them informally by checking whether the model's prediction falls within the empirical confidence interval. The IV goodness-of-fit test of \citet{adao2023putting} provides a complementary statistic using observed duty-inclusive prices. We instrument model-implied price changes with statutory tariff changes and test whether the model reproduces the average tariff-induced price response in the data. The test is a natural complement to Table~\ref{tab:table5_new}, with the caveat that average pass-through is the moment least able to discriminate among the variants, as discussed above. Figure~\ref{fig:IVtest_fit} in Appendix~\ref{sec:model_fit_robustness} rejects neither the baseline model nor the restricted variants, consistent with that caveat. Table~\ref{tab:Modelfit_IV_altparam_panel} shows that these conclusions hold across alternative calibrations and fixed-effects specifications.

\paragraph{Quantities and expenditure.}

We extend the out-of-sample evaluation to bilateral quantities and expenditures. Appendix~\ref{sec:model_fit_robustness} describes the exercise. Table~\ref{tab:quantity_response_combined} examines quantity pass-through and its heterogeneity, while Table~\ref{tab:PT_IV_q_sales_combined} reports formal IV tests for quantity and expenditure responses under the baseline and alternative model variants. Because the model is estimated on price levels alone, these are untargeted predictions.

Table~\ref{tab:quantity_response_combined} shows that the model gets the average direction right: tariff increases reduce traded quantities in both the model and the data, with larger declines under more stringent fixed effects. It overstates the magnitude, however, predicting quantity declines two to three times larger than those in the data. For heterogeneity, the model predicts smaller quantity declines for high-share buyers. The estimated interactions are too imprecise to establish their sign, but they are substantially below the model’s predictions.

The IV comparisons in Table~\ref{tab:PT_IV_q_sales_combined} indicate a weaker fit for quantities and expenditures across the model variants, with the predictions rejected in most specifications. We interpret this result cautiously. Physical quantities in customs data are noisier than prices, and the mapping from predicted prices to quantities relies on the calibrated elasticity of substitution, $\rho=10$, so price-prediction errors can be magnified in the quantity predictions. The model also abstracts from quantity adjustments induced by price changes elsewhere in the trading network.

Taken together, these results indicate that matching prices, quantities, and expenditures would require a richer framework that allows trade to reallocate across the network. They do not, by themselves, overturn the model's account of price adjustment within existing relationships.

\section{Aggregate Pass-Through in Firm-to-Firm Trade}\label{sec:Implications}

The estimated model matches tariff pass-through within buyer--supplier relationships out of sample. This section uses it to quantify the incidence of the 2018 tariffs within continuing relationships and to decompose it into its markup and cost components. 

We compute model-implied bilateral price changes and aggregate them into a single statistic by regressing predicted price changes on the tariff shock, with observations weighted by initial import shares. The coefficient is a measure of aggregate pass-through within our sample of continuing relationships. We then decompose predicted price changes into their cost and markup components using the model's expressions for the two elasticities: for each channel, we recompute counterfactual price changes holding the other fixed and re-estimate the weighted regression.

\begin{table}[t]
\caption{Aggregate Tariff Pass-Through and Decomposition}
\label{tab:tariff_pass_through}
\centering
{\small
\renewcommand{\arraystretch}{1.2}
\arrayrulecolor{gray!50}
\begin{tabular}{
  >{\raggedright\arraybackslash}p{7.2cm}
  >{\centering\arraybackslash}p{2.5cm}
  >{\centering\arraybackslash}p{2.5cm}
}
\toprule
& Baseline FE & Stringent FE \\
\cmidrule(lr){2-3}
& (1) & (2) \\
\midrule
\addlinespace[0.8em]
 \multicolumn{3}{c}{\textit{Panel A}: Aggregate Pass-through (\% )} \\
 \midrule
Agg. pass-through: $1/(1+ \Lambda_{ij}+\Gamma_{ij})$   & 72.7 & 66.5 \\
\hspace{1em} {Cost channel only}: $1/(1+\Lambda_{ij})$                 &  71.5 & 64.8  \\
\addlinespace[0.8em]
\midrule
\addlinespace[0.8em]
 \multicolumn{3}{c}{\textit{Panel B}: Variance Decomposition of $\Lambda_{ij}+\Gamma_{ij}$} \\
 \midrule
 Cost Elasticity: $\Lambda_{ij}$ & 1.023 & 0.997 \\
 Markup Elasticity: $\Gamma_{ij}$ & -0.023 & 0.003 \\
 
\bottomrule
\end{tabular}
}
\begin{minipage}{1\textwidth}\vspace{0.5em}
\scriptsize
\textit{Notes}: This table reports model-implied aggregate pass-through estimates following the 2018 U.S. tariff increases. Column (1) includes product-time and country-sector fixed effects. Column (2) includes product-time, country-time, and buyer-time fixed effects. The overall pass-through elasticity is computed as \(1 +\) the estimated coefficient on \(\Delta \ln(1 + \tau)\), and decomposed into contributions from the cost channel ($\Lambda_{ij}$) and the markup channel ($\Gamma_{ij}$). The counterfactual ``Cost channel only'' row shows the predicted pass-through when markup elasticities are set to zero. For Panel~B, we first remove each column's fixed effects from the cost and markup elasticities using the same projection. Denote the residuals by $\widetilde\Lambda_{ij}$ and $\widetilde\Gamma_{ij}$ and their sum by $\widetilde D_{ij}$. The reported shares are $\operatorname{Cov}(\widetilde\Lambda_{ij},\widetilde D_{ij})/\operatorname{Var}(\widetilde D_{ij})$ and $\operatorname{Cov}(\widetilde\Gamma_{ij},\widetilde D_{ij})/\operatorname{Var}(\widetilde D_{ij})$. They sum to one and decompose the variation in total elasticity remaining after removing fixed effects; different projections therefore yield different shares across columns. \textit{Source}: FSRDC Project Number 2109 (CBDRB-FY25-P2109-R12520).
\end{minipage}
\end{table}

Panel~A of Table~\ref{tab:tariff_pass_through} reports aggregate pass-through under the full model and under a counterfactual that shuts down markup adjustments. The full model predicts pass-through of 73\% under the baseline specification (Column~(1)) and 67\% under the more stringent fixed effects (Column~(2)). The corresponding unweighted regression estimates are 77.8\% under the baseline fixed effects (Table~\ref{tab:table5_new}) and 77.4\% under the stringent fixed effects (Table~\ref{tab:table5_new_stringentFE}). The lower weighted estimates indicate that pass-through is lower for matches accounting for a larger share of imports.

Setting $\Gamma_{ij}=0$ yields pass-through of 72\% under the baseline fixed effects and 65\% under the stringent fixed effects—within two percentage points of the full-model predictions. Markup adjustments therefore change aggregate pass-through only modestly. They raise pass-through relative to the cost-only counterfactual, as expected when importers hold substantial bargaining power, but their net contribution is small.

Panel~B decomposes the variance of total elasticity after removing the fixed effects in each column. In both columns, the cost elasticity accounts for more than 99\% of the residual variance of total elasticity. Thus, after netting out the fixed effects, nearly all remaining variation in total elasticity is associated with the cost channel.

This exercise addresses the question left open in Section~\ref{sec: Theory}: whether export supply is sufficiently steep at the estimated parameters for the cost channel to dominate. Proposition~\ref{prop:phi_theta} and Corollary~\ref{cor:cov} establish cost-channel dominance in the limit as $\theta\to0$. Our estimates show that this dominance already holds at the finite estimated value $\hat{\theta}$. Together, Panels~A and~B show that the cost channel accounts for nearly all of both the aggregate response and the residual cross-sectional variation.

\section{Conclusions}\label{sec:conclusions}

The incidence of tariffs is usually studied at the level of products and countries. Yet part of the burden is settled within individual trading relationships, through the price an importer negotiates with its foreign supplier, and these adjustments matter in the aggregate. This paper studies pricing and pass-through at the relationship level under two-sided market power. It offers a novel view of importers' oligopsony power as arising from two features: short-run decreasing returns in the exporter's technology, which generate quasi-rents that grow with the buyer share, and bargaining power, which lets the importer capture part of them. In U.S. import data, we estimate that importers hold most of the bargaining power and that short-run export supply is steep. Oligopsony power is thus a feature of U.S. imports, and an important determinant of price differences across the buyers of a given exporter.

A theory of prices is also a theory of pass-through. Because oligopsony power in our framework is rooted in decreasing returns, adjustments in the exporter's marginal cost enter pass-through alongside changes in the negotiated markup. This channel is absent from standard models of exporter market power, which assume constant returns and therefore attribute pass-through primarily to markup adjustment. In our framework, the same supply curvature that generates oligopsony power implies that marginal costs decline with tariffs, and more so when buyers account for a large share of shipments. As export supply becomes sufficiently steep, the cost channel dominates pass-through regardless of how bargaining power is allocated. By our estimates, this is the case for the 2018 tariffs: foreign exporters absorbed about 30 percent of the tariff within continuing relationships, almost entirely through lower marginal costs.

The broader lesson concerns how market power and shock transmission interact in production networks. Market power can be large in levels and still contribute little to how prices respond at the margin. Inferring the division of rents from pass-through, or pass-through from the division of rents, can therefore be misleading. The same logic should apply to other cost shocks that travel through buyer--supplier relationships, such as exchange rate movements and input price changes, and to other settings, such as domestic production networks. 

Two features of the analysis define the scope of our conclusions. The first is the sample: our estimates describe repricing within selected continuing relationships, and broadening the sample substantially raises measured pass-through. The second is the horizon: ours is a short-run analysis of prices within an existing network, whereas tariffs also restructure the network itself \citep[e.g.,][]{GrossmanHelpmanRedding2024}, as relationships end, new ones form, and exporters adjust capacity over longer horizons. The gap between our estimates and the near-complete pass-through found in product-level data, together with the weaker fit of our model to observed quantity responses, which are where reallocation across suppliers first shows up, suggests that these extensive-margin adjustments could matter considerably for aggregate incidence. Combining two-sided market power with a model of how trading relationships form and end is the natural next step.
\clearpage
\bibliographystyle{ecta}
\bibliography{library.bib}
\clearpage{}

\appendix

\pagenumbering{arabic} %
\setcounter{page}{1} %

\numberwithin{equation}{section}

\numberwithin{figure}{section}

\numberwithin{table}{section}
\begin{center}
\textsc{\large{}Online Appendix\medskip{}
}{\large\par}
\par\end{center}

\begin{center}
\textbf{\Large{}``Two-Sided Market Power in Firm-to-Firm Trade''}\textsc{\large{}\medskip{}
}{\large\par}
\par\end{center}

\begin{center}
{\large{}Vanessa Alviarez, Michele Fioretti, Ken Kikkawa, Monica Morlacco}\textsc{\vspace{2cm}
}
\par\end{center}

This Appendix is organized as follows. 
Section \ref{subsec: Theory Appendix} contains the derivations and proofs for the
theoretical results.
Section \ref{sec: Theory Extensions} contains the discussion of theory extensions.
Section \ref{sec:Data Appendix} contains details about the data and sample selection criteria. 
Section \ref{sec: Additional Tables and Figures} provides additional empirical results referenced in the text. 
Section \ref{sec:Estimation Appendix} provides additional details regarding the estimation.

\section{Mathematical Derivations\label{subsec: Theory Appendix}}

\subsection{Demand Elasticity\label{subsec: Demand Elasticity}}

The residual demand elasticity enters both the oligopoly markup and the bilateral bargaining solution. We derive it here from the importer’s cost-minimization problem.

\paragraph{Input demand.} The importer's nested CES production structure (equations~\eqref{nested_ces_1}–\eqref{nested_ces_2}) yields the demand for input $q_{ij}$ in \eqref{eq:demand function q_ij}:
\begin{equation*}
q_{ij} = q_{j}^{f} \varsigma_{ij}^{\rho} \left( \frac{p_{ij}}{p_{j}^{f}} \right)^{-\rho}, \label{eq:q_ij_eq}
\end{equation*}
where $p_{j}^{f}= \left( \sum_{i} \varsigma_{ij}^{\rho} p_{ij}^{1-\rho} \right)^{\frac{1}{1-\rho}}$ is the price index (i.e., shadow cost) of imported inputs. Solving the outer CES aggregator for total input demand yields:
\begin{align}
q_{j}^{f} &= \gamma MC_j q_j \left(p_{j}^{f}\right)^{-1}, \label{eq:q^f} \\
q_{j}^{d} &= (\varrho - \gamma) MC_j q_j \left(p_{j}^{d}\right)^{-1}, \label{eq:q^d}
\end{align}
where $MC_j$ is the marginal cost of output $q_j$, given by:
\begin{equation}
MC_j = \left[ \varphi_j^{-1} \left( \frac{p_j^f}{\gamma} \right)^{\gamma} \left( \frac{p_j^d}{\varrho - \gamma} \right)^{\varrho - \gamma} \right]^{\frac{1}{\varrho}} q_j^{\frac{1 - \varrho}{\varrho}} \equiv k_j q_j^{\frac{1 - \varrho}{\varrho}}. \label{eq:unit cost}
\end{equation}

This form illustrates symmetry with exporter-side technology. Finally, note that the cost share of foreign inputs is constant and equal to
\[
\frac{p_{j}^{f} q_{j}^{f}}{p_{j}^{f} q_{j}^{f} + p_{j}^{d} q_{j}^{d}} = \frac{\gamma}{\varrho}.
\]

\paragraph{Residual demand elasticity.} We first compute the response of the importer’s marginal cost to the imported-input price. Using downstream demand, write $q_j$ as
\begin{equation}
q_{j}=\left(\frac{\nu}{\nu-1}\right)^{-\nu}MC_{j}^{-\nu}D_{j}^\nu,\label{eq:demand}
\end{equation}
where $D_j$ is revenue shifter in equation~\eqref{eq:revenue_j}, so inverse demand is $p_j=D_jq_j^{-1/\nu}$. 
Substituting equation (\ref{eq:demand}) into equation (\ref{eq:unit cost}) and rearranging, we can write:
\begin{align*}
MC_{j} & =\left[\varphi_{j}^{-1}\left(\frac{p_{j}^{f}}{\gamma}\right)^{\gamma}\left(\frac{p_{j}^{d}}{\varrho-\gamma}\right)^{(\varrho-\gamma)}\right]^{\frac{1}{\varrho+\nu-\nu\varrho}}\left(\frac{\nu}{\nu-1}\right)^{-\nu\frac{1-\varrho}{\varrho+\nu-\nu\varrho}}\left(D_{j}\right)^{\frac{\nu(1-\varrho)}{\varrho+\nu-\nu\varrho}},
\end{align*}
which implies 
\[
\frac{d\ln MC_{j}}{d\ln p_{j}^{f}}=\frac{\gamma}{\varrho+\nu-\nu\varrho}.
\]
Combining this result and equation~\eqref{eq:q^f} with the log input demand
\[
\ln q_{ij}=\ln q_{j}^{f}+\rho\ln\varsigma_{ij}-\rho\left(\ln p_{ij}-\ln p_{j}^{f}\right),
\]
gives
\begin{align}
\varepsilon_{ij}=-\frac{d\ln q_{ij}}{d\ln p_{ij}} & =-\left(\frac{d\ln q_{j}^{f}}{d\ln p_{j}^{f}}+\rho\right)\frac{d\ln p_{j}^{f}}{d\ln p_{ij}}+\rho \notag\\
 & =\left((\nu-1)\frac{d\ln c_{j}}{d\ln p_{j}^{f}}+1-\rho\right)s_{ij}+\rho\nonumber \\
 & =\eta s_{ij}+(1-s_{ij})\rho,\label{eq:varepsilon_}
\end{align}
where we defined 
\begin{equation}
\eta \equiv-\frac{d\ln q_{j}^{f}}{d\ln p_{j}^{f}} =\frac{(\varrho-\gamma)+\nu(1-(\varrho-\gamma))}{\varrho+\nu(1-\varrho)}.
\end{equation}

\subsection{Proof of Proposition \ref{prop:markup}\label{subsec: Derivations Markup}}

Using the demand elasticity derived in Section~\ref{subsec: Demand Elasticity}, we now solve the bilateral bargaining problem.

\paragraph{Bargaining first-order condition.} The $i-j$ pair chooses the bilateral price $p_{ij}$ to solve
\begin{equation}
\max_{p}\Big(\underbrace{\pi_{i}\left(p\right)-\tilde{\pi}_{i\left(-j\right)}}_{GFT_{ij}^{i}\left(p\right)}\Big)^{1-\phi}\Big(\underbrace{\pi_{j}\left(p\right)-\tilde{\pi}_{j\left(-i\right)}}_{GFT_{ij}^{j}\left(p\right)}\Big)^{\phi},\label{eq:nash_problem-1}
\end{equation}
where $\phi\in(0,1)$ is $j$'s bargaining power, and the terms inside parentheses are the gains from trade for exporter $i$ ($GFT_{ij}^{i}\left(p\right)$) and importer $j$ ($GFT_{ij}^{j}\left(p\right)$), written as a function of $p_{ij}$. 

The FOC associated with the problem (\ref{eq:nash_problem-1}) can be written as:

\begin{align}
0 & =\frac{d\ln\pi_{i}}{d\ln p_{ij}}+\frac{\phi}{1-\phi}\cdot\frac{GFT_{ij}^{i}}{\pi_{i}}\frac{\pi_{j}}{GFT_{ij}^{j}}\cdot\frac{d\ln\pi_{j}}{d\ln p_{ij}},\label{eq:FOC}
\end{align}
where we used the fact that $\frac{dGFT_{ij}^{k}}{dp}=\frac{d\pi_{k}}{dp}$
for $k=\{i,j\}.$ In what follows, we derive expressions for $\frac{d\ln\pi_{i}}{d\ln p_{ij}},\frac{d\ln\pi_{j}}{d\ln p_{ij}},GFT_{ij}^{i}$
and $GFT_{ij}^{j}.$

\paragraph{Exporter profits and gains from trade.} Firm $i$'s profits under a successful negotiation are
\[
\pi_{i}=p_{ij}q_{ij}+\sum_{k\neq j}p_{ik}q_{ik}-\theta MC_iq_{i},
\]
where $MC_i=k_{i}q_{i}^{\frac{1-\theta}{\theta}}.$ The elasticity
of the profit $\pi_{i}$ with respect to $p_{ij}$ is:
\begin{align}
\frac{d\ln\pi_{i}}{d\ln p_{ij}} & =\frac{q_{ij}}{\pi_{i}}(\varepsilon_{ij}-1)\left(\mu_{ij}^{\text{oligopoly}}MC_i-p_{ij}\right),\label{eq:dpi_i / dp_ij}
\end{align}
where $\varepsilon_{ij}\equiv-\frac{d\ln q_{ij}}{d\ln p_{ij}}$ is
defined in equation (\ref{eq:varepsilon_}) and
\[
\mu^{\text{oligopoly}}:=\frac{\varepsilon_{ij}}{\varepsilon_{ij}-1}.
\]

The ``outside'' profits of firm $i$ are
\[
\tilde{\pi}_{i\left(-j\right)}=\sum_{k\neq j}p_{ik}q_{ik}-\theta\widetilde{MC_iq_{i}}.
\]
For a constant $k_{i}$ and given $q_{i}=\sum_{j\in\mathcal{Z}_{i}}q_{ij}$, the total cost in case of failed agreement is: 
\begin{align*}
\theta\widetilde{MC_iq_{i}} & =\theta MC_iq_{i}\left(\frac{\widetilde{q_{i}}}{q_{i}}\right)^{\frac{1}{\theta}}.\\
 & =\theta MC_iq_{i}\left(1-x_{ij}\right)^{\frac{1}{\theta}}.
\end{align*}

The exporter's gains from trade $GFT_{ij}^{i}$ are thus given by:
\begin{equation}
\begin{aligned}
GFT_{ij}^{i}\left(p_{ij}\right) & \equiv\pi_{i}\left(p\right)-\tilde{\pi}_{i\left(-j\right)}\label{eq:GFT_i}\\
 & =q_{ij}\left(p_{ij}-MC_i\mu_{ij}^{\text{oligopsony}}\right),
\end{aligned}
\end{equation}
where \(\mu_{ij}^{\text{oligopsony}}:=\theta\left[\frac{\Delta_{ij}^{x}}{x_{ij}}\right]\) and  \(\Delta_{ij}^{x}:=\left[1-\left(1-x_{ij}\right)^{\frac{1}{\theta}}\right].\)

\paragraph{Importer profits and gains from trade.} Firm $j$'s profits under a successful negotiation are
as 
\[ \pi_{j} =p_{j}q_{j}-\varrho MC_{j}q_{j} = \frac{\gamma}{\eta-1}MC_{j}q_{j}, 
\]
where $MC_{j}$ is defined in equation (\ref{eq:unit cost}), and $q_{j}$ in equation (\ref{eq:demand}). The elasticity of the profit $\pi_{j}$ with respect to $p_{ij}$ is:

\begin{equation}
-\frac{d\ln\pi_{j}}{d\ln p_{ij}}=(\eta-1)s_{ij}.\label{eq:eladlnp_ij}
\end{equation}
The outside profits of firm $j$ under a failed negotiation are
\[
\tilde{\pi}_{j\left(-i\right)}=\frac{\gamma}{\eta-1}\widetilde{MC_{j}q_{j}},
\]
where $\varrho\widetilde{MC_{j}q_{j}}$ denotes the total ``outside''
cost of importer $j$. The assumptions on technology and demand downstream
imply: 
\begin{align*}
\widetilde{MC_{j}q_{j}} & =\left(1-s_{ij}\right)^{\frac{\eta-1}{\rho-1}}MC_{j}q_{j}.
\end{align*}

Putting things together, the importer's gains from trade can be written
as: 
\begin{align}
GFT_{j}(p_{ij}) & =\pi_{j}\Delta_{ij}^{s}\label{eq:GFT_j}
\end{align}
where we defined \(\Delta_{ij}^{s}:=\left(1-\left(1-s_{ij}\right)^{\frac{\eta-1}{\rho-1}}\right).\)

\paragraph{Bilateral price.} Substituting equations~\eqref{eq:dpi_i / dp_ij}, \eqref{eq:GFT_i}, \eqref{eq:eladlnp_ij}, and \eqref{eq:GFT_j} into \eqref{eq:FOC} yields
\begin{equation}
\mu_{ij}:=\frac{p_{ij}}{MC_i}=(1-\omega_{ij})\mu_{ij}^{\text{oligopoly}}+\omega_{ij}\mu_{ij}^{\text{oligopsony}},\label{eq:bilateral markup APP}
\end{equation}
where
\begin{equation}
\omega_{ij}:=\frac{\frac{\phi}{1-\phi}\lambda_{ij}}{1+\frac{\phi}{1-\phi}\lambda_{ij}}\in(0,1)\label{eq:omega_ij}
\end{equation}
 and where 
\begin{align}
\lambda_{ij} & :=\frac{\pi_{j}}{GFT_{ij}^{j}}\cdot\left(-\frac{d\ln\pi_{j}}{d\ln p_{ij}}\cdot(\varepsilon_{ij}-1)^{-1}\right)\label{eq:lambda_ij}\\
 & =\underbrace{\frac{\pi_{j}}{GFT_{ij}^{j}}}_{\lambda_{ij}^{N}}\cdot\underbrace{-\frac{d\ln\pi_{j}}{d\ln p_{ij}q_{ij}}}_{\lambda_{ij}^{I}} \notag\\
 & =\left(\Delta_{ij}^{s}\right)^{-1}\cdot\frac{\eta-1}{\varepsilon_{ij}-1}s_{ij}>0. \tag*{\qedsymbol}
\end{align}

\clearpage

\subsection{Proof of Proposition \ref{prop:2}}
\label{appendix:proof-prop2}

Proposition~\ref{prop:2} characterizes how $\mu_{ij}$ varies with
bilateral market shares. Part~I states that a positive relationship between $\mu_{ij}$ and $s_{ij}$ requires $\phi<1$. Part~II states that a negative
relationship between $\mu_{ij}$ and $x_{ij}$ can arise only if $\phi>0$ and $\theta<1$.

\paragraph{Part I: Supplier share.}

Differentiating \eqref{eq:bilateral markup APP} gives
\begin{equation}
\frac{\partial \mu_{ij}}{\partial s_{ij}}
=(1-\omega_{ij})\frac{\partial \mu^{\text{oligopoly}}_{ij}}{\partial s_{ij}}
+\frac{\partial \omega_{ij}}{\partial s_{ij}}
\bigl(\mu^{\text{oligopsony}}_{ij}-\mu^{\text{oligopoly}}_{ij}\bigr).
\label{eq:dmu_ds}
\end{equation}
If $\phi=1$, then $\omega_{ij}=1$ and \eqref{eq:bilateral markup APP} reduces to
\[
\mu_{ij}=\mu^{\text{oligopsony}}_{ij}(x_{ij}),
\]
which is independent of $s_{ij}$. Hence $\partial\mu_{ij}/\partial s_{ij}=0$ when
$\phi=1$, so $\partial\mu_{ij}/\partial s_{ij}>0$ requires $\phi<1$. 

\paragraph{Part II: Buyer share.}

The buyer share enters \eqref{eq:bilateral markup APP} only through
$\mu^{\text{oligopsony}}_{ij}(x_{ij})$. Therefore,
\begin{equation}
\frac{\partial \mu_{ij}}{\partial x_{ij}}
=\omega_{ij}\frac{\partial \mu^{\text{oligopsony}}_{ij}}{\partial x_{ij}},
\qquad
\frac{\partial \mu^{\text{oligopsony}}_{ij}}{\partial x_{ij}}
=
\frac{x_{ij}(1-x_{ij})^{\frac{1}{\theta}-1}
-\theta\bigl[1-(1-x_{ij})^{\frac{1}{\theta}}\bigr]}
{x_{ij}^{2}}.
\label{eq:dmu_dx}
\end{equation}
If $\phi=0$, then $\omega_{ij}=0$; if $\phi>0$, then $\omega_{ij}>0$. It remains only
to sign the numerator in \eqref{eq:dmu_dx}.

\begin{lemma}\label{lem:sign}
For any $x\in(0,1)$ and $\theta>0$,
\[
\operatorname{sign}\!\left(
x(1-x)^{\frac{1}{\theta}-1}
-\theta\bigl[1-(1-x)^{\frac{1}{\theta}}\bigr]
\right)
=\operatorname{sign}(\theta-1).
\]
\end{lemma}

\begin{proof}
Let $t=(1-x)^{1/\theta}\in(0,1)$. The expression above equals
$f_t(\theta)=t^{1-\theta}-t-\theta+t\theta$. For fixed $t$,
\[
f_t''(\theta)=(\ln t)^2t^{1-\theta}>0,
\qquad f_t(0)=f_t(1)=0.
\]
Strict convexity implies $f_t(\theta)<0$ for $0<\theta<1$. Moreover,
$f_t'(1)=t-1-\ln t>0$, so $f_t(\theta)>0$ for $\theta>1$.
\end{proof}

Because $x_{ij}^{2}>0$, the lemma and \eqref{eq:dmu_dx} imply
\[
\frac{\partial \mu_{ij}}{\partial x_{ij}}<0
\quad\Longleftrightarrow\quad
\phi>0 \text{ and } \theta<1,
\]
which proves Part~II.
\qed

\clearpage
\subsection{Proof of Proposition \ref{prop:3}}
\label{appendix:proof-prop3}

The log tariff-inclusive price is
\[
\ln p_{ij}=\ln\mu_{ij}+\ln MC_i+\ln T_{c},
\]
where $MC_i$ and $\mu_{ij}$ are given by \eqref{eq:MC_i} and
\eqref{eq:bilateral_markup_prop}. Using the definitions of $\Gamma_{ij}$ and $\Lambda_{ij}$
in Proposition~\ref{prop:3}, a full log-differential gives
\[
d\ln p_{ij}
=-\Gamma_{ij}\,d\ln p_{ij}-\Lambda_{ij}\,d\ln p_{ij}+d\ln T_c.
\]
Rearranging,
\[
\frac{d\ln p_{ij}}{d\ln T_c}=\frac{1}{1+\Gamma_{ij}+\Lambda_{ij}}.
\]
\qed

\subsection{Derivation of the Cost and Markup Elasticities}
\label{subsec: Derivations Pass-through}

We now derive the two elasticities that enter the pass-through formula.

\subsubsection*{The Cost Elasticity.}

Taking the logarithm of equation (\ref{eq:MC_i}), we obtain:
\begin{align*}
\ln MC_i & =\ln k_{i}+\frac{1-\theta}{\theta}\ln q_{i}.
\end{align*}
It immediately follows that:
\begin{align*}
\Lambda_{ij}\equiv-\frac{d\ln MC_i}{d\ln p_{ij}} & =\frac{1-\theta}{\theta}\frac{d\ln q_{i}}{d\ln q_{ij}}\left(-\frac{d\ln q_{ij}}{d\ln p_{ij}}\right)\\
 & =\frac{1-\theta}{\theta}\cdot x_{ij}\cdot\varepsilon_{ij}\geq0.
\end{align*}
Moreover, the comparative statics with respect to the bilateral shares
are easy to compute as:
\[
\frac{d\Lambda_{ij}}{dx_{ij}}=\frac{1-\theta}{\theta}\cdot\varepsilon_{ij}\geq0,
\]
with strict inequality whenever $\theta<1,$ whereas:
\[
\frac{d\Lambda_{ij}}{ds_{ij}}=\frac{1-\theta}{\theta}\cdot x_{ij}\cdot(\eta-\rho)<0.
\]
Thus, the cost elasticity weakly increases with the importer's buyer share $x_{ij},$ and it decreases with the exporter's supplier share $s_{ij}.$ 

\newpage
\subsubsection*{Markup Elasticity}

Taking logs of equation~\eqref{eq:bilateral_markup_prop} and differentiating, we obtain:
\begin{align*}
d \ln \mu_{ij} 
&= \frac{(1 - \omega_{ij}) \mu_{ij}^{\text{oligopoly}}}{\mu_{ij}} \, d \ln \mu_{ij}^{\text{oligopoly}} 
+ \frac{\omega_{ij} \mu_{ij}^{\text{oligopsony}}}{\mu_{ij}} \, d \ln \mu_{ij}^{\text{oligopsony}} \\
&\  \ + \frac{\omega_{ij} \left( \mu_{ij}^{\text{oligopsony}} - \mu_{ij}^{\text{oligopoly}} \right)}{\mu_{ij}} \, d \ln \omega_{ij}.
\end{align*}

Rearranging terms, the price elasticity of the bilateral markup can be expressed as:
\begin{align*}
\Gamma_{ij} \equiv -\frac{d \ln \mu_{ij}}{d \ln p_{ij}} 
= \frac{(1 - \omega_{ij}) \mu_{ij}^{\text{oligopoly}}}{\mu_{ij}} \, \Gamma_{ij}^{\text{oligopoly}} 
+ \frac{\omega_{ij} \mu_{ij}^{\text{oligopsony}}}{\mu_{ij}} \, \Gamma_{ij}^{\text{oligopsony}} 
+ \left(1 - \frac{\mu_{ij}^{\text{oligopoly}}}{\mu_{ij}}\right) \Gamma_{ij}^{\omega},
\end{align*}
where: \(\Gamma_{ij}^{\text{oligopoly}}\equiv -\frac{d \ln \mu_{ij}^{\text{oligopoly}}}{d \ln p_{ij}}, \ \  \Gamma_{ij}^{\text{oligopsony}} \equiv -\frac{d \ln \mu_{ij}^{\text{oligopsony}}}{d \ln p_{ij}},\) and \(\Gamma_{ij}^{\omega} \equiv -\frac{d \ln \omega_{ij}}{d \ln p_{ij}}. \)

\vspace{0.5cm}
\textbf{Oligopoly Markup Elasticity.} The oligopoly markup elasticity is given by:
\begin{align*}
\Gamma_{ij}^{\text{oligopoly}}\equiv & -\frac{d\ln\mu_{ij}^{\text{oligopoly}}}{d\ln p_{ij}}=-\frac{d\ln\mu_{ij}^{\text{oligopoly}}}{d\ln s_{ij}}\cdot\frac{d\ln s_{ij}}{d\ln p_{ij}}.
\end{align*}

From the definition of $s_{ij},$ the last term is:
\[
\frac{d\ln s_{ij}}{d\ln p_{ij}}=-\left(\rho-1\right)\left(1-s_{ij}\right).
\]

Given $\mu_{ij}^{oligopoly}=\frac{\varepsilon_{ij}}{\varepsilon_{ij}-1}$
, we find: 
\begin{align*}
-\frac{d\ln\mu_{ij}^{\text{oligopoly}}}{d\ln s_{ij}} & =-\frac{1}{(\varepsilon_{ij}-1)}\cdot\frac{\rho-\varepsilon_{ij}}{\varepsilon_{ij}},
\end{align*}
which implies: 
\[
\Gamma_{ij}^{\text{oligopoly}}=-\frac{d\ln\mu_{ij}^{\text{oligopoly}}}{d\ln s_{ij}}\cdot\frac{d\ln s_{ij}}{d\ln p_{ij}}=\frac{1}{\varepsilon_{ij}-1}\frac{\rho-\varepsilon_{ij}}{\varepsilon_{ij}}\left(\rho-1\right)\left(1-s_{ij}\right)\geq0.
\]

\vspace{0.5cm}
\textbf{Oligopsony Markup Elasticity.} The oligopsony markup elasticity is given by:
\begin{align*}
\Gamma_{ij}^{\text{oligopsony}}\equiv & -\frac{d\ln\mu_{ij}^{\text{oligopsony}}}{d\ln p_{ij}}=-\left(\frac{d\ln\mu_{ij}^{\text{oligopsony}}}{d\ln x_{ij}}\right)\left(\frac{d\ln x_{ij}}{d\ln p_{ij}}\right).
\end{align*}
From the definition of $x_{ij},$ the last term is:
\[
\frac{d\ln x_{ij}}{d\ln p_{ij}}=-(1-x_{ij})\varepsilon_{ij}.
\]
Given $\mu_{ij}^{oligopsony}:=\theta\left(\frac{1-\left(1-x_{ij}\right)^{\frac{1}{\theta}}}{x_{ij}}\right),$
we find:
\[
\frac{d\ln\mu_{ij}^{\text{oligopsony}}}{d\ln x_{ij}}=\left(\frac{x_{ij}\left(1-x_{ij}\right)^{\frac{1}{\theta}-1}}{\theta\left(1-\left(1-x_{ij}\right)^{\frac{1}{\theta}}\right)}-1\right)
\]
which implies: 
\[
\Gamma_{ij}^{\text{oligopsony}}=-\left(\frac{d\ln\mu_{ij}^{\text{oligopsony}}}{d\ln x_{ij}}\right)\left(\frac{d\ln x_{ij}}{d\ln p_{ij}}\right)=\left(\frac{x_{ij}\left(1-x_{ij}\right)^{\frac{1}{\theta}-1}}{\theta\left(1-\left(1-x_{ij}\right)^{\frac{1}{\theta}}\right)}-1\right)(1-x_{ij})\varepsilon_{ij}.
\]

\subsubsection*{Omega Elasticity}

The elasticity of the weight $\omega_{ij}$ with respect to price
is:
\begin{align*}
\Gamma_{ij}^{\omega}\equiv-\frac{d\ln\omega_{ij}}{d\ln p_{ij}}= & \frac{d\ln\omega_{ij}}{d\ln s_{ij}}\left(-\frac{d\ln s_{ij}}{d\ln p_{ij}}\right).
\end{align*}
Given
\[
\frac{d\ln\omega_{ij}}{d\ln s_{ij}}=\left(1-\omega_{ij}\right)\frac{d\ln\lambda_{ij}}{d\ln s_{ij}},
\]
the above elasticity becomes:
\[
\Gamma_{ij}^{\omega}=\frac{d\ln\lambda_{ij}}{d\ln s_{ij}}\left(1-\omega_{ij}\right)\left(\rho-1\right)\left(1-s_{ij}\right),
\]

Since $\lambda_{ij}=\frac{(\eta-1)s_{ij}}{\varepsilon_{ij}-1}\cdot\left(1-(1-s_{ij})^{\frac{\eta-1}{\rho-1}}\right)^{-1}$
, we find:
\begin{align*}
\frac{d\ln\lambda_{ij}}{d\ln s_{ij}} & =1-\frac{\varepsilon_{ij}-\rho}{\varepsilon_{ij}-1}-\frac{\eta-1}{\rho-1}\cdot\frac{(1-s_{ij})^{\frac{\eta-1}{\rho-1}}}{1-(1-s_{ij})^{\frac{\eta-1}{\rho-1}}}\cdot\frac{s_{ij}}{(1-s_{ij})}.
\end{align*}

Thus:
\begin{align*}
\Gamma_{ij}^{\omega} & =\left(1-\frac{\varepsilon_{ij}-\rho}{\varepsilon_{ij}-1}-\frac{\eta-1}{\rho-1}\cdot\frac{(1-s_{ij})^{\frac{\eta-1}{\rho-1}}}{1-(1-s_{ij})^{\frac{\eta-1}{\rho-1}}}\cdot\frac{s_{ij}}{(1-s_{ij})}\right)\left(1-\omega_{ij}\right)\left(\rho-1\right)\left(1-s_{ij}\right).
\end{align*}

\clearpage

\subsection{Proof of Proposition \ref{prop:4}}
\label{appendix:proof-prop4}

Fix the supplier share $s$ and suppress the $ij$ subscripts. We prove the results for an interior buyer share, $x\in(0,1)$; the boundary cases $x=0$ and $x=1$ follow by continuity.

\paragraph{Part 1.}
Let $\theta<1$, $\beta\equiv1/\theta>1$, and $y\equiv1-x$. Write
$m(x)\equiv\mu^{\text{oligopsony}}(x)$ for the oligopsony markdown and define the
auxiliary elasticity term
\[
r(x)\equiv1+\frac{d\ln m(x)}{d\ln x}
=
\frac{x(1-x)^{\frac{1}{\theta}-1}}
{\theta\left[1-(1-x)^{\frac{1}{\theta}}\right]}
=\frac{\beta(1-y)y^{\beta-1}}{1-y^{\beta}}.
\]
Using $d\ln x/d\ln p=-(1-x)\varepsilon$ from
Appendix~\ref{subsec: Derivations Pass-through}, we obtain
$\Gamma^{\text{oligopsony}}=-(1-r)(1-x)\varepsilon$. Moreover,
\begin{equation}
1-r=\frac{h(y)}{1-y^{\beta}},
\qquad
h(y)\equiv1-\beta y^{\beta-1}+(\beta-1)y^{\beta}
=(\beta-1)\int_y^1\beta t^{\beta-1}\frac{1-t}{t}\,dt>0.
\label{eq:h}
\end{equation}
Thus $r\in(0,1)$, as in Lemma~\ref{lem:sign}, and
$\Gamma^{\text{oligopsony}}\le0$. For the lower bound,
Chebyshev's integral inequality and the fact that $\beta t^{\beta-1}$ is increasing while
$(1-t)/t$ is decreasing give
\begin{align}
\frac{1-r}{\beta-1}
&=\frac{\int_y^1\beta t^{\beta-1}\frac{1-t}{t}\,dt}
{\int_y^1\beta t^{\beta-1}\,dt}
\le\frac{1}{1-y}\int_y^1\frac{1-t}{t}\,dt \notag\\
&=\frac{-\ln y-(1-y)}{1-y}
\le\frac{1-y}{2y}.
\label{eq:cheb}
\end{align}
The last inequality sets $u=1-y$ and uses
$-\ln(1-u)-u\le u^2/[2(1-u)]$. The difference between the right- and left-hand
sides vanishes at $u=0$ and has derivative $u^2/[2(1-u)^2]\ge0$. Multiplying
\eqref{eq:cheb} by $(\beta-1)y\varepsilon>0$ and using $y=1-x$ therefore gives
\[
(1-r)(1-x)\varepsilon
\le\frac{\beta-1}{2}x\varepsilon
=\frac{1}{2}\Lambda.
\]
Since $\Gamma^{\text{oligopsony}}=-(1-r)(1-x)\varepsilon$, it follows that
$\Gamma^{\text{oligopsony}}\ge-\Lambda/2$, as claimed.

\paragraph{Part 2.}
Write $M\equiv\mu^{\text{oligopoly}}$, retain
$m\equiv\mu^{\text{oligopsony}}$, let $\mu=(1-\omega)M+\omega m$, and
$w\equiv\omega m/\mu=\omega^{\Gamma}$. The net response generated by decreasing returns is
$D\equiv\Lambda+w\,\Gamma^{\text{oligopsony}}$. Since $w\in[0,1]$, Part~1 gives
$D\ge\Lambda-\tfrac{1}{2}w\Lambda\ge\tfrac{1}{2}\Lambda>0$.

To sign $\partial D/\partial x$, set $f(x)\equiv-(1-r)(1-x)$, so that
$\Gamma^{\text{oligopsony}}=f(x)\varepsilon$. Direct differentiation gives
\begin{align}
\frac{m'(x)}{m(x)}&=-\frac{1-r}{x},\label{eq:P1}\\
\frac{r'(x)}{r(x)}&=\frac{1-r}{x}-\frac{1-\theta}{\theta(1-x)},\label{eq:P2}\\
f'(x)+\frac{1-\theta}{\theta}
&=(1-r)\left[1+\frac{r(1-x)}{x}+\frac{1-\theta}{\theta}\right].\label{eq:P3}
\end{align}
Neither $\omega$ nor $M$ depends on $x$, so equation~\eqref{eq:P1} implies
$\partial w/\partial x=-w(1-w)(1-r)/x$. Differentiating
$D=wf\varepsilon+(1-\theta)x\varepsilon/\theta$ and using \eqref{eq:P3},
\begin{equation}
\frac{\partial D}{\partial x}
=w\,\frac{1-r}{x}\,\varepsilon\big[1-w(1-r)(1-x)\big]
+\varepsilon\,(1-wr)\,\frac{1-\theta}{\theta}.
\label{eq:dD}
\end{equation}
Both terms are nonnegative and the second is strictly positive, because $w\in[0,1]$, $r\in(0,1)$, and $x\in(0,1)$. Hence $D$ is strictly increasing in $x$. \qed

\paragraph{Constant returns.}
If $\theta=1$, then $\Lambda=0$, $m=r=1$, and $\Gamma^{\text{oligopsony}}=0$. The weight
$w=\omega/[(1-\omega)M+\omega]$ is also independent of $x$. The buyer share therefore drops out of the pricing equation, and $\Phi=1/(1+\Gamma+\Lambda)$ does not vary with $x$.

\paragraph{Buyer-share gradient of the full pass-through elasticity.}
With $c\equiv(d\ln\lambda/d\ln s)(\rho-1)(1-s)$, the markup elasticity is
\begin{equation}
\Gamma=(1-w)\Gamma^{\text{oligopoly}}
+w\Gamma^{\text{oligopsony}}+(w-\omega)c,
\qquad \Gamma^{\omega}=(1-\omega)c.
\label{eq:Gamma_clean}
\end{equation}
Only $w$ and $\Gamma^{\text{oligopsony}}$ vary with $x$. Relative to $D$, the sum $\Gamma+\Lambda$ contains the additional terms $(1-w)\Gamma^{\text{oligopoly}}+(w-\omega)c$, which depend on $x$ through $w$. Using \eqref{eq:dD},
\begin{equation}
\frac{\partial(\Gamma+\Lambda)}{\partial x}
=w\,\frac{1-r}{x}\Big\{\varepsilon\big[1-w(1-r)(1-x)\big]+(1-w)\big(\Gamma^{\text{oligopoly}}-c\big)\Big\}
+\varepsilon\,(1-wr)\,\frac{1-\theta}{\theta}.
\label{eq:dGL}
\end{equation}
The last term is positive, while the expression in braces is bounded below
by $(1-w)(\varepsilon+\Gamma^{\text{oligopoly}}-c)$. Hence
$\partial(\Gamma+\Lambda)/\partial x>0$ provided
\begin{equation}
c\;\le\;\varepsilon+\Gamma^{\text{oligopoly}}.
\label{eq:cond_prop4}
\end{equation}
Since $\partial\Phi/\partial x=-\Phi^2\,\partial(\Gamma+\Lambda)/\partial x$, this condition
is sufficient for pass-through to decrease with $x$ when $\theta<1$. If $\phi\in\{0,1\}$ (full oligopoly or full oligopsony), then
$w(1-w)=0$ and no condition is required. Using the expressions in Appendix~\ref{subsec: Derivations Pass-through},
\[
c-\Gamma^{\text{oligopoly}}=(\rho-1)(1-s)\left[\frac{\rho}{\varepsilon}-\frac{ks(1-s)^{k-1}}{1-(1-s)^{k}}\right],
\qquad k\equiv\frac{\eta-1}{\rho-1},
\]
so the condition depends only on $(s,\rho,\eta)$. 

\clearpage

\subsection{Proof of Proposition \ref{prop:phi_theta} and Corollary \ref{cor:cov}}
\label{appendix:proof-prop-phi-theta}

\paragraph{Proof of Proposition \ref{prop:phi_theta}.}
Fix $(s_{ij},x_{ij})\in(0,1)^2$ and suppress the subscript $ij$. Let $\zeta\equiv(1-\theta)/\theta$, so that $\zeta\to\infty$ as $\theta\to0$. The cost elasticity
is $\Lambda=\zeta x\varepsilon$, so it grows linearly with $\zeta$. We first show
that the markup elasticity remains bounded as supply steepens, uniformly over
bargaining power.

Using the notation from the preceding proof, equation~\eqref{eq:Gamma_clean}
decomposes the markup elasticity as
\[
\Gamma
=(1-w)\Gamma^{\text{oligopoly}}
+w\Gamma^{\text{oligopsony}}
+(w-\omega)c.
\]
The quantities $\Gamma^{\text{oligopoly}}$, $M$, $\varepsilon$, $c$, and $\omega$
do not depend on $\theta$. The bargaining parameter $\phi$ enters only through
$\omega\in[0,1]$ and hence $w$.

It remains to bound the terms that vary with $\theta$. Recall that
\[
m=\theta\frac{1-(1-x)^{1/\theta}}{x},
\qquad
w=\frac{\omega m}{(1-\omega)M+\omega m}.
\]
Because $1/\theta>1$, $(1-x)^{1/\theta}\le1-x$, while Bernoulli's inequality
gives $(1-x)^{1/\theta}\ge1-x/\theta$. These two inequalities imply
$\theta\le m\le1$. Since $M\ge1$, it follows that
$0\le w\le\omega\le1$. In addition, the proof of
Proposition~\ref{prop:4} established that $r\in(0,1)$. Because
$\Gamma^{\text{oligopsony}}=-(1-r)(1-x)\varepsilon$, we therefore have
\[
-(1-x)\varepsilon
\le
\Gamma^{\text{oligopsony}}
\le0.
\]
The first two terms in the decomposition of $\Gamma$ are therefore a weighted
average of two bounded objects, and $|w-\omega|\le1$. Hence
\[
|\Gamma|
\le
B
\equiv
\max\left\{\Gamma^{\text{oligopoly}},(1-x)\varepsilon\right\}+|c|,
\]
where $B$ is finite and independent of $\phi$ and $\theta$. Using $\Lambda=\zeta x\varepsilon$,
\[
\sup_{\phi\in[0,1]}
\frac{|\Gamma|}{\Lambda}
\le
\frac{B}{\zeta\,x\varepsilon},
\]
which converges to zero at rate $1/\zeta=\theta/(1-\theta)$, that is, at rate $\theta$. For pass-through, note that
\[
\Phi\,(1+\Lambda)-1
=
-\frac{\Gamma}{1+\Lambda+\Gamma}.
\]
The numerator is bounded by $B$, while the denominator is at least $1+\zeta x\varepsilon-B$. The right-hand side is therefore $O(\zeta^{-1})$ uniformly in $\phi$. \qed

\paragraph{Proof of Corollary \ref{cor:cov}.}
Consider the finite cross-section $\mathcal I$ and let
$z_{ij}\equiv x_{ij}\varepsilon_{ij}$. Proposition~\ref{prop:phi_theta} supplies a finite bound for each
relationship. Because $\mathcal I$ is finite, these bounds have a finite maximum;
hence $\Gamma_{ij}=O(1)$ uniformly over the cross-section and over bargaining power.
In particular, $\operatorname{Var}_{\mathcal I}(\Gamma_{ij})=O(1)$ and
$\operatorname{Cov}_{\mathcal I}(z_{ij},\Gamma_{ij})=O(1)$.

Since
\[
\Lambda_{ij}=\zeta z_{ij},
\qquad
D_{ij}=\zeta z_{ij}+\Gamma_{ij},
\]
expanding the relevant variances and covariances gives
\begin{align*}
\operatorname{Var}_{\mathcal I}(D_{ij})
&=
\zeta^2\operatorname{Var}_{\mathcal I}(z_{ij})
+2\zeta\operatorname{Cov}_{\mathcal I}(z_{ij},\Gamma_{ij})
+\operatorname{Var}_{\mathcal I}(\Gamma_{ij})\\
&=
\zeta^2\operatorname{Var}_{\mathcal I}(z_{ij})+O(\zeta),\\
\operatorname{Cov}_{\mathcal I}(\Lambda_{ij},D_{ij})
&=
\zeta^2\operatorname{Var}_{\mathcal I}(z_{ij})
+\zeta\operatorname{Cov}_{\mathcal I}(z_{ij},\Gamma_{ij})\\
&=
\zeta^2\operatorname{Var}_{\mathcal I}(z_{ij})+O(\zeta),\\
\operatorname{Cov}_{\mathcal I}(\Gamma_{ij},D_{ij})
&=
\zeta\operatorname{Cov}_{\mathcal I}(\Gamma_{ij},z_{ij})
+\operatorname{Var}_{\mathcal I}(\Gamma_{ij})
=O(\zeta).
\end{align*}
By assumption, $\operatorname{Var}_{\mathcal I}(z_{ij})>0$. The terms of order
$\zeta^2$ therefore dominate, and division by
$\operatorname{Var}_{\mathcal I}(D_{ij})$ yields, writing
$S_{\Lambda}\equiv\operatorname{Cov}_{\mathcal I}(\Lambda_{ij},D_{ij})/\operatorname{Var}_{\mathcal I}(D_{ij})$ and
$S_{\Gamma}\equiv\operatorname{Cov}_{\mathcal I}(\Gamma_{ij},D_{ij})/\operatorname{Var}_{\mathcal I}(D_{ij})$,
\[
S_{\Lambda}=1+O(\zeta^{-1}),
\qquad
S_{\Gamma}=O(\zeta^{-1}).
\]
Finally, the two covariance contributions sum to one exactly because
$D_{ij}=\Lambda_{ij}+\Gamma_{ij}$:
\[
S_{\Lambda}+S_{\Gamma}
=
\frac{\operatorname{Cov}_{\mathcal I}(\Lambda_{ij}+\Gamma_{ij},D_{ij})}
{\operatorname{Var}_{\mathcal I}(D_{ij})}
=1.
\]
All bounds are uniform over bargaining power. \qed

\clearpage

\section{Theory: Extensions\label{sec: Theory Extensions}}

\subsection{Generalized Outside Option ~\label{subsec:Generalized-outside-option}}

In the baseline model, each party's disagreement payoff is its profit from trading with all current partners except the other party, at unchanged prices and, for the exporter, with unchanged sales to its other buyers. This rules out two forces that may matter in richer environments: after a failed negotiation, firms may form or reallocate relationships, and broader equilibrium adjustments may change costs or demand elsewhere in the network. This section shows that the structure of the pricing equation survives arbitrary disagreement payoffs, but that the price-cost ratio then depends on objects that bilateral shares do not summarize.

Let $\tilde{\pi}_{i(-j)}$ and $\tilde{\pi}_{j(-i)}$ denote arbitrary disagreement payoffs, restricted only to be independent of the negotiated price $p_{ij}$ and to leave both parties with positive gains from trade. Define
\[
\mu_{ij}^{\text{oligopsony},G}\equiv\frac{p_{ij}q_{ij}-\left(\pi_{i}-\tilde{\pi}_{i(-j)}\right)}{MC_i\,q_{ij}},
\qquad
\Delta_{ij}^{s,G}\equiv 1-\frac{\tilde{\pi}_{j(-i)}}{\pi_{j}} .
\]
The first object is the exporter's net incremental cost of serving importer $j$, per unit and relative to current marginal cost. The second is the share of the importer's profit that is lost if the negotiation fails. By construction,
\begin{align*}
GFT_{ij}^{i}(p_{ij}) &= q_{ij}\left(p_{ij}-MC_i\,\mu_{ij}^{\text{oligopsony},G}\right), \\
GFT_{ij}^{j}(p_{ij}) &= \pi_j\,\Delta_{ij}^{s,G} .
\end{align*}
Because the disagreement payoffs do not depend on $p_{ij}$, we still have $dGFT_{ij}^{k}/dp_{ij}=d\pi_{k}/dp_{ij}$ for $k\in\{i,j\}$, and the first-order condition \eqref{eq:FOC} is unchanged. Following the steps in Appendix~\ref{subsec: Derivations Markup},
\begin{equation}
\mu_{ij}\equiv\frac{p_{ij}}{MC_i}=(1-\omega^G_{ij})\,\mu_{ij}^{\text{oligopoly}}+\omega^G_{ij}\,\mu_{ij}^{\text{oligopsony},G},
\label{mu_outside_general}
\end{equation}
with
\[
\omega^G_{ij}
=
\frac{\phi\,\xi^G(GFT_{ij}^{j},p_{ij})}
{\phi\,\xi^G(GFT_{ij}^{j},p_{ij})+(1-\phi)(\varepsilon_{ij}-1)}
,
\qquad
\xi^G(GFT_{ij}^{j},p_{ij})
\equiv
-\frac{d\ln GFT_{ij}^{j}}{d\ln p_{ij}}
=
\frac{(\eta-1)s_{ij}}{\Delta_{ij}^{s,G}} .
\]
The baseline model is the special case $\mu_{ij}^{\text{oligopsony},G}=\theta\,[1-(1-x_{ij})^{1/\theta}]/x_{ij}$ and $\Delta_{ij}^{s,G}=1-(1-s_{ij})^{\frac{\eta-1}{\rho-1}}$.

\paragraph{Example: cost changes under disagreement.}
Suppose a failed negotiation changes the two parties' marginal costs by the factors $\Delta_{ij}^{MC_i}\equiv\widetilde{MC}_i/MC_i$ and $\Delta_{ij}^{c_j}\equiv\tilde{c}_j/c_j$, where the exporter's sales to its other buyers are unchanged and, for the importer, $\varrho=1$, so that $\eta-1=(\nu-1)\gamma$. Then
\[
\mu_{ij}^{\text{oligopsony},G}=\theta\,\frac{1-\Delta_{ij}^{MC_i}(1-x_{ij})}{x_{ij}},
\qquad
\Delta_{ij}^{s,G}=1-\left(\Delta_{ij}^{c_j}\right)^{1-\nu},
\]
and the baseline corresponds to $\Delta_{ij}^{MC_i}=(1-x_{ij})^{\frac{1-\theta}{\theta}}$ and $\Delta_{ij}^{c_j}=(1-s_{ij})^{-\frac{\gamma}{\rho-1}}$. A change in the exporter's cost conditions, or the importer's access to alternative suppliers, moves these factors away from their baseline values.

\paragraph{Implication.}
Equation \eqref{mu_outside_general} retains the weighted-average structure of equation~\eqref{eq:bilateral_markup_prop}, but its components are no longer functions of the bilateral shares alone.  Without additional data or stronger assumptions, the model is not identified in a way that is useful for our empirical exercise. The baseline specification imposes more structure precisely so that the components of the price-cost ratio can be recovered from observable shares and a small set of parameters.

\subsection{Pass-Through with Network Spillovers}
\label{subsec:Generalized-pass-through-elasticity}

Proposition~\ref{prop:3} holds other link prices and aggregate variables fixed, while allowing the importer to re-optimize quantities. Here we illustrate selected spillovers using a truncated approximation. The exercise simply aims to illustrate possible wedges between reduced-form pass-through and the direct elasticity $\Phi_{ij}$, in a simplified setting. We allow a tariff on exporter $i$ to affect all its buyers and allow its rivals to respond to the initial supplier-share changes. We omit rival-price-induced substitution in the quantity equations, rivals' own-price and further cross-rival feedback in their pricing response, direct tariff exposure of rivals, and changes in their marginal costs and buyer shares. These omissions can be first-order in the tariff shock. 

Let $\Psi_{ij}$ denote the pass-through response when the tariff applies to all relationships of exporter $i$.
Define
\[
\Gamma_{ij}^{s}\equiv\frac{\partial\ln\mu_{ij}}{\partial\ln s_{ij}},
\qquad
\Gamma_{ij}^{x}\equiv\frac{\partial\ln\mu_{ij}}{\partial\ln x_{ij}},
\]
where each derivative holds the other share fixed. When all other match prices are fixed, $d\ln s_{ij}=-(\rho-1)(1-s_{ij})\,d\ln p_{ij}$ and $d\ln x_{ij}=-\varepsilon_{ij}(1-x_{ij})\,d\ln p_{ij}$, so the markup elasticity of Proposition~\ref{prop:3} is $\Gamma_{ij}=(\rho-1)(1-s_{ij})\Gamma_{ij}^{s}+\varepsilon_{ij}(1-x_{ij})\Gamma_{ij}^{x}$. Because the bargaining weight $\omega_{ij}$ does not depend on $x_{ij}$, $\Gamma_{ij}^{x}=\omega_{ij}^{\Gamma}\,\partial\ln\mu_{ij}^{\text{oligopsony}}/\partial\ln x_{ij}\le 0$.

\paragraph{Exporter side.} Retaining only own-price demand responses, the approximation gives $d\ln q_i=-\sum_{\ell\in\mathcal{Z}_i}x_{i\ell}\varepsilon_{i\ell}\Psi_{i\ell}\,d\ln T_{c(i)}$. The buyer share and marginal cost then respond according to
\begin{align*}
\frac{d\ln x_{ij}}{d\ln T_{c(i)}} &= -\varepsilon_{ij}(1-x_{ij})\Psi_{ij}+\sum_{\ell\in\mathcal{Z}_i\setminus\{j\}}x_{i\ell}\varepsilon_{i\ell}\Psi_{i\ell},\\
\frac{d\ln MC_i}{d\ln T_{c(i)}} &= -\frac{1-\theta}{\theta}\Big(\varepsilon_{ij}x_{ij}\Psi_{ij}+\sum_{\ell\in\mathcal{Z}_i\setminus\{j\}}x_{i\ell}\varepsilon_{i\ell}\Psi_{i\ell}\Big).
\end{align*}

\paragraph{Importer side.} A higher $p_{ij}$ raises the supplier shares of importer $j$'s other suppliers, which adjust their prices in turn. For this step, we retain only the share change induced by the initial movement in $p_{ij}$, holding each rival's marginal cost and buyer share fixed and omitting its own-price and cross-rival feedback: $d\ln s_{rj}=(\rho-1)s_{ij}\,d\ln p_{ij}$ and $d\ln p_{rj}=\Gamma_{rj}^{s}\,d\ln s_{rj}$ for $r\in\mathcal{Z}_j\setminus\{i\}$. The supplier share of exporter $i$ then satisfies
\[
\frac{d\ln s_{ij}}{d\ln T_{c(i)}}=-(\rho-1)\Big((1-s_{ij})-s_{ij}(\rho-1)\sum_{r\in\mathcal{Z}_j\setminus\{i\}}s_{rj}\Gamma_{rj}^{s}\Big)\Psi_{ij}.
\]

\paragraph{Approximate elasticity.} Substituting the three responses into $d\ln p_{ij}=\Gamma_{ij}^{s}\,d\ln s_{ij}+\Gamma_{ij}^{x}\,d\ln x_{ij}+d\ln MC_i+d\ln T_{c(i)}$ and collecting terms in $\Psi_{ij}$ gives
\begin{equation*}
\Psi_{ij}=\tilde{\Phi}_{ij}\Big[1+\Big(\Gamma_{ij}^{x}-\frac{1-\theta}{\theta}\Big)\sum_{\ell\in\mathcal{Z}_i\setminus\{j\}}x_{i\ell}\varepsilon_{i\ell}\Psi_{i\ell}\Big],
\end{equation*}
where
\begin{equation*}
\tilde{\Phi}_{ij}\equiv\Big[1+(\rho-1)\Gamma_{ij}^{s}\Big((1-s_{ij})-s_{ij}(\rho-1)\sum_{r\in\mathcal{Z}_j\setminus\{i\}}s_{rj}\Gamma_{rj}^{s}\Big)+\varepsilon_{ij}(1-x_{ij})\Gamma_{ij}^{x}+\frac{1-\theta}{\theta}\varepsilon_{ij}x_{ij}\Big]^{-1}.
\end{equation*}
The system is linear in the elasticities $\{\Psi_{i\ell}\}_{\ell\in\mathcal{Z}_i}$. With other match prices fixed, the sums vanish and $\Psi_{ij}=\tilde{\Phi}_{ij}=\Phi_{ij}$.

Within the approximation, the two retained indirect effects work in opposite directions, provided the relevant pass-through elasticities are positive. On the importer side, when the focal and rival markups increase with their supplier shares ($\Gamma^{s}>0$), rival suppliers raise their prices as they gain share. This limits the loss of share by exporter $i$ and hence the fall in its markup, so that $\tilde{\Phi}_{ij}>\Phi_{ij}$. On the exporter side, price increases on the other links reduce the exporter's total quantity further. Marginal cost falls by more, and the buyer share $x_{ij}$ rises relative to the direct case, which deepens the markdown. Both forces lower $p_{ij}$: the coefficient $\Gamma_{ij}^{x}-(1-\theta)/\theta$ is negative whenever $\theta<1$, so $\Psi_{ij}<\tilde{\Phi}_{ij}$ as long as pass-through on the other links is positive. Under constant returns, this coefficient is zero and the exporter-side effect disappears. Whether $\Psi_{ij}$ lies above or below $\Phi_{ij}$ therefore depends on the relative strength of the two effects. In the main text, we focus on the direct elasticity $\Phi_{ij}$, which follows from bilateral shares and the model's parameters alone.

\subsection{Efficient Bargaining} \label{sec:APP_efficient-bargaining}

Under efficient bargaining, the importer and exporter solve:
\[
\max_{p_{ij},\, q_{ij}} \left(GFT_{ij}^{i}\right)^{1-\phi} \left(GFT_{ij}^{j}\right)^{\phi}
\quad \text{s.t.} \quad GFT_{ij}^{i} \geq 0,\quad GFT_{ij}^{j} \geq 0,
\]
where gains are measured relative to the disagreement payoffs in Section~\ref{sec: Theory}, holding other link prices and the exporter's sales to its other buyers fixed. The importer re-optimizes its other inputs conditional on the negotiated quantity. The pair first selects $q_{ij}$ to maximize joint surplus, then negotiates over $p_{ij}$ to split it.

Let $V_j(q_{ij})$ denote the importer's optimized downstream revenue net of all other input costs, excluding payment for $q_{ij}$, and let $\Delta V_j(q_{ij})=V_j(q_{ij})-V_j(0)$. By the envelope theorem, $MRP_j(q_{ij})\equiv V_j'(q_{ij})$ is the marginal revenue product of this input at the optimized other-input choices. The joint-surplus-maximizing quantity, conditional on the other relationships, solves:
\[
q_{ij}^{*} : \; MRP_j(q^*_{ij}) = MC_i(q_i^{*}),
\qquad \text{where} \qquad
q_i^{*} \equiv q_{i(-j)} + q_{ij}^{*}.
\]

Given $q_{ij}^*$, the price $p_{ij}$ solves:
\[
\max_{p_{ij}} \left(p_{ij} q^*_{ij} - \Delta C_i(q^*_{ij})\right)^{1-\phi}
\left(\Delta V_j(q^*_{ij}) - p_{ij} q^*_{ij} \right)^{\phi},
\]
Here $\Delta V_j(q_{ij})$ is the incremental value net of other input costs defined above, and $\Delta C_i(q_{ij})\equiv TC_i(q_{i(-j)}+q_{ij})-TC_i(q_{i(-j)})$ is the exporter's incremental production cost. Thus $GFT_{ij}^j=\Delta V_j-p_{ij}q_{ij}$ and $GFT_{ij}^i=p_{ij}q_{ij}-\Delta C_i$.

Standard derivations lead to the equilibrium price:
\begin{equation*}
p_{ij} = (1 - \phi) \cdot \frac{{\Delta V_j}(q_{ij})}{q_{ij}} 
+ \phi \cdot \frac{\Delta C_i(q_{ij})}{q_{ij}}
\end{equation*}
a weighted average of the importer's per-unit incremental value net of other input costs and the exporter's per-unit incremental production cost, evaluated at $q_{ij}=q_{ij}^*$, with bargaining weights given by $\phi$.

\begin{figure}[t]
\caption{Efficient Bargaining\label{fig:eff_bgn_fig}}
\begin{centering}
\includegraphics[width=0.60\textwidth]{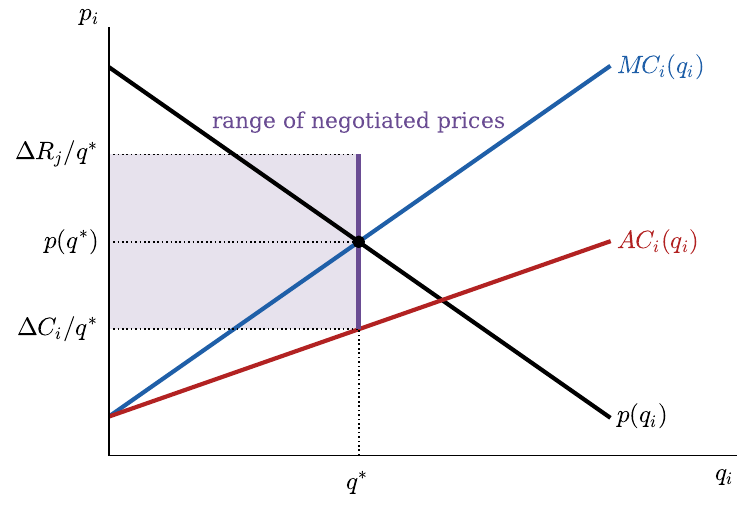}
\par\end{centering}
\footnotesize \textit{Notes}{: The figure illustrates the efficient bargaining protocol for an exporter with a single buyer, so that $q_i=q_{ij}$. The black schedule is importer $j$'s marginal revenue product $p(q_i)$, the blue schedule the exporter's marginal cost $MC_i(q_i)$, and the red schedule its average cost $AC_i(q_i)$. The efficient quantity $q^{*}$ equates marginal cost to marginal revenue product. The shaded band is the set of prices at which both parties have nonnegative gains from trade. {The figure's label $\Delta R_j/q^*$ denotes $\Delta V_j/q^*$, the importer's incremental value net of other input costs.}}
\end{figure}

Figure~\ref{fig:eff_bgn_fig} illustrates this result for an exporter with a single buyer, so that $q_i = q_{ij}$. The efficient quantity $q^*$ is determined where the exporter's marginal cost $\mathrm{MC}(q_i)$ equals the importer's marginal revenue product, drawn as the downward-sloping schedule $p(q_i)$. At $q^*$, the negotiated price lies between the exporter's average incremental cost $\Delta C_i/q^*$, which equals average cost $\mathrm{AC}(q^*)$ with a single buyer and $\mathrm{AIC}_{ij}$ in general, and the importer's average incremental value net of other input costs ${\Delta V_j}/q^*$. The weight $\phi$ determines the exact price within this interval: as $\phi \to 0$ the price approaches ${\Delta V_j}/q^*$, and as $\phi \to 1$ it approaches $\Delta C_i/q^*$.

\subsection{Supply-Driven Quantity Bargaining} \label{sec:APP_supply-bargaining}

In supply-driven quantity bargaining, the exporter chooses quantity at a given price. Holding its other buyers' quantities fixed, its supply schedule solves
\[
\max_{q_{ij}\geq 0}\left\{p_{ij}q_{ij}-\Delta C_i(q_{ij})\right\},
\qquad
\Delta C_i(q_{ij})=TC_i(q_{i(-j)}+q_{ij})-TC_i(q_{i(-j)}).
\]
For $0<\theta<1$, the interior first-order condition gives the inverse supply schedule
\[
p_{ij}(q_{ij})=MC_i(q_{i(-j)}+q_{ij})
=k_i(q_{i(-j)}+q_{ij})^{\frac{1-\theta}{\theta}}.
\]
The parties then bargain over quantity along this schedule. Let $V_j(q_{ij})$ be importer $j$'s optimized downstream revenue net of all other input costs, excluding payment for $q_{ij}$, with other input prices and aggregate demand shifters fixed. Define $\Delta V_j(q_{ij})=V_j(q_{ij})-V_j(0)$ and $MRP_j=V_j'(q_{ij})$. The gains from trade under this protocol are
\[
GFT_{ij}^{i}=p_{ij}(q_{ij})q_{ij}-\Delta C_i(q_{ij}),
\qquad
GFT_{ij}^{j}=\Delta V_j(q_{ij})-p_{ij}(q_{ij})q_{ij}.
\]
The Nash problem maximizes $(GFT_{ij}^{i})^{1-\phi}(GFT_{ij}^{j})^{\phi}$ over quantities with nonnegative gains. For an interior solution with $0<\phi<1$ and strictly positive gains, its first-order condition is
\[
\psi_{ij}p_{ij}-MRP_j
=\frac{1-\omega_{ij}^{S}}{\omega_{ij}^{S}}(\psi_{ij}-1)p_{ij},
\qquad
\psi_{ij}=1+\frac{q_{ij}p_{ij}'(q_{ij})}{p_{ij}(q_{ij})}
=1+\frac{1-\theta}{\theta}x_{ij},
\]
where
\[
\omega_{ij}^{S}=\frac{\phi GFT_{ij}^{i}}
{(1-\phi)GFT_{ij}^{j}+\phi GFT_{ij}^{i}}\in(0,1).
\]
This follows from $(GFT_{ij}^{i})'=(\psi_{ij}-1)p_{ij}$ and $(GFT_{ij}^{j})'=MRP_j-\psi_{ij}p_{ij}$. Unlike the baseline bargaining weight, $\omega_{ij}^{S}$ depends on gains-from-trade levels not summarized by the bilateral shares alone.

At $\phi=1$, the importer maximizes its gains along the supply schedule, giving $\psi_{ij}p_{ij}=MRP_j$ at an interior solution, the standard oligopsony condition. The price is then a factor $\psi_{ij}^{-1}$ below marginal revenue product. The $\phi=0$ endpoint must instead be obtained from the exporter's maximization subject to importer participation, rather than by substituting zero into the interior ratio. Since exporter gains increase with quantity along an upward-sloping supply curve, the endpoint, when finite, exhausts importer surplus: $\Delta V_j(q_{ij})=p_{ij}(q_{ij})q_{ij}$. It need not coincide with the efficient quantity.

\begin{figure}[t]
    \centering
    \includegraphics[width=0.5\linewidth]{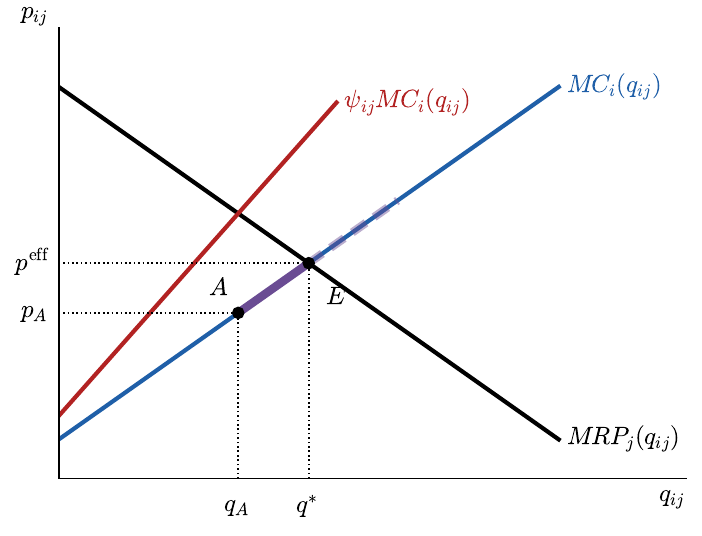}
    \caption{\small Supply-Driven Quantity Bargaining}
    \label{fig:supp_bgn_fig}
\end{figure}

Figure~\ref{fig:supp_bgn_fig} illustrates the equilibrium. The blue curve is the exporter's marginal cost $MC_i$ (equivalently, the supply curve), and the black curve is the importer's marginal revenue product $MRP_j$. The red curve is the importer's marginal factor cost $\psi_{ij}\,MC_i$, which exceeds $MC_i$ whenever $\theta < 1$. At $\phi = 1$ the equilibrium lies at point~A, where $MRP_j = \psi_{ij}\,MC_i$: the importer acts as a pure monopsonist, restricting quantity below the efficient level $q^*$ (where $MC_i = MRP_j$). The shaded segment between point~A and the efficient benchmark is illustrative, not the full feasible bargaining range: exporter-favored outcomes can lie beyond $q^*$, up to the importer-participation boundary.

\section{Data Appendix\label{sec:Data Appendix}}

\subsection{Related-Party Trade Measured via Ownership Linkages}
\label{subsec:Related-party-trade}
A key advantage of the ORBIS dataset is the breadth and detail of its ownership information. It provides comprehensive listings of both direct and indirect shareholders and subsidiaries, along with indicators of each firm's independence, global ultimate ownership, and group affiliations. This enables us to identify corporate structures at the firm level, including ownership links between firms located in different countries. We define a parent–subsidiary relationship as one in which the parent firm holds at least a 50\% ownership stake in the affiliate.

\paragraph{Linking US Importers to Multinational Ownership}
To identify US-based multinational firms, we match firms in the Census Business Register to their ORBIS counterparts using names, addresses, and GPS coordinates. This linkage combines probabilistic record matching with manual validation, producing a high match rate. As a result, we can flag US establishments that are either majority-owned affiliates of foreign multinationals or parent firms with majority-owned affiliates abroad. This information allows us to identify multinationals with operations in the US without relying solely on the Related Party Trade (RPT) indicator reported in the LFTTD.

\paragraph{Identifying Cross-Border Ownership Links} \label{subsec:Merging-foreign-exporter-id-orbis}
To assess whether the foreign exporter also belongs to the same corporate group, we match the Manufacturer ID (MID) reported in LFTTD to firm records in ORBIS. The MID is constructed by US Customs based on the exporter’s name, address, and country of origin using a set of formatting rules. The MID begins with a two-character country code (or a province code for Canada), followed by a name-based segment derived from the first three letters of the first and second words in the company’s name. If the company name consists of only one word, the first six letters are used. The next segment contains the first four digits from the address number, and the final three characters are the first three alphabetic characters of the city name. Standard formatting conventions apply, including the exclusion of punctuation, one-letter initials, and common stop words such as “the,” “and,” or “of.” Country-specific prefixes (e.g., “OAO” or “ZAO” in Russia or “PT” in Indonesia) are also omitted when constructing the MID.

Using these same rules, we replicate the MID structure for foreign firms in the ORBIS database. We then match each MID in the customs data to candidate firms in ORBIS based on the reconstructed name segment. We assess the quality of each potential match using two dimensions: location and product alignment. A location score is computed based on the match between city names in the MID and in ORBIS. A product match score is constructed at the MID–firm level by comparing the foreign firm’s NAICS6 industry code in ORBIS to the distribution of HS6 products shipped under that MID in the customs data, using the concordance developed by \citet{Pierce2009}. We retain only those matches where both the location and product match scores exceed 90\%. In addition, we drop from the matched LFTTD–ORBIS dataset any ORBIS-linked firm that is associated with fewer than five customs transactions. This filter is based on the number of shipment records matched to each ORBIS identifier in the merged data and is intended to eliminate weak or potentially noisy MID–firm matches. %

Another concern is that the MID may sometimes refer to intermediaries rather than manufacturers. Although US customs rules require that the MID correspond to the producer or manufacturer, not to wholesalers or freight forwarders, compliance with this rule is imperfect. To mitigate this concern, we use ORBIS industry codes to exclude retailers, wholesalers, and logistics providers from the matched dataset.

Finally, another challenge with the MID is that it is not a unique firm identifier: a given MID can correspond to multiple legal entities. In our matched data, we address this issue directly by checking whether a MID maps to more than one firm in ORBIS. If multiple firms share the same MID but belong to the same corporate group based on majority ownership links reported in ORBIS, we retain the match. Otherwise, we exclude the ambiguous MID from the analysis.

Taken together, these steps yield a linked dataset that offers a more transparent and conservative definition of related-party trade, based on majority ownership (at least 50\%). In contrast to the standard related-party trade (RPT) flag in customs data, which applies a lower threshold of 5\% ownership for imports, this approach reduces false positives and more precisely captures transactions where ownership ties are likely to influence pricing. By combining MID-based matching with firm-level ownership structures, the final dataset is well-suited for analyzing pricing behavior in cross-border transactions.

\clearpage

\subsection{Data Cleaning, Sample Construction and Summary Statistics} 
\label{subsec:Sample-Selection}

We construct the analysis sample in several steps to align the data with the model and the identification strategy. As in the main text, it is useful to distinguish between restrictions that are essential for identification and those that primarily ensure consistency with the model’s environment.

\paragraph{Preliminary restrictions.}
We begin by aggregating transaction-level customs records to the annual importer–exporter–product (HS10) level. Because relatively few relationships appear in adjacent months or in the same month across years, annual aggregation provides a natural unit of observation for relationship-level price changes. 

To ensure reliable price measurement, we apply several data-quality filters. We drop transactions with missing or zero import value or quantity, invalid exporter identifiers, or US\ importers that cannot be linked to the Longitudinal Business Database (LBD). We exclude HS codes 98–99 (special provisions), remove observations with log unit values below the 1st or above the 99th percentile of the HS10–country distribution, and trim extreme year-on-year log price changes outside the interval $[-4,4]$.\footnote{These filters are applied to the estimation sample but retained when computing tenure and relationship length.} We also exclude energy goods (HS chapter 27), which fall outside the scope of the model.

\paragraph{Identification-driven restrictions.}

First, we require importer–exporter–product triplets to be observed in at least two consecutive calendar years. This panel structure is necessary to compute year-on-year price changes and to estimate pass-through.

Second, our identification strategy relies on cross-sectional price variation across buyers sourcing the same product from the same supplier. We therefore restrict attention to supplier–product pairs that transact with at least two distinct US\ buyers in consecutive years. This condition is essential for identifying the bargaining and returns-to-scale parameters, as outlined in Section~\ref{sec:Identification-and-estimation}.

\paragraph{Model-alignment restrictions.}

The remaining restrictions ensure that the empirical setting matches the model’s focus on decentralized bargaining over intermediate inputs. We exclude products classified as consumption goods under the Broad Economic Categories (BEC) system, restricting attention to capital and intermediate inputs.

We also exclude related-party trade, as such transactions may reflect intra-firm transfer pricing rather than arm’s-length bargaining.\footnote{\citet{bernard2006transfer} document systematic differences between related-party and arm’s-length prices.} We classify a pair as related if ORBIS identifies a common corporate parent. The baseline excludes all such pairs. For robustness, we consider alternative definitions based on the LFTTD related-party flag and on multinational status (see Appendix~\ref{subsec:Related-party-trade}).

\medskip

With these restrictions, the final sample isolates repeated, arm’s-length buyer–supplier relationships in intermediate goods with sufficient cross-sectional variation for identification; it retains about 8\% of US imports by value and 5\% of all buyer--supplier--product relationships in the LFTTD.

Table~\ref{tab:sample_stats_bec1} summarizes the cumulative impact of these steps. Panel~A focuses on the 2001–2016 sample used for structural estimation. Panel~B reports the corresponding summary for the 2017–2018 sample used in the pass-through analysis. The first row of each panel (“All Imports”) includes all US import records for the relevant years. Subsequent rows show the effect of each restriction in turn, including the requirement that buyer–supplier pairs trade the same product in two consecutive calendar years, the exclusion of consumption and energy goods, and the restriction to arm’s-length relationships with sufficient variation for identification.

The final rows report the estimation samples used in the analysis. For 2001–2016 (Panel~A), the sample includes approximately \$880 billion in import value, 480,000 buyer–supplier pairs, and 630,000 buyer–supplier–HS10 triplets. For 2017–2018 (Panel~B), the corresponding sample comprises \$160 billion in imports, 190,000 pairs, and 250,000 triplets.\footnote{Panel~A omits intermediate sample steps and reports only the full and final samples for 2001–2016, as intermediate counts are neither used in the analysis nor releasable under Census disclosure policies. The same applies to Table~\ref{tab:sample_stats_bec0}.} These samples correspond exactly to those used in the structural estimation and post-estimation analysis. Table~\ref{tab:summary_stats_main} reports summary statistics for the main estimation sample, analogous to Table~\ref{tab:summary_stats} in the main text.

Tables~\ref{tab:sample_stats_bec0} and~\ref{tab:summary_stats_bec0} present analogous results for a broader sample that includes capital, intermediate, and consumption goods (“BEC – All (Consec.)”), rather than restricting to non-consumption goods.

\begin{table}[htbp]
\caption{Sample Composition by Period – Excluding BEC-Classified Consumption Goods}
\label{tab:sample_stats_bec1}
\centering
{\small
\renewcommand{\arraystretch}{1.2}
\arrayrulecolor{gray!40}
\begin{tabular}{
  >{\raggedright\arraybackslash}p{5cm} 
  >{\raggedright\arraybackslash}p{1.7cm} 
  >{\raggedright\arraybackslash}p{1.7cm} 
  >{\raggedright\arraybackslash}p{1.7cm} 
  >{\raggedright\arraybackslash}p{1.7cm} 
  >{\raggedright\arraybackslash}p{1.7cm}
}
\specialrule{.05em}{0em}{0em}
\makecell{Sample} & 
\makecell{Import\\value\\(bn USD)} & 
\makecell{Importers\\(th)} & 
\makecell{Exporters\\(th)} & 
\makecell{Pairs\\(th)} & 
\makecell{Triplets\\(th)} \\
\specialrule{.05em}{0em}{0.3em}
\\[-0.6em]

\multicolumn{6}{c}{\textit{Panel A}: 2001--2016} \\
\cmidrule(lr){1-6}
\\[-0.6em]
All Imports                 & 22{,}000 & 1{,}000 & 6{,}900 & 17{,}000 & 40{,}000 \\
BEC – Non-Cons. (Consec.)       &    -   &    -  &   -  &     -  &    - \\
+ No Energy/Outliers/RPT       &    -   &   -   &   -   &     -  &    - \\
+ Supplier Multi-Buyer        &    880   &    70  &   100  &     480  &    630 \\

\cmidrule(lr){1-6} 
\\[-0.3em]

\multicolumn{6}{c}{\textit{Panel B}: 2017--2018} \\
\cmidrule(lr){1-6}
\\[-0.6em]
All Imports                 &  2{,}000 &   330   & 1{,}300 &   2{,}500 &  5{,}300 \\
BEC – All (Consec.)   &  1{,}600 &   160  &   530  &     890  &  1{,}800 \\
BEC – Non-Cons. (Consec.)        &  1{,}000 &   120   &   320   &     540  &    950 \\
+ No Energy/Outliers/RPT       &    420   &   110   &   270   &     470  &    730 \\
+ Supplier Multi-Buyer       &    160 &   71   &   43   &     190  &    250 \\
\\[-0.9em]
\cmidrule(lr){1-6}
\end{tabular}
\arrayrulecolor{black}
}
\vspace{1em}
\noindent
\begin{minipage}{16cm}\vspace{0.5em}
\footnotesize \textit{Notes}: This table reports sample characteristics for a series of progressively restricted datasets used in the empirical analysis. The first row (“All Imports”) includes all US import records in the sample period. All subsequent rows restrict the sample to buyer–supplier pairs that trade the same HS-10 product in two consecutive calendar years. The “BEC – excl. Cons. (Consec.)” sample includes only capital and intermediate goods, excluding consumption goods as defined by the Broad Economic Categories (BEC) classification. The next sample (“+ No Energy/Outliers”) adds four filters: (i) transactions involving energy-sector goods are excluded; (ii) observations with price levels below the 1st percentile or above the 99th percentile of the within-product price distribution are removed; (iii) extreme log price changes (above 4 or below –4) are excluded; and (iv) related-party transactions, defined as pairs flagged as related by the ORBIS ownership links, are dropped. “+ Supplier Multi-Buyer” restricts to suppliers that trade with at least two different buyers in consecutive years for the same product. “Import value” denotes the total annual value of imports in billions of US dollars. “Importers” and “Exporters” correspond to distinct US buyers and foreign suppliers, respectively. All figures are reported separately for the 2001–2016 and 2017–2018 periods and are rounded to four significant digits in accordance with US Census Bureau disclosure guidelines. These samples are the exact ones used in the empirical analysis. \textit{Source}: FSRDC Project Number 2109 (CBDRB-FY25-P2109-R12520).
\end{minipage}
\end{table}

\vspace{1ex}
\begin{table}[!htbp]
\caption{Summary Statistics for Main Estimation Sample (2001--2018)}
\label{tab:summary_stats_main}
\centering
{\small
\renewcommand{\arraystretch}{1.2}
\arrayrulecolor{gray!50}
\begin{tabular}{
  >{\raggedright\arraybackslash}p{6.8cm} 
  >{\centering\arraybackslash}p{1.4cm} 
  >{\centering\arraybackslash}p{1.4cm} 
  >{\centering\arraybackslash}p{1.4cm} 
  >{\centering\arraybackslash}p{1.4cm} 
  >{\centering\arraybackslash}p{1.4cm}
}
\toprule
\text{Variable} & \text{Mean} & \text{Std. Dev.} & \text{P25} & \text{Median} & \text{P75} \\
\midrule
\addlinespace
\multicolumn{6}{c}{\textit{Panel A}: Trade Relationships} \\
\midrule
\addlinespace
\addlinespace
\quad $s_{ijh}$: Supplier share & 0.32 & 0.35 & 0.03 & 0.15 & 0.57 \\
\quad $x_{ijh}$: Buyer share & 0.25 & 0.29 & 0.02 & 0.10 & 0.40 \\
\quad Relationship length (product $h$) & 4.00 & 2.80 & 2.50 & 3.50 & 5.50 \\
\quad Relationship length (all products) & 4.80 & 3.30 & 2.50 & 4.50 & 6.50 \\
\quad \# Transactions (product $h$) & 120 & 1100 & 6.50 & 16 & 50 \\
\quad \# Transactions (all products) & 360 & 3000 & 11 & 36 & 140 \\
\quad \# Products per pair & 3.80 & 7.30 & 1.50 & 2.50 & 4.50 \\
\quad Multi-HS10 dummy & 0.59 & 0.49 & 0.00 & 1.00 & 1.00 \\
\quad \# Suppliers per buyer (HS10) & 1.80 & 3.20 & 1.50 & 2.50 & 5.50 \\
\quad Buyer tenure (all products) & 9.90 & 5.00 & 6.50 & 10.00 & 14.00 \\
\quad Buyer tenure (product $h$) & 6.90 & 4.40 & 3.50 & 6.50 & 10.00 \\
\quad \# Buyers per supplier (HS10) & 3.20 & 3.90 & 2.50 & 3.50 & 7.50 \\
\quad Supplier tenure (all products) & 8.00 & 4.60 & 4.50 & 8.50 & 12.00 \\
\quad Supplier tenure (product $h$) & 6.40 & 4.00 & 3.50 & 6.50 & 9.50 \\
\quad Corr.\ between $s_{ijh}$ and $x_{ijh}$ & 0.041 & --- & --- & --- & --- \\
\bottomrule
\addlinespace
\multicolumn{6}{c}{\textit{Panel B}: Prices (log unit values)} \\
\midrule
\addlinespace
\quad $\log p$ (pre-duty, excl. charges) & 3.40 & 2.80 & 1.30 & 3.00 & 5.40 \\
\quad $\log p$ (pre-duty) & 3.50 & 2.80 & 1.40 & 3.10 & 5.40 \\
\quad $\log p^{\text{duty}}$ (post-duty) & 3.50 & 2.80 & 1.40 & 3.10 & 5.40 \\

\bottomrule
\end{tabular}
}
\vspace{2em}
\begin{minipage}{17cm}\vspace{0.5em}
\footnotesize \textit{Notes}: This table extends Table~\ref{tab:summary_stats} with additional variables for the main estimation sample at the importer-supplier-product level. Full details on sample construction are provided in Appendix~\ref{subsec:Sample-Selection}. Panel~A reports trade relationship characteristics. $s_{ijh}$ is exporter $i$'s share in buyer $j$'s imports of product $h$; $x_{ijh}$ is buyer $j$'s share in exporter $i$'s US exports of product $h$. Relationship length (product $h$) is the number of years between the supplier-buyer pair's first transaction in product $h$ and year~$t$; Relationship length (all products) is defined analogously, using the pair's first transaction in any product. \# Transactions (product $h$) is the cumulative number of transactions between the pair in product $h$, from their first transaction through year~$t$; \# Transactions (all products) is defined analogously, summing across all products. \# Products per pair is the number of distinct HS10 products traded by the pair in year~$t$, and Multi-HS10 dummy equals one when this number exceeds one. \# Suppliers per buyer (HS10) is the number of distinct exporters from which buyer $j$ sources product $h$ in year~$t$; \# Buyers per supplier (HS10) is defined analogously, counting distinct buyers per exporter. Buyer tenure (product $h$) is the number of years between the buyer's first recorded transaction in product $h$ (with any supplier) and year~$t$; Buyer tenure (all products) is defined analogously, using the buyer's first recorded transaction in any product. Supplier tenure is defined analogously for the exporter. Corr.\ between $s_{ijh}$ and $x_{ijh}$ is the sample correlation between the two shares. Panel~B reports log unit values (value divided by quantity). $\log p$ (pre-duty, excl.\ charges) corresponds to the FOB (Free on Board) unit value, excluding shipping and insurance charges; $\log p$ (pre-duty) includes these charges to obtain the CIF (Cost, Insurance, and Freight) unit value; and $\log p^{\text{duty}}$ (post-duty) further incorporates import duties. Statistics are based on confidential LFTTD data and rounded according to US Census Bureau disclosure guidelines. \textit{Source}: FSRDC Project Number 2109 (CBDRB-FY25-P2109-R12520).
\end{minipage}
\end{table}

\begin{table}[htbp]
\caption{Sample Composition by Period – All BEC Categories}
\label{tab:sample_stats_bec0}
\centering
{\small
\renewcommand{\arraystretch}{1.2}
\arrayrulecolor{gray!40}
\begin{tabular}{
  >{\raggedright\arraybackslash}p{5cm} 
  >{\raggedright\arraybackslash}p{1.7cm} 
  >{\raggedright\arraybackslash}p{1.7cm} 
  >{\raggedright\arraybackslash}p{1.7cm} 
  >{\raggedright\arraybackslash}p{1.7cm} 
  >{\raggedright\arraybackslash}p{1.7cm}
}
\specialrule{.05em}{0em}{0em}
\makecell{Sample} & 
\makecell{Import\\value\\(bn USD)} & 
\makecell{Importers\\(th)} & 
\makecell{Exporters\\(th)} & 
\makecell{Pairs\\(th)} & 
\makecell{Triplets\\(th)} \\
\specialrule{.05em}{0em}{0.3em}
\\[-0.6em]

\multicolumn{6}{c}{\textit{Panel A}: 2001--2016} \\
\cmidrule(lr){1-6}
\\[-0.6em]
All Imports     & 22{,}000 & 1{,}000 & 6{,}900 & 17{,}000 & 40{,}000 \\
BEC – All (Consec.)      &    -   &   -   &  -    &   -    & - \\
+ No Energy/Outliers/RPT    &   -    &      &   -   &   -    & - \\
+ Supplier Multi-Buyer   &  1{,}600 &   110   &   210 &     990  &  1{,}500 \\

\cmidrule(lr){1-6} 
\\[-0.3em]

\multicolumn{6}{c}{\textit{Panel B}: 2017--2018} \\
\cmidrule(lr){1-6}
\\[-0.6em]
All Imports                       &  2{,}000 &   330   & 1{,}300 &   2{,}500 &  5{,}300 \\
BEC – All (Consec.)   &  1{,}600 &   160  &   530  &     890  &  1{,}800 \\
+ No Energy/Outliers/RPT             &    710   &   150   &   470   &     800  &  1{,}500 \\
+ Supplier Multi-Buyer             &    260   &    99   &    79   &     330  &    470 \\
\\[-0.9em]
\cmidrule(lr){1-6}
\end{tabular}
\arrayrulecolor{black}
}
\vspace{1em}
\noindent
\begin{minipage}{16cm}\vspace{0.5em}
\footnotesize 
\textit{Notes}: This table reports sample characteristics for a series of progressively restricted datasets used in the empirical analysis. The first row (“All Imports”) includes all US import records in the sample period. All subsequent rows restrict the sample to buyer–supplier pairs that trade the same HS-10 product in two consecutive calendar years. The “BEC – All Categories (Consec.)” sample retains all transactions in capital, intermediate, and consumption goods as defined by the Broad Economic Categories (BEC) system, subject to the consecutive-year condition. The next sample (“+ No Energy/Outliers”) adds four filters: (i) transactions involving energy-sector goods are excluded; (ii) observations with price levels below the 1st percentile or above the 99th percentile of the within-product price distribution are removed; (iii) extreme log price changes (above 4 or below –4) are excluded; and (iv) related-party transactions are dropped following ORBIS ownership links. “+ Supplier Multi-Buyer” restricts to suppliers that trade with at least two different buyers in consecutive years for the same product. “Import value” denotes the total annual value of imports in billions of US dollars. “Importers” and “Exporters” correspond to distinct US buyers and foreign suppliers, respectively.  All figures are reported separately for the 2001–2016 and 2017–2018 periods and are rounded to four significant digits in accordance with US Census Bureau disclosure guidelines. \textit{Source}: FSRDC Project Number 2109 (CBDRB-FY25-P2109-R12520).
\end{minipage}
\end{table}

\begin{table}[htbp]
\caption{Summary Statistics -- All BEC Categories (2001--2018)}
\label{tab:summary_stats_bec0}
\centering
{\small
\renewcommand{\arraystretch}{1.2}
\arrayrulecolor{gray!50}
\begin{tabular}{
  >{\raggedright\arraybackslash}p{6.8cm} 
  >{\centering\arraybackslash}p{1.4cm} 
  >{\centering\arraybackslash}p{1.4cm} 
  >{\centering\arraybackslash}p{1.4cm} 
  >{\centering\arraybackslash}p{1.4cm} 
  >{\centering\arraybackslash}p{1.4cm}
}
\toprule
\text{Variable} & \text{Mean} & \text{Std. Dev.} & \text{P25} & \text{Median} & \text{P75} \\
\midrule
\addlinespace
\multicolumn{6}{c}{\textit{Panel} A: Trade Relationships} \\
\midrule
\addlinespace
\addlinespace
\quad $s_{ijh}$: Supplier share & 0.27 & 0.32 & 0.02 & 0.10 & 0.43 \\
\quad $x_{ijh}$: Buyer share & 0.27 & 0.30 & 0.03 & 0.13 & 0.45 \\
\quad Relationship length (product $h$) & 3.90 & 2.60 & 2.50 & 3.50 & 5.50 \\
\quad Relationship length (all products) & 4.60 & 3.10 & 2.50 & 4.50 & 6.50 \\
\quad \# Transactions (product $h$) & 100 & 890 & 6.50 & 16 & 50 \\
\quad \# Transactions (all products) & 410 & 3200 & 13 & 45 & 180 \\
\quad \# Products per pair & 5.60 & 12.00 & 1.50 & 2.50 & 5.50 \\
\quad Multi-HS10 dummy & 0.68 & 0.47 & 0.00 & 1.00 & 1.00 \\
\quad \# Suppliers per buyer (HS10) & 2.00 & 3.60 & 1.50 & 2.50 & 6.50 \\
\quad Buyer tenure (all products) & 9.50 & 4.90 & 5.50 & 9.50 & 14.00 \\
\quad Buyer tenure (product $h$) & 6.80 & 4.30 & 3.50 & 6.50 & 10.00 \\
\quad \# Buyers per supplier (HS10) & 3.00 & 3.40 & 2.50 & 3.50 & 5.50 \\
\quad Supplier tenure (all products) & 7.70 & 4.40 & 4.50 & 7.50 & 11.00 \\
\quad Supplier tenure (product $h$) & 6.00 & 3.80 & 3.50 & 5.50 & 9.50 \\
\quad Corr.\ between $s_{ijh}$ and $x_{ijh}$ & 0.053 & --- & --- & --- & --- \\
\bottomrule
\addlinespace
\multicolumn{6}{c}{\textit{Panel B}: Prices (log unit values)} \\
\midrule
\addlinespace
\quad $\log p$ (pre-duty,  excl. charges) & 3.30 & 2.50 & 1.40 & 3.00 & 5.00 \\
\quad $\log p$ (pre-duty) & 3.40 & 2.50 & 1.50 & 3.10 & 5.10 \\
\quad $\log p^{\text{duty}}$ (post-duty) & 3.40 & 2.50 & 1.50 & 3.10 & 5.20 \\

\bottomrule
\end{tabular}
}
\begin{minipage}{17cm}\vspace{0.5em}
\footnotesize \textit{Notes}: This table reports summary statistics for the pooled sample, defined analogously to the main estimation sample (Table~\ref{tab:summary_stats}) but without restricting on BEC end-use category, at the importer-supplier-product level. Full details on sample construction are provided in Appendix~\ref{subsec:Sample-Selection}. Panel~A reports trade relationship characteristics. $s_{ijh}$ is exporter $i$'s share in buyer $j$'s imports of product $h$; $x_{ijh}$ is buyer $j$'s share in exporter $i$'s US exports of product $h$. Relationship length (product $h$) is the number of years between the supplier-buyer pair's first transaction in product $h$ and year~$t$; Relationship length (all products) is defined analogously, using the pair's first transaction in any product. \# Transactions (product $h$) is the cumulative number of transactions between the pair in product $h$, from their first transaction through year~$t$; \# Transactions (all products) is defined analogously, summing across all products. \# Products per pair is the number of distinct HS10 products traded by the pair in year~$t$, and Multi-HS10 dummy equals one when this number exceeds one. \# Suppliers per buyer (HS10) is the number of distinct exporters from which buyer $j$ sources product $h$ in year~$t$; \# Buyers per supplier (HS10) is defined analogously, counting distinct buyers per exporter. Buyer tenure (product $h$) is the number of years between the buyer's first recorded transaction in product $h$ (with any supplier) and year~$t$; Buyer tenure (all products) is defined analogously, using the buyer's first recorded transaction in any product. Supplier tenure is defined analogously for the exporter. Corr.\ between $s_{ijh}$ and $x_{ijh}$ is the sample correlation between the two shares. Panel~B reports log unit values (value divided by quantity). $\log p$ (pre-duty, excl.\ charges) corresponds to the FOB (Free on Board) unit value, excluding shipping and insurance charges; $\log p$ (pre-duty) includes these charges to obtain the CIF (Cost, Insurance, and Freight) unit value; and $\log p^{\text{duty}}$ (post-duty) further incorporates import duties. Statistics are based on confidential LFTTD data and rounded according to US Census Bureau disclosure guidelines. \textit{Source}: FSRDC Project Number 2109 (CBDRB-FY25-P2109-R12520).
\end{minipage}
\end{table}

\section{Additional Empirical Results\label{sec: Additional Tables and Figures}}

\subsection{Decomposition of Price Dispersion}\label{sec:price_dispersion}

To explore the sources of price heterogeneity, we report in Table \ref{fact_2} the results from OLS regressions decomposing price variation using the specification:
\[
\ln p_{ijht} = FE_i + FE_j + FE_{ht} + \beta \boldsymbol{X}_{ijht} + \varepsilon_{ijht},
\]
estimated over the period 2001-2016.
We consider three alternative prices: prices that exclude duties ($\ln p_{ijht}$), prices that exclude duties and charges ($\ln p^{c}_{ijht}$), and prices that include duties and charges ($\ln p^{\text{duty}}_{ijht}$). 

Table~\ref{fact_2} finds that controlling for product and year fixed effects explains approximately 50\% of the overall price dispersion, while 4\% is attributed to match-specific residuals. Notably, this figure changes substantially when isolating variation within supplier–product–year combinations (Panel B), with the buyer–supplier match accounting for 77\% of the price variance. This emphasizes that a significant share of price heterogeneity stems from bilateral characteristics that are not solely attributable to either buyer or supplier individually.

\begin{table}[H]
\caption{Fixed-Effect Decomposition of Price Dispersion}
\label{fact_2}
\centering
{\small
\renewcommand{\arraystretch}{1.2}
\arrayrulecolor{gray!50}
\begin{tabular}{
  >{\raggedright\arraybackslash}p{6.2cm} 
  >{\centering\arraybackslash}p{2.1cm} 
  >{\centering\arraybackslash}p{2.1cm} 
  >{\centering\arraybackslash}p{2.1cm}
}
\toprule
Source of Variation &  $\ln p_{ijht}$ & $\ln p^{c}_{ijht}$ & $\ln p^{\text{duty}}_{ijht}$ \\
\midrule
\addlinespace
\addlinespace

\multicolumn{4}{c}{\textit{Panel A}: Overall price dispersion} \\
\midrule
\addlinespace
\addlinespace
$FE_{ht}$              & 0.485 & 0.483   &   0.486            \\
$FE_{i}$               & 0.424 & 0.427   &   0.423            \\
$FE_{j}$               & 0.0464 & 0.0452 &   0.0463            \\
Match residual       & 0.0441  & 0.0444  &   0.0442            \\
\midrule
\addlinespace
\addlinespace
\multicolumn{4}{c}{\textit{Panel B}: Within exporter–product-year dispersion} \\
\midrule
\addlinespace
\addlinespace
$FE_{j}$               & 0.233 & 0.231     &  0.233             \\
Match residual         & 0.765  & 0.768   &    0.765           \\
\bottomrule
\end{tabular}
}
\medskip
\begin{minipage}{16cm}\vspace{0.5em}
\footnotesize
\textit{Notes}: The columns correspond to alternative price definitions: $\ln p_{ijht}$ excludes duties; $\ln p^{c}_{ijht}$ exclude both duties and charges; $\ln p^{\text{duty}}_{ijht}$ includes duties. The estimation sample includes importer–exporter–product matches observed in two consecutive calendar years, and applies the following restrictions: (i) excludes transactions involving consumption goods (based on the BEC classification), energy-sector products, statistical outliers, and related-party trade; and (ii) retains only suppliers that trade with at least two distinct US buyers in consecutive years. This corresponds to the final sample in Panel A of Table~\ref{tab:sample_stats_bec1}. The control vector $\boldsymbol{X}_{ijht}$ includes the log of transaction value, the log of relationship longevity (years since the exporter first supplied the buyer with the given HS10 product), and the log of the relative number of partners (the supplier’s number of HS10-level buyers divided by the buyer’s number of HS10-level suppliers). The sample includes 1,200,000 importer–exporter–product–year observations, which have been rounded to four significant digits per US Census Bureau
disclosure guidelines. $ R^2 = 0.956$. All coefficients in a regression model are significantly different from zero at the 1\% significance level. \textit{Source}: FSRDC Project Number 2109 (CBDRB-FY25-P2109-R12520).
\end{minipage}
\end{table}

\clearpage

\subsection{Pass-Through Heterogeneity}\label{sec:PT_heterogeneity}

Table \ref{Table_PT_strictFE} examines heterogeneity in tariff pass-through using specification \eqref{eq:pass_through} with a more demanding set of fixed effects. In addition to product–time fixed effects ($FE_{ht}$), the specification includes importer–time and exporting country–time fixed effects ($FE_{jt} + FE_{ct}$), accounting for time-varying shocks at the buyer level while flexibly controlling for exporter-side conditions. 

Findings replicate the incomplete tariff pass-through estimates reported in Table~\ref{Table_PT_FEhtcs}, consistent with exporters adjusting marginal costs in response to demand shifts from large buyers and thereby absorbing a substantial fraction of tariff shocks. In line with Proposition \ref{prop:4}, pass-through declines with the buyer share $x_{ij}$, highlighting the central role of the cost channel.

\begin{table}[H]
\caption{Pass-Through and Relationship Heterogeneity, Stringent Fixed Effects\label{Table_PT_strictFE}}
\centering
{\small
\renewcommand{\arraystretch}{1.2}
\begin{tabular}{
  >{\raggedright\arraybackslash}p{5cm} 
  >{\centering\arraybackslash}p{1.5cm} 
  >{\centering\arraybackslash}p{1.5cm} 
  >{\centering\arraybackslash}p{1.5cm} 
  >{\centering\arraybackslash}p{1.5cm} 
  >{\centering\arraybackslash}p{1.5cm} 
  >{\centering\arraybackslash}p{1.5cm} 
}
\toprule
Dependent variable: & \multicolumn{6}{c}{$\Delta\ln p_{ijht}$} \\
\cmidrule(lr){2-7}
& (1) & (2) & (3) & (4) & (5) & (6) \\
\midrule
$\Delta\ln(1+\tau_{cht})$ & -0.184 & -0.306 & -0.183 & -0.113 & -0.114 & -0.247 \\
                          & (0.107) & (0.155) & (0.093) & (0.126) & (0.104) & (0.145) \\
$\Delta\ln(1+\tau_{cht}) \cdot \ln \text{longevity}_{ijht}$ 
                          &        & 0.088   &         &         &         & 0.101   \\
                          &        & (0.047) &         &         &         & (0.047) \\
$\Delta\ln(1+\tau_{cht}) \cdot s_{ijht-1}$ 
                          &        &         & -0.003   &         & 0.004   & -0.007   \\
                          &        &         & (0.149) &         & (0.165) & (0.165) \\
$\Delta\ln(1+\tau_{cht}) \cdot x_{ijht-1}$ 
                          &        &         &         & -0.282  & -0.280  & -0.289  \\
                          &        &         &         & (0.125) & (0.134) & (0.128) \\
\midrule
$FE_{ht} + FE_{ct} + FE_{jt}$ & Yes & Yes & Yes & Yes & Yes & Yes \\
\midrule
Observations              & 249,000 & 249,000 & 249,000 & 249,000 & 249,000 & 249,000 \\
R-squared                 & 0.31   & 0.31    & 0.31    & 0.31    & 0.31    & 0.31    \\
\bottomrule
\end{tabular}
}
\begin{minipage}{17cm}\vspace{0.5em}
\footnotesize
\textit{Notes:} This table reports estimates of the pass-through of statutory tariffs, $\Delta \ln (1 + \tau_{cht})$, to duty-exclusive CIF prices at the exporter–importer–product–year level, $\Delta \ln p_{ijht}$. Columns (2) and (6) interact tariffs with the log of relationship longevity, measured as the number of years that buyer $j$ and supplier $i$ have transacted in product $h$. Columns (3) and (5) interact tariffs with the lagged supplier share, $s_{ijht-1}$, defined as supplier $i$'s share in buyer $j$'s imports of product $h$. Columns (4) and (5) interact tariffs with the lagged buyer share, $x_{ijht-1}$, defined as buyer $j$'s share in supplier $i$'s exports of product $h$. All regressions include product–year, exporter country–year, and importer–year fixed effects ($FE_{ht} + FE_{ct} + FE_{jt}$). Controls include: (i) $\ln \text{longevity}_{ijht}$; (ii) $\Delta \ln q_{i(-j)ht}$, the change in exporter $i$'s total sales of $h$ to US buyers other than $j$; and (iii) $\Delta \ln p_{(-i)jht}$, the weighted average price change charged by other suppliers of $h$ to buyer $j$, using lagged shares as weights. Standard errors are clustered at the HS8 product and exporter-country level. The sample corresponds to the "+ Supplier Multi-Buyer" definition in Table~\ref{tab:sample_stats_bec1}. Observation counts are rounded to four significant digits per US Census Bureau disclosure guidelines. \textit{Source}: FSRDC Project Number 2109 (CBDRB-FY25-P2109-R12520).
\end{minipage}
\end{table}

\paragraph{Nonlinear Effects.}

\begin{figure}[H]
    \centering
    \caption{Pass-Through by Bilateral Market Share Quartiles}
    \label{fig:PT_nonlinearities}
    \vspace{1em}

    \textbf{(A) Fixed Effects: $FE_{ht} + FE_{cs}$} \par
    \vspace{0.5em}
    \includegraphics[width=0.8\linewidth]{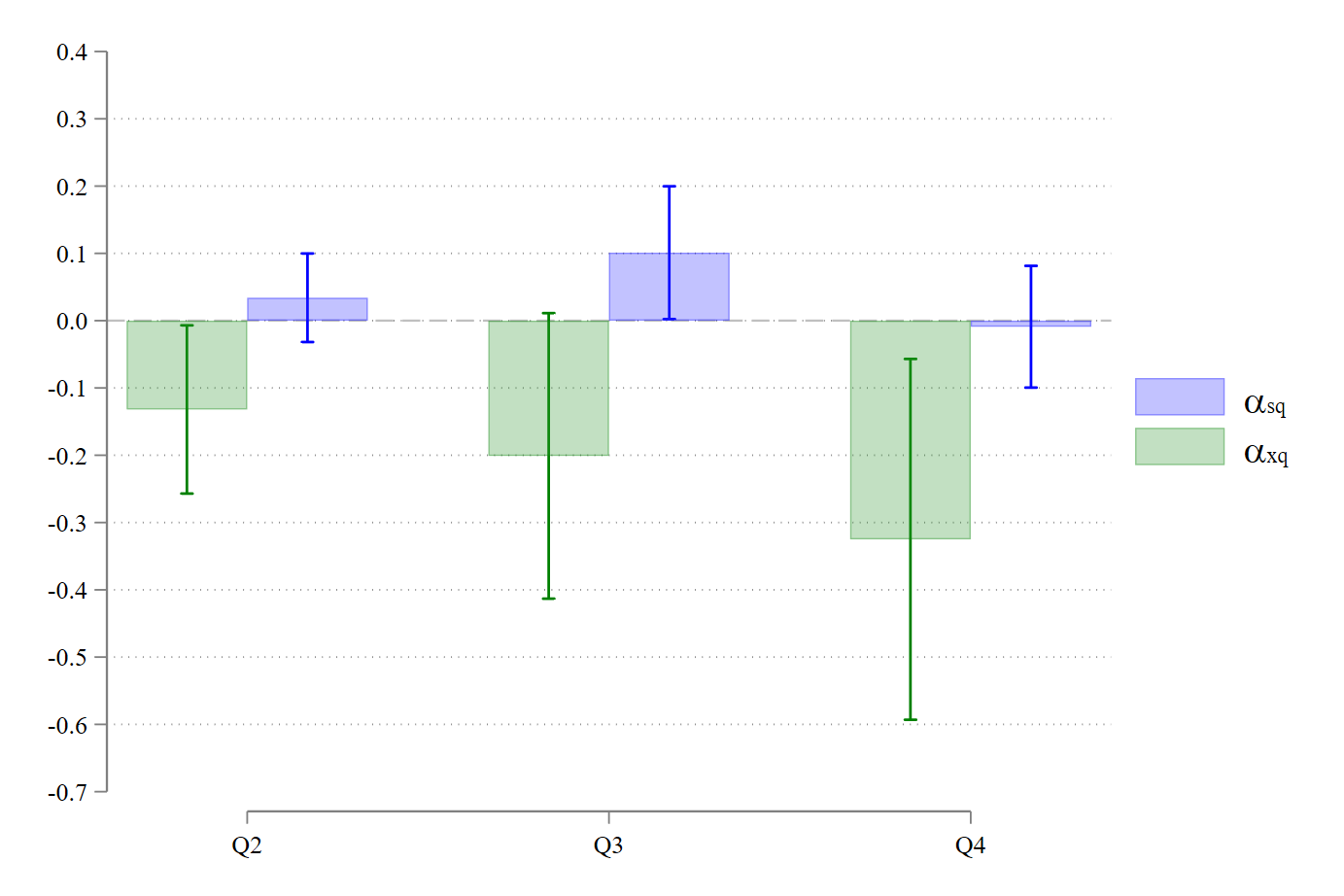}

    \vspace{1em}

    \textbf{(B) Fixed Effects: $FE_{ht} + FE_{ct} + FE_{jt}$} \par
    \vspace{0.5em}
    \includegraphics[width=0.83\linewidth]{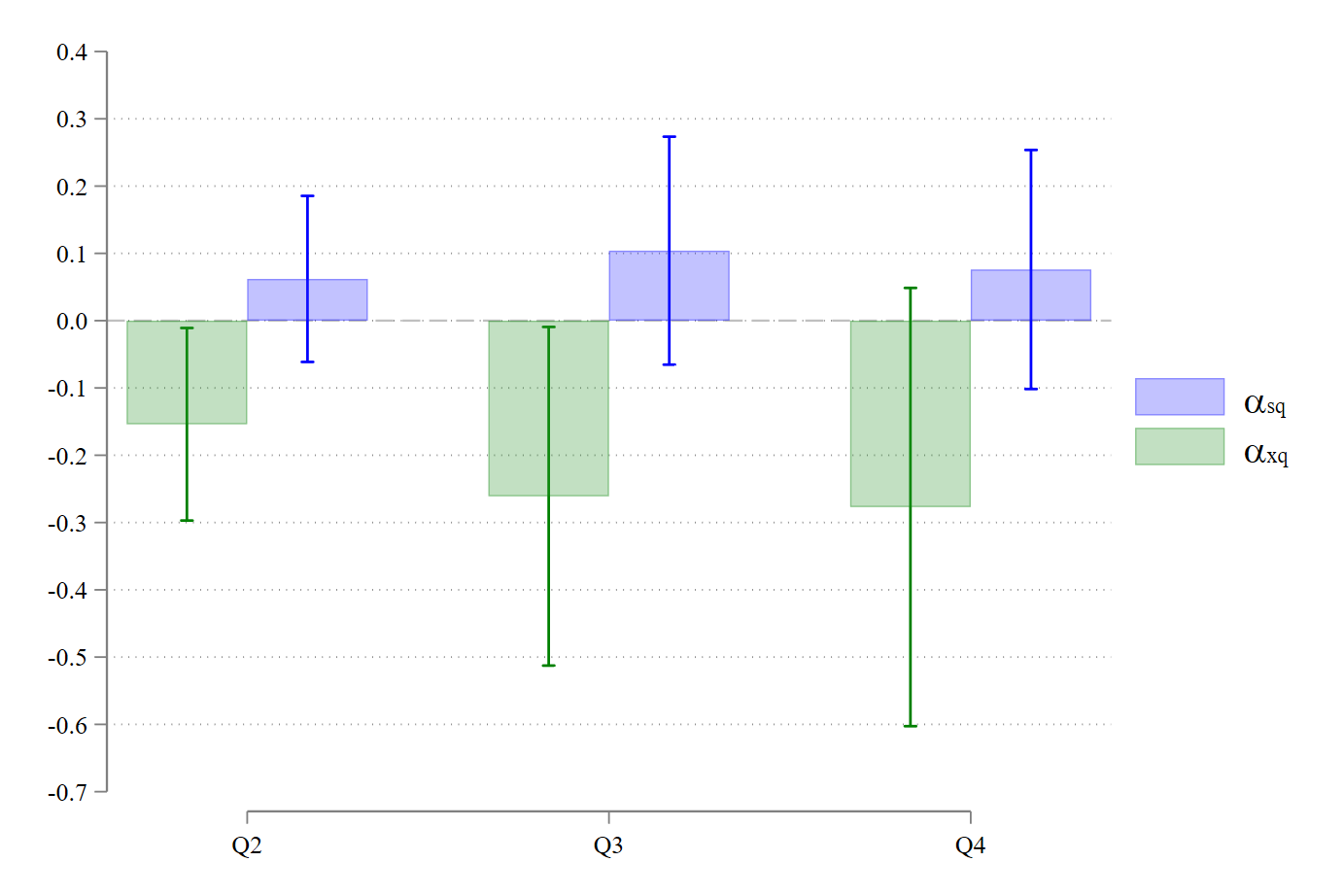}

    \vspace{1em}

    \begin{minipage}{1.0\linewidth}
    \footnotesize
    \textit{Notes}: The figure plots estimated coefficients from regressions of bilateral price changes on tariff changes interacted with quartiles of supplier share ($s_{ijht}$) and buyer share ($x_{ijht}$). The first quartile serves as the omitted category. The estimated coefficients correspond to equation~\eqref{eq:pass_through_quartiles}, where $\alpha_{s,q}$ and $\alpha_{x,q}$ capture the interaction of the tariff term with the $q$th quartile of supplier and buyer shares, respectively. The top panel includes product-year and exporting country-sector fixed effects. The bottom panel includes product-year, importer-year, and exporting country-year fixed effects. Bars indicate 90\% confidence intervals. \textit{Source}: FSRDC Project Number 2109 (CBDRB-FY25-P2109-R12520).
    \end{minipage}

\end{figure}

To explore nonlinearities in tariff pass-through along the distribution of bilateral concentration, we interact the tariff change with quartile dummies of the lagged supplier share (\(s_{ijh,t-1}\)) and buyer share (\(x_{ijh,t-1}\)). Specifically, we estimate \eqref{eq:pass_through_quartiles}, where \(\mathbf{1}\{s_{ijh,t-1} \in Q_q\}\) and \(\mathbf{1}\{x_{ijh,t-1} \in Q_q\}\) are indicator variables for quartiles \(q=2,3,4\), with the first quartile serving as the omitted category. To separate level and interaction effects, the regression also includes the shares \(s_{ijh,t-1}\) and \(x_{ijh,t-1}\) themselves as covariates. This specification allows us to test whether pass-through varies nonlinearly across the concentration distribution, while flexibly controlling for underlying differences in market structure.
Panel (A) of Figure~\ref{fig:PT_nonlinearities} includes product–time and exporting country–sector fixed effects, while Panel (B) features product–time, importer–time, and exporting country–time fixed effects. \begin{align}
\Delta \ln p_{ijht} =\ & \alpha_0 + \alpha_1 \Delta \ln(1 + \tau_{cht}) 
+ \sum_{q=2}^{Q} \alpha_{s,q} \cdot \Delta \ln(1 + \tau_{cht}) \cdot \mathbf{1}\{s_{ijh,t-1} \in Q_q\} \notag \\
& + \sum_{q=2}^{Q} \alpha_{x,q} \cdot \Delta \ln(1 + \tau_{cht}) \cdot \mathbf{1}\{x_{ijh,t-1} \in Q_q\} 
+ \gamma' \textbf{X}_{ijht} + \textbf{FE} + \epsilon_{ijht}. \label{eq:pass_through_quartiles}
\end{align}

The results reveal no systematic pattern in pass-through across supplier-share quartiles (in purple): the interaction coefficients are small, of either sign, and statistically insignificant. Buyer shares, by contrast, display a monotonic pattern (in green): the interaction coefficients fall from the second to the fourth quartile under both sets of fixed effects, although the individual coefficients are imprecisely estimated and only the top quartile under the baseline fixed effects is significant at the 5 percent level. The pattern is consistent with the model's prediction that pass-through declines with the buyer share.

\clearpage

\paragraph{Additional Robustness: GE Controls and Price Definitions.}

Table~\ref{tab:robustness_pass_through} presents robustness checks using alternative price definitions and specifications. Columns (1)-(2) exclude the general equilibrium controls included in the baseline model. Columns (3)-(4) use duty-exclusive prices that \textit{exclude} charges, while Columns (5)-(6) use tariff-inclusive prices. In all cases, tariff changes are interacted with lagged supplier and buyer shares, and the set of fixed effects is held constant to ensure comparability across specifications.  The negative buyer-share interaction is stable across these variants, but the main tariff coefficient differs with the price definition. Recorded-duty-inclusive prices are not constructed by adding the statutory tariff change to duty-exclusive prices, so the main coefficients need not differ by exactly one.

\begin{table}[H]
\caption{Additional Robustness: General Equilibrium Controls and Price Definitions}
\label{tab:robustness_pass_through}
\centering
{\small
\renewcommand{\arraystretch}{1.2}
\begin{tabular}{
  >{\raggedright\arraybackslash}p{4.6cm} 
  >{\centering\arraybackslash}p{1.5cm} 
  >{\centering\arraybackslash}p{1.5cm} 
  >{\centering\arraybackslash}p{1.5cm} 
  >{\centering\arraybackslash}p{1.5cm} 
  >{\centering\arraybackslash}p{1.5cm} 
  >{\centering\arraybackslash}p{1.5cm}
}
\toprule
Dependent variable: 
& \multicolumn{2}{c}{\makecell{$\Delta \ln p_{ijht}$ \\ (excl. GE \\controls)}} 
& \multicolumn{2}{c}{\makecell{$\Delta \ln p^{c}_{ijht}$ \\ (before duty, \\ excl. charges)}} 
& \multicolumn{2}{c}{\makecell{$\Delta \ln p^{\text{duty}}_{ijht}$ \\ (tariff inclusive)}} \\
\cmidrule(lr){2-3} \cmidrule(lr){4-5} \cmidrule(lr){6-7}
& (1) & (2) & (3) & (4) & (5) & (6) \\
\midrule
$\Delta\ln(1+\tau_{cht})$ & -0.025 & -0.103 & -0.066 & -0.163 & 0.506 & 0.401 \\
                          & (0.090) & (0.103) & (0.093) & (0.107) & (0.099) & (0.126) \\
$\Delta\ln(1+\tau_{cht}) \cdot s_{ijht-1}$ 
                          & 0.027 & 0.011 & 0.054 & 0.029 & 0.017 & 0.007 \\
                          & (0.076) & (0.170) & (0.072) & (0.157) & (0.068) & (0.143) \\
$\Delta\ln(1+\tau_{cht}) \cdot x_{ijht-1}$ 
                          & -0.403 & -0.287 & -0.403 & -0.271 & -0.421 & -0.287 \\
                          & (0.112) & (0.135) & (0.113) & (0.135) & (0.121) & (0.145) \\
\midrule
$FE_{ht} + FE_{cs}$         & Yes & No  & Yes & No  & Yes & No  \\
$FE_{ht} + FE_{ct} + FE_{jt}$ & No  & Yes & No  & Yes & No  & Yes \\
\midrule
Observations                & 249,000 & 249,000 & 249,000 & 249,000 & 249,000 & 249,000 \\
R-squared                   & 0.04 & 0.31 & 0.04 & 0.31 & 0.05 & 0.31 \\
\bottomrule
\end{tabular}
}
\vspace{2em}
\begin{minipage}{17.3cm}\vspace{0.5em}
\footnotesize
\textit{Notes:} This table reports robustness checks on tariff pass-through specifications using alternative price definitions and control sets. Columns (1)–(2) exclude general equilibrium controls; Columns (3)–(4) use pre-duty prices excluding charges; and Columns (5)–(6) use tariff-inclusive prices. In each case, we report specifications using either baseline fixed effects ($FE_{ht} + FE_{cs}$) or a more stringent set of fixed effects ($FE_{ht} + FE_{ct} + FE_{jt}$). Standard errors are clustered at the HS8 product and exporter-country level. The number of observations is rounded to four significant digits in accordance with US Census Bureau disclosure guidelines. \textit{Source}: FSRDC Project Number 2109 (CBDRB-FY25-P2109-R12520).
\end{minipage}
\end{table}

\clearpage
\subsection{Additional Pass-Through Results}\label{sec:additional_PT}

\subsubsection{Pass-Through Across Samples}\label{sec:PT_duke}
Table \ref{tab:price_changes_sample_CTCS}  examines how tariff pass-through varies across sample restrictions under the baseline fixed effects (Panel~A) and the preferred stringent fixed effects (Panel~B). In each panel, Column~(1) reports the model-implied response and Column~(2) reports the observed response in the baseline sample. Panel~A, Columns~(1)--(2), reports the average model and data moments used in the main-text out-of-sample evaluation (Table~\ref{tab:table5_new}), in reverse order; Panel~B reports the same moments under the more stringent fixed effects (Table~\ref{tab:table5_new_stringentFE}).

As the sample is progressively broadened--from including consumption goods (column 3), to incorporating single-buyer matches (column 4), and finally to the most inclusive specification, which adds related-party transactions, energy goods, and outliers (column 5)--estimated pass-through rises, reaching 94–96 percent. Under stringent fixed effects, the increase is monotonic, although pairwise comparisons across columns are not statistically significant.

These results show that sample composition importantly affects measured pass-through. At the 5\% level, complete pass-through cannot be rejected in the baseline sample under either fixed-effects specification, although the stringent fixed-effects estimate rejects it at the 10\% level. Adding single-buyer suppliers improves precision and leads us to reject complete pass-through in Column~(4) under both specifications. In the broadest sample, however, complete pass-through cannot be rejected in Column~(5) under either specification. To reconcile these results with the broader evidence, we next aggregate the data to the product level and exploit monthly variation.

\begin{table}[H]
\caption{Tariff Pass-Through Across Different Samples\label{tab:price_changes_sample_CTCS}}
\centering
{\small
\renewcommand{\arraystretch}{1.2}
\arrayrulecolor{gray!50}
\begin{tabular}{
  >{\raggedright\arraybackslash}p{3.1cm} 
  >{\centering\arraybackslash}p{2cm} 
  >{\centering\arraybackslash}p{2cm} 
  >{\centering\arraybackslash}p{2.8cm} 
  >{\centering\arraybackslash}p{2cm} 
  >{\centering\arraybackslash}p{2cm} 
}
\toprule

Dependent variable:  & Model & Baseline &  + Consumption &   +Suppliers  & + Energy/  \\
$\Delta \ln p_{ijht}$ &  &  & Goods & w/  & RPT/   \\
&  & & & $<$ 2 Buyers &  Outliers   \\

\midrule
& (1) & (2) & (3) & (4) & (5) \\
\midrule
\addlinespace[1.5em]
\multicolumn{6}{c}{\textit{Panel A}: Baseline Fixed Effects} \\
\midrule
$\Delta \ln(1+\tau_{cht})$ & 
-0.222 & -0.126 & -0.151 & -0.138 & -0.060 \\
& (0.008) & (0.091) & (0.089) & (0.034) & (0.044) \\
\midrule
$FE_{ht} + FE_{cs}$ & Yes & Yes & Yes & Yes & Yes \\
R-squared & 0.31 & 0.04 & 0.04 & 0.02 & 0.02 \\
\midrule
\addlinespace[1.5em]
\multicolumn{6}{c}{\textit{Panel B}: Stringent Fixed Effects} \\
\midrule
$\Delta \ln(1+\tau_{cht})$ & -0.226 & -0.184 & -0.118 & -0.083 & -0.036 \\
& (0.009) & (0.107) & (0.075) & (0.032) & (0.042) \\
\midrule
$FE_{ht} + FE_{ct} + FE_{jt}$ & Yes & Yes & Yes & Yes & Yes \\
R-squared & 0.48 & 0.31 & 0.26 & 0.14 & 0.12 \\
\midrule
\addlinespace[1em]
Observations & 249,000 & 249,000 & 473,000   & 1,490,000 & 1,768,000 \\
\bottomrule
\end{tabular}
}
\begin{minipage}{\textwidth} \vspace{0.75em}
\footnotesize
\textit{Notes}: This table reports tariff pass-through estimates to duty-exclusive CIF prices at the exporter–importer–product level. Panel~A uses baseline fixed effects: product–time and country–sector ($FE_{ht} + FE_{cs}$). Panel~B uses a more stringent specification: product–time, country–time, and buyer–time ($FE_{ht} + FE_{ct} + FE_{jt}$). Column (1) uses the model-predicted price change as the dependent variable. Column (2) uses the observed price change in the baseline sample; in Panel~A it is identical to Column (1) of Table~\ref{Table_PT_FEhtcs}, while Panel~B uses the stringent fixed effects stated above. Column (3) expands the baseline sample to include consumption goods. Column (4) further includes relationships involving suppliers that trade with fewer than two US buyers. Column (5) additionally includes energy commodities, related-party transactions, and observations with extreme price levels or changes. Standard errors are clustered at the HS8 product and country level. Observation counts are rounded per US Census Bureau disclosure guidelines. \textit{Source}: FSRDC Project Number 2109 (CBDRB-FY25-P2109-R12520).
\end{minipage}
\end{table}

\subsubsection{Product-Level Pass-Through Estimates \label{sec: pass-through_and_literature}}

To complement our main analysis, we replicate a standard pass-through specification using data aggregated at the product-country-month level.
We construct the data directly from the same underlying buyer-supplier-product triplets that form the basis of our firm-level regressions. 

While much of the literature analyzes monthly price and tariff changes at the product–country level, such approaches reflect both intensive and extensive margin adjustments, including changes in trading partners or the entry and exit of relationships. 
In contrast, our aggregation focuses exclusively on consecutive transactions between the same buyer and supplier for a given product. 
This restricts the product-level aggregate to transactions within ongoing relationships, so that the comparison with Table~\ref{tab:reconciliation}, Panel~A, changes only the unit of measurement and the frequency, not the set of relationships.

\paragraph{Data} We measure price changes at the HS10–source-country–month level using the same set of buyer–supplier–product links as in the baseline analysis, restricted to relationships with at least one transaction in both 2017 and 2018. For each relationship, we retain the month in which transactions occur in both years. We then aggregate values and quantities across these transactions to the product–source-country–month level, construct unit values, and compute month-to-month  price changes.

Relative to Table~\ref{tab:price_changes_sample_CTCS}, this procedure introduces two changes: prices are aggregated across trading relationships to the product–source-country level, and price changes are measured at the monthly frequency rather than annually. The objective of Table~\ref{tab:price_changes_sample_product_level} is to evaluate how pass-through estimates change when prices are aggregated across firms within a product–source-country–month cell, while maintaining the same underlying set of buyer–supplier links observed in both 2017 and 2018. 

Column~(1) reports results for the baseline sample. The sample is then progressively broadened along the same margins as in Table~\ref{tab:price_changes_sample_CTCS}: Column~(2) adds consumption goods, Column~(3) allows suppliers serving only one buyer, Column~(4) further includes related-party transactions, energy goods, and extreme price changes, and Column~(5) additionally incorporates non-consecutive buyer–supplier relationships.

\paragraph{Results} When the aggregation is performed using the same set of repeated buyer–supplier relationships as in the baseline analysis, pass-through is well below one (35 percent) and significantly so. As the sample is progressively broadened to include consumption goods, single-buyer suppliers, related-party transactions, energy goods, and eventually non-consecutive relationships, estimated pass-through increases steadily, to 65, 94, 95, and 98 percent. From Column~(3) onward, the estimates are statistically indistinguishable from complete pass-through, in line with the literature.

Differences between match-level and product-level pass-through therefore primarily reflect the set of relationships included in the data rather than aggregation itself. Because the columns are cumulative, the exercise cannot isolate the contribution of any single restriction.

\begin{table}[H]
\caption{Product-Level Tariff Pass-Through Across Alternative Samples\label{tab:price_changes_sample_product_level}}
\centering
{\small
\renewcommand{\arraystretch}{1.2}
\arrayrulecolor{gray!50}
\begin{tabular}{
  >{\raggedright\arraybackslash}p{3.6cm} 
  >{\centering\arraybackslash}p{2.1cm} 
  >{\centering\arraybackslash}p{2.8cm} 
  >{\centering\arraybackslash}p{2.1cm} 
  >{\centering\arraybackslash}p{2.1cm} 
  >{\centering\arraybackslash}p{2.5cm} 
}
\toprule

Dependent variable:  & Baseline &  + Consumption   &  + Suppliers &  + Energy/ & + Non- \\
$\Delta \ln p_{cht}$ &    & Goods  & w/& RPT/  &Consecutive  \\
&   & & $<$ 2 Buyers &  Outliers  & Relationships \\

\midrule
& (1) & (2) & (3) & (4) & (5) \\
\midrule
\addlinespace[1.5em]
\
$\Delta \ln(1+\tau_{cht})$ & 
   -0.654 & -0.353 & -0.063 &  -0.054 & -0.017   \\
  &  (0.245) & (0.156) & (0.073) & (0.068) & (0.068)   \\
\midrule

$FE_{ht} + FE_{cs}$ & Yes & Yes & Yes & Yes & Yes \\
R-squared &    0.23 & 0.22 & 0.14 & 0.13 & 0.11 \\
\midrule

\addlinespace[1em]
Observations &  124,000 & 213,000 & 665,000 & 805,000 &  1,071,000 \\
\bottomrule
\end{tabular}
}
\begin{minipage}{1.05\textwidth} \vspace{0.75em}
\footnotesize
\textit{Notes}: This table reports product-level pass-through estimates. The outcome is $\Delta\ln p_{cht}$ for product $h$, source country $c$, and time $t$ at monthly frequency. In Columns (1)--(4), product-level observations are constructed from firm-buyer-HS relationships that transact in two consecutive calendar years. Values and quantities are aggregated to the product–source-country–month level, unit values are constructed at that frequency, and month-to-month changes in log CIF prices are computed. Column (1) aggregates the baseline sample of Table~\ref{Table_PT_FEhtcs}, restricted to repeated arm's-length relationships in intermediate and capital goods among multi-buyer suppliers, to the product--source-country level. Column (2) broadens the sample to include consumption goods (all BEC categories). Column (3) further adds suppliers that trade with only one US importer. Column (4) further includes related-party transactions, energy commodities, and observations with extreme price levels or changes. Column (5) removes the consecutive-year restriction and includes all exporter–importer–product combinations. %
Standard errors are clustered at the HS8 product and country level. Observation counts are rounded in accordance with US Census Bureau disclosure guidelines. \textit{Source}: FSRDC Project Number 2109 (CBDRB-FY25-P2109-R12520).
\end{minipage}
\end{table}
\pagebreak

\subsection{Additional Results on Model Estimation}\label{sec:model_estimation_robustness}

This section complements Section~\ref{subsec:Estimation-results}. We first report an additional diagnostic of the model's mechanism. We then report the full estimates underlying Table~\ref{tab:robust_model_estimates_d} in the main text.

\paragraph{Simulated Markups and Bilateral Concentration.} Table~\ref{Table_simulated_markups} repeats the OLS price regressions of Table~\ref{Table 2 - Prices} using model-implied markups. With exporter--product--year fixed effects, the buyer-share coefficient is $-0.584$ in the model and $-0.617$ in the data, while the supplier-share coefficient is $0.023$ versus $0.237$. The model thus reproduces the buyer-share coefficient. The supplier-share coefficient is positive in both, though smaller in the model; Section~\ref{subsec:ModelTest} explains why its size is a poor guide. At $\phi=0$, the markup has no structural dependence on buyer share; small nonzero buyer-share coefficients can nevertheless arise in a linear projection of the nonlinear markup function.

\paragraph{Robustness of Structural Estimates}
We assess the robustness of our structural estimates in Section~\ref{subsec:Estimation-results}.
We first re-estimate the main GMM specification using an expanded sample that includes all Broad Economic Categories (BECs). We also estimate the structural parameters on a sample restricted to consumption goods only. Results are reported in Tables~\ref{tab:model_primitives_robust_bec0} and \ref{tab:model_primitives_robust_bec2}, respectively

Estimated returns to scale are systematically higher for consumption goods. Restricting the estimation to consumption goods raises $\hat{\theta}$ from 0.42–0.51 in the baseline to 0.54–0.66. Expanding the sample to include all BEC categories also increases estimates, with $\hat{\theta}$ ranging from 0.48 to 0.60, consistent with the higher values observed for consumption goods. This pattern is consistent with our interpretation of supply curvature as short-run capacity constraints, which are more severe for intermediate inputs than for consumption goods.

Finally, Table \ref{tab:model_primitives_robust_RPTcensus} examines the robustness of our structural estimates to an alternative definition of related-party trade. Instead of using our baseline ORBIS-based classification, we re-estimate the model excluding transactions identified as related-party trade using the standard RPT indicator in the US Census data. Appendix~\ref{subsec:Related-party-trade} explains the difference between the two indicators. Reassuringly, the estimated bargaining-power and returns-to-scale parameters remain very close to the baseline results, indicating that our main conclusions are not driven by the particular way related-party relationships are classified.

\begin{table}[H]
\caption{Simulated Markups and Bilateral Concentration}
\label{Table_simulated_markups}
\centering
\makebox[\textwidth][c]{%
{\small
\renewcommand{\arraystretch}{1.2}
\arrayrulecolor{gray!50}
\begin{tabular}{
  >{\raggedright\arraybackslash}p{5.5cm}
  >{\centering\arraybackslash}p{2cm}
  >{\centering\arraybackslash}p{2cm}
  >{\centering\arraybackslash}p{2cm}
  >{\centering\arraybackslash}p{2cm}
}
\toprule
Dependent variable: & \multicolumn{4}{c}{$\ln \hat{\mu}_{ijht}$} \\
\cmidrule(lr){2-5}
& (1) & (2) & (3) & (4) \\
\midrule
\multicolumn{5}{l}{\textit{\textbf{Panel A}: Estimated bargaining power} ($\hat{\phi}$)} \\[0.3em]
$s_{ijht}$   & 0.016     & 0.020     &    0.023       &    0.025       \\
 & (0.005) & (0.005) & (0.005) & (0.005) \\
$x_{ijht}$   & -0.588    & -0.590    &   -0.592   &   -0.584        \\
           & (0.007) & (0.007) & (0.007) & (0.007) \\[0.3em]
R-squared    & 0.995     & 0.997     &    0.998       &     1.000      \\
\midrule
\multicolumn{5}{l}{\textit{\textbf{Panel B}: No bargaining power} ($\phi = 0$)} \\[0.3em]
$s_{ijht}$   & 0.532     & 0.591     &  0.600         &  0.457         \\
           & (0.022) & (0.025) & (0.024) & (0.025) \\
$x_{ijht}$   & -0.006    & -0.034    &  -0.042   &  -0.037        \\
           & (0.003) & (0.003) & (0.003) & (0.003) \\[0.3em]
R-squared    & 0.537     & 0.651     &    0.735       &   0.972        \\
\midrule
$\text{FE}_{ht}$                             & Yes & No  & No  & No  \\
$\text{FE}_i + \text{FE}_j + \text{FE}_{ht}$ & No  & Yes & No  & No  \\
$\text{FE}_{iht} + \text{FE}_j$              & No  & No  & Yes & No  \\
$\text{FE}_{iht} + \text{FE}_{jht}$          & No  & No  & No  & Yes \\
\midrule
Observations & 1,200,000 & 1,200,000 & 1,200,000 & 1,200,000 \\
\bottomrule
\end{tabular}
}}
\vspace{0.5em}
\makebox[\textwidth][c]{%
\begin{minipage}{15.5cm}
\scriptsize \textit{Notes:} This table reports estimates of equation~\eqref{eq:fact_3} applied to model-implied bilateral markups. The dependent variable is the log price--cost ratio $\hat{\mu}_{ijht}$ obtained from  equation~\eqref{eq:bilateral_markup_prop}, evaluated at the observed supplier and buyer shares and at the estimated structural parameters. Column~(1) includes product--year fixed effects; Columns~(2)--(4) use the same fixed-effects specifications as  Table~\ref{Table 2 - Prices}. Panel~A uses the baseline estimates $\hat{\phi}$ and $\hat{\theta}$ from Column~(1) of  Table~\ref{tab:model_primitives}. Panel~B sets $\phi = 0$, shutting down importer bargaining power while holding $\hat{\theta}$ fixed. All regressions control for log relationship length (in years) within HS10 products. The simulated markups contain no independently simulated pricing shock, but the linear regressions have approximation residuals because the markup function is nonlinear in the shares and also depends on calibrated importer--product parameters. Standard errors in parentheses combine the sampling variance of the coefficients, clustered by HS8 product and exporter country, with the variance contributed by the estimation of $(\hat{\phi},\hat{\theta})$, computed by a parametric bootstrap with 1,000 replications. In Panel~B, $\phi$ is fixed at zero; since the oligopoly markup is independent of $\theta$, redrawing only $\theta$ should contribute zero parameter-estimation variance. The number of observations is rounded to four significant digits in accordance with US Census Bureau disclosure guidelines. \textit{Source}: FSRDC Project Number 2109 (CBDRB-FY25-P2109-R12520).
\end{minipage}}
\end{table}

\begin{table}[H]
\caption{Estimated Model Primitives - All BEC Categories }
\label{tab:model_primitives_robust_bec0}
\centering
{\small
\renewcommand{\arraystretch}{1.2}
\arrayrulecolor{gray!50}
\begin{adjustbox}{max width=\textwidth,center}
\begin{tabular}{
  >{\raggedright\arraybackslash}p{6cm}
  >{\centering\arraybackslash}p{2.2cm}
  >{\centering\arraybackslash}p{2.2cm}
  >{\centering\arraybackslash}p{2.2cm}
  >{\centering\arraybackslash}p{2.2cm}
}
\toprule
\multicolumn{5}{c}{\text{Panel A: Calibrated Parameters}} \\
\midrule
\multicolumn{1}{c}{$\hat{\nu}$} & \multicolumn{2}{c}{$\hat{\gamma}$} & \multicolumn{2}{c}{$\hat{\rho}$} \\
\multicolumn{1}{c}{4} & \multicolumn{2}{c}{0.5} & \multicolumn{2}{c}{10} \\
\midrule
\multicolumn{5}{c}{\text{Panel B: Estimated Parameters (GMM)}} \\
\midrule
& (1) & (2) & (3) & (4) \\
\midrule
Rel. bargaining power: $\ln \widehat{\frac{\phi}{1-\phi}}$ & 1.180 &  & 0.940 &  \\
 & (0.097)&  & (0.086) &  \\
\addlinespace
Returns to scale ($\hat{\theta}$) & 0.508 & 0.585 & 0.483 & 0.595 \\
& (0.009) & (0.010) & (0.010) & (0.010) \\
Constant &   & 3.355 &   & 1.299 \\
&  & (0.610) &  & (0.395) \\
Longevity &  & 0.115 &  & 0.902 \\
&  & (0.111) &  & (0.287) \\
Number of HS10 transactions &  & -0.315 &  & -0.090 \\
&  & (0.080) &  & (0.087) \\
Multiple HS10 dummy &  & 0.137 &  & 0.199 \\
&  & (0.116) &  & (0.108) \\
Lagged outside option &  & -0.267 &  & -0.272 \\
&  & (0.063) &  & (0.078) \\
\midrule
None & \multicolumn{1}{c}{Yes} & \multicolumn{1}{c}{Yes} & \multicolumn{1}{c}{No} & \multicolumn{1}{c}{No} \\
$FE_{h} + FE_{t} + FE_{j}$ & \multicolumn{1}{c}{No} & \multicolumn{1}{c}{No} & \multicolumn{1}{c}{Yes} & \multicolumn{1}{c}{Yes} \\
\midrule
Observations & \multicolumn{4}{c}{6,143,000} \\
\midrule
\multicolumn{5}{c}{\text{Panel C: Implied Bargaining Powers ($\hat{\phi}$)}} \\
\midrule
Mean & 0.765 & 0.923 & 0.719 & 0.899 \\
& (0.017) & (0.075) & (0.017) & (0.095) \\
Median & -- & 0.946 & --  & 0.928 \\
& -- & (0.075) & -- & (0.095) \\
\bottomrule
\end{tabular}
\end{adjustbox}
}
\vspace{0.75em}
\begin{minipage}{17cm}\vspace{0.5em}
\footnotesize
\textit{Notes}: This table presents model estimates based on a sample that includes all Broad Economic Categories (BEC), including consumption goods, for the period 2001-2016. Panel A reports calibrated parameters: demand elasticity ($\nu$), cost elasticity to foreign input prices ($\gamma$), and the elasticity of substitution across foreign varieties ($\rho$). 
Panel B reports GMM estimates. Columns (1) and (3) impose a constant $\phi$, while Columns (2) and (4) allow heterogeneous bargaining power via $\boldsymbol{\delta}$. Specifications differ in fixed effects. Controls include relationship longevity, transaction intensity, the relative outside option (lagged), and a multi-product indicator. Panel C reports the mean and median implied bargaining power. Standard errors are clustered at the exporter-product level; Panel C uses the delta method. Instruments are the numbers of importers and exporters in each HS10 product-year, and lagged leave-pair-out means and medians of the supplier- and buyer-share distributions. All estimations use CIF duty-exclusive unit values. The number of observations is rounded to four significant digits in accordance with US Census Bureau disclosure guidelines. \textit{Source}: FSRDC Project Number 2109 (CBDRB-FY25-P2109-R12520).
\end{minipage}
\end{table}

\begin{table}[H]
\caption{Estimated Model Primitives - BEC Consumption Categories}
\label{tab:model_primitives_robust_bec2}
\centering
{\small
\renewcommand{\arraystretch}{1.2}
\arrayrulecolor{gray!50}
\begin{adjustbox}{max width=\textwidth,center}
\begin{tabular}{
  >{\raggedright\arraybackslash}p{6cm}
  >{\centering\arraybackslash}p{2.2cm}
  >{\centering\arraybackslash}p{2.2cm}
  >{\centering\arraybackslash}p{2.2cm}
  >{\centering\arraybackslash}p{2.2cm}
}
\toprule
\multicolumn{5}{c}{\text{Panel A: Calibrated Parameters}} \\
\midrule
\multicolumn{1}{c}{$\hat{\nu}$} & \multicolumn{2}{c}{$\hat{\gamma}$} & \multicolumn{2}{c}{$\hat{\rho}$} \\
\multicolumn{1}{c}{4} & \multicolumn{2}{c}{0.5} & \multicolumn{2}{c}{10} \\
\midrule
\multicolumn{5}{c}{\text{Panel B: Estimated Parameters (GMM)}} \\
\midrule
& (1) & (2) & (3) & (4) \\
\midrule
Rel. bargaining power: $\ln \widehat{\frac{\phi}{1-\phi}}$ & 0.936 &  & 0.729 &  \\
 & (0.082)&  & (0.077) &  \\
\addlinespace
Returns to scale ($\hat{\theta}$) & 0.558 & 0.646 & 0.539 & 0.664 \\
& (0.009) & (0.013) & (0.010) & (0.016) \\
Constant &   & 2.501 &   & 0.568 \\
&  & (0.727) &  & (0.375) \\
Longevity &  & 0.628 &  & 1.414 \\
&  & (0.283) &  & (0.771) \\
Number of HS10 transactions &  & -0.386 &  & -0.137 \\
&  & (0.157) &  & (0.135) \\
Multiple HS10 dummy &  & 0.452 &  & 0.344 \\
&  & (0.243) &  & (0.224) \\
Lagged outside option &  & -0.266 &  & -0.299 \\
&  & (0.096) &  & (0.149) \\
\midrule
None & \multicolumn{1}{c}{Yes} & \multicolumn{1}{c}{Yes} & \multicolumn{1}{c}{No} & \multicolumn{1}{c}{No} \\
$FE_{h} + FE_{t} + FE_{j}$ & \multicolumn{1}{c}{No} & \multicolumn{1}{c}{No} & \multicolumn{1}{c}{Yes} & \multicolumn{1}{c}{Yes} \\
\midrule
Observations & \multicolumn{4}{c}{3,024,000} \\
\midrule
\multicolumn{5}{c}{\text{Panel C: Implied Bargaining Powers ($\hat{\phi}$)}} \\
\midrule
Mean & 0.718 & 0.919 & 0.674 & 0.891 \\
& (0.017) & (0.077) & (0.017) & (0.111) \\
Median & -- & 0.942 & --  & 0.928 \\
& -- & (0.077) & -- & (0.111) \\
\bottomrule
\end{tabular}
\end{adjustbox}
}
\vspace{0.75em}
\begin{minipage}{17cm}\vspace{0.5em}
\footnotesize
\textit{Notes}: This table presents model estimates based on a sample that only includes consumption goods based on Broad Economic Categories (BEC), for the period 2001-2016. Panel A reports calibrated parameters: demand elasticity ($\nu$), cost elasticity to foreign input prices ($\gamma$), and the elasticity of substitution across foreign varieties ($\rho$). 
Panel B reports GMM estimates. Columns (1) and (3) impose a constant $\phi$, while Columns (2) and (4) allow heterogeneous bargaining power via $\boldsymbol{\delta}$. Specifications differ in fixed effects. Controls include relationship longevity, transaction intensity, the relative outside option (lagged), and a multi-product indicator. 
Panel C reports the mean and median implied bargaining power. Standard errors are clustered at the exporter-product level; Panel C uses the delta method. Instruments are the numbers of importers and exporters in each HS10 product-year, and lagged leave-pair-out means and medians of the supplier- and buyer-share distributions. All estimations use CIF duty-exclusive unit values. The number of observations is rounded to four significant digits in accordance with US Census Bureau disclosure guidelines. \textit{Source}: FSRDC Project Number 2109 (CBDRB-FY25-P2109-R12520).
\end{minipage}
\end{table}

\begin{table}[H]
\caption{Estimated Model Primitives under Alternative Related-Party Classification}
\label{tab:model_primitives_robust_RPTcensus}
\centering
{\small
\renewcommand{\arraystretch}{1.2}
\arrayrulecolor{gray!50}
\begin{adjustbox}{max width=\textwidth,center}
\begin{tabular}{
  >{\raggedright\arraybackslash}p{6.4cm}
  >{\centering\arraybackslash}p{1.8cm}
  >{\centering\arraybackslash}p{1.8cm}
  >{\centering\arraybackslash}p{1.8cm}
  >{\centering\arraybackslash}p{1.8cm}
}
\toprule
& (1) & (2) & (3) & (4) \\
\midrule
\addlinespace
Implied Mean Bargaining Power ($\hat{\phi}$) & 0.837 & 0.928 & 0.761 & 0.899 \\
& (0.034) & (0.078) & (0.034) & (0.082) \\

Returns to scale ($\hat{\theta}$) & 0.455 & 0.500 & 0.410 & 0.504 \\
& (0.017) & (0.017) & (0.021) & (0.019) \\

\midrule
None & \multicolumn{1}{c}{Yes} & \multicolumn{1}{c}{Yes} & \multicolumn{1}{c}{No} & \multicolumn{1}{c}{No} \\
$FE_{h} + FE_{t} + FE_{j}$ & \multicolumn{1}{c}{No} & \multicolumn{1}{c}{No} & \multicolumn{1}{c}{Yes} & \multicolumn{1}{c}{Yes} \\
\bottomrule
\end{tabular}
\end{adjustbox}
}
\vspace{0.75em}
\begin{minipage}{17cm}\vspace{0.5em}
\footnotesize
\textit{Notes}: Model estimates are based on US arm’s-length imports of intermediate inputs and capital goods over 2001–2016, excluding transactions identified as related-party trade using the RPT indicator in the US Census data; see Appendix~\ref{subsec:Related-party-trade} for details. Columns (1) and (3) impose a constant $\phi$, while Columns (2) and (4) allow for heterogeneous bargaining power through $\boldsymbol{\delta}$, parameterized as a function of relationship longevity, transaction intensity, the lagged relative outside option, and a multi-product indicator. The demand elasticity is set to $\nu = 4$, the cost elasticity to foreign input prices to $\gamma = 0.5$, and the elasticity of substitution across foreign varieties to $\rho = 10$. Specifications differ in the fixed effects included. Standard errors are clustered at the exporter-product level. All estimations use CIF duty-exclusive unit values. All regressions include 3,120,000 observations. The number of observations is rounded to four significant digits in accordance with US Census Bureau disclosure guidelines. \textit{Source}: FSRDC Project Number 2109 (CBDRB-FY25-P2109-R12520).
\end{minipage}
\end{table}

\newpage 
\subsection{Prices and Pass-Through in Consumption Goods}
\label{sec:appendix_consumption}
Tables~\ref{tab:prices_consumption} and~\ref{tab:PT_consumption} re-estimate, respectively, the price-level regression in Table~\ref{Table 2 - Prices} and the pass-through regression in Table~\ref{tab:pass_through_main} using the consumption-goods sample.

In the price-level regression, the buyer-share coefficient shrinks in magnitude from $-0.56$ to $-0.31$, while the supplier-share coefficient changes little. This pattern is consistent with the lower $\hat{\phi}$ estimated for consumption goods (Table~\ref{tab:robust_model_estimates_d}, Column~(2)). In the pass-through regression, the buyer-share interaction moves from $-0.41$ to $0.06$ (s.e.~$0.11$). This attenuation is consistent with the weaker cost channel implied by the higher $\hat{\theta}$, although the estimate is imprecise and the model still predicts a negative gradient at $\hat{\theta}=0.56$.

\begin{figure}[h]
\caption{Pass-Through Elasticity at the Estimated Parameters: Baseline and Consumption Goods}
\label{fig:hetmap_consumption}
\begin{centering}
\includegraphics[width=0.85\columnwidth]{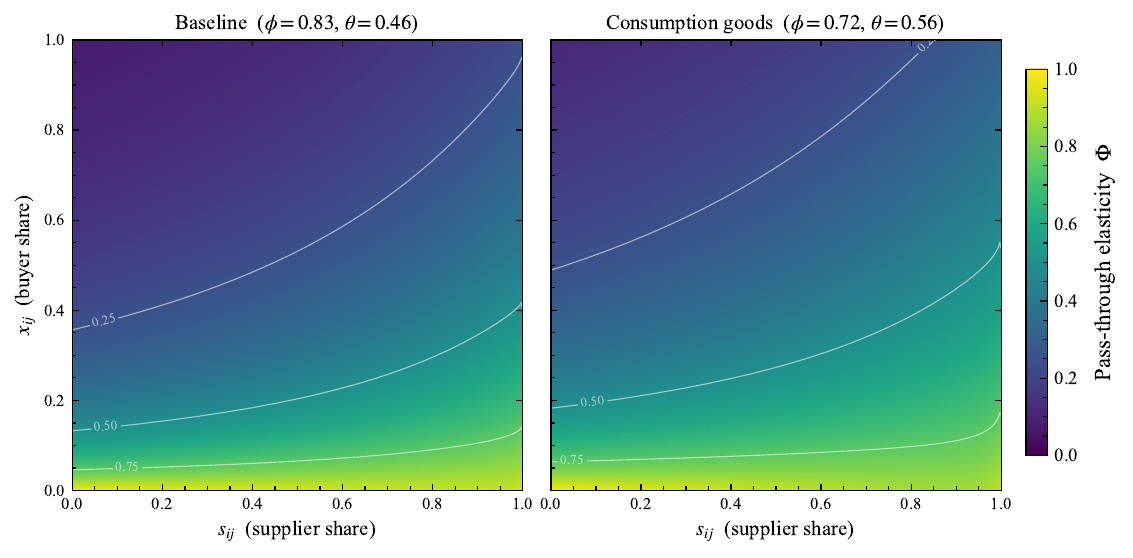}
\par\end{centering}
\scriptsize
\textit{Notes:} The figure presents heatmaps of the pass-through elasticity $\Phi_{ij}$ as a function of the supplier share $s_{ij}$ (horizontal axis) and the buyer share $x_{ij}$ (vertical axis), as in Figure~\ref{fig:hetmap}. The left panel uses the baseline estimates in Column~(1) of Table~\ref{tab:robust_model_estimates_d}; the right panel uses the consumption-goods estimates in Column~(2). The remaining parameters are fixed at $\gamma=0.5$, $\varrho=1$, $\nu=4$, and $\rho=10$.
\end{figure}

Figure~\ref{fig:hetmap_consumption} compares model-implied pass-through under the consumption-goods and baseline parameter estimates. The buyer-share gradient is flatter for consumption goods but remains clearly negative. At the baseline median shares, $(s,x)=(0.15,0.10)$, and mean shares, $(s,x)=(0.32,0.25)$, the consumption-goods estimates raise predicted pass-through by about 7 and 8 percentage points, respectively. The reduced-form average tariff coefficient, however, is $-0.48$ (s.e.~$0.32$), too imprecise to establish whether average pass-through differs from the baseline.

\vspace{1ex}
\begin{table}[H]
\caption{Prices and Bilateral Concentration --- Consumption Goods}
\label{tab:prices_consumption}
\centering
{\small
\renewcommand{\arraystretch}{1.2}
\arrayrulecolor{gray!50}
\begin{tabular}{
  >{\raggedright\arraybackslash}p{4.3cm}
  >{\centering\arraybackslash}p{1.6cm}
  >{\centering\arraybackslash}p{1.6cm}
  >{\centering\arraybackslash}p{1.6cm}
  >{\centering\arraybackslash}p{1.6cm}
  >{\centering\arraybackslash}p{1.6cm}
  >{\centering\arraybackslash}p{1.6cm}
}
\toprule
\small{Dependent Variable:} & \multicolumn{6}{c}{$\ln p_{ijht}$} \\
\cmidrule(lr){2-7}
& (1) & (2) & (3) & (4) & (5) & (6) \\
& OLS & IV & OLS & IV & OLS & IV \\
\midrule
$s_{ijht}$ & 0.148 & 0.260 & 0.196 & 0.235 & 0.220 & 0.512 \\
& (0.024) & (0.090) & (0.021) & (0.049) & (0.031) & (0.053) \\
$x_{ijht}$ & -0.306 & -0.119 & -0.347 & -0.171 & -0.306 & -0.439 \\
& (0.019) & (0.020) & (0.023) & (0.051) & (0.022) & (0.024) \\
\midrule
$\text{FE}_i + \text{FE}_j + \text{FE}_{ht}$     & Yes & Yes & No  & No  & No  & No \\
$\text{FE}_{iht} + \text{FE}_j$           & No  & No  & Yes & Yes & No  & No \\
$\text{FE}_{iht} + \text{FE}_{jht}$       & No  & No  & No  & No  & Yes & Yes \\
\midrule
Observations & 1,617,000 &1,108,000 & 1,617,000 & 1,108,000 & 1,617,000 & 1,108,000 \\
R-squared & 0.967 & --- & 0.981 & --- & 0.994 & --- \\
First-stage F stat. & --- & 471 & --- & 1,022 & --- & 259 \\
\makecell[l]{SW F stat  ($s_{ijht}$)} & --- & 1,003 & --- & 2,769 & --- & 1,828 \\
\makecell[l]{SW F stat  ($x_{ijht}$)} & --- & 455 & --- & 1,653 & --- & 2,746 \\
\bottomrule
\end{tabular}
}
\begin{minipage}{16.8cm}\vspace{0.5em}
\footnotesize \textit{Notes:} This table reports OLS and IV estimates of equation~\eqref{eq:fact_3} for the sample of consumption goods, where the dependent variable is the log CIF duty-exclusive unit value of product $h$ imported by buyer $j$ from supplier $i$ in year $t$. Columns alternate between OLS and IV specifications. All regressions control for log relationship length (in years) within HS10 products.
Columns (1)--(2) include buyer ($\text{FE}_j$), supplier ($\text{FE}_i$), and product-year ($\text{FE}_{ht}$) fixed effects. Columns (3)--(4) use supplier--product--year ($\text{FE}_{iht}$) and buyer ($\text{FE}_j$) fixed effects. Columns (5)--(6) include fully interacted buyer--product--year and supplier--product--year fixed effects ($\text{FE}_{jht}$, $\text{FE}_{iht}$). IV estimates (even-numbered columns) use leave-one-out instruments: $s_{ijht}$ is instrumented with exporter $i$'s average supplier share across its other buyers (excluding $j$), and $x_{ijht}$ with importer $j$'s average buyer share across its other suppliers (excluding $i$), within product $h$ and year $t$. Because the model includes multiple endogenous regressors, we report both first-stage and conditional F-statistics from \citet{SANDERSON2016212}. Standard errors are clustered at the HS8 product and exporter-country level. The number of observations is rounded to four significant digits in accordance with US Census Bureau disclosure guidelines. \textit{Source}: FSRDC Project Number 2109 (CBDRB-FY25-P2109-R12520).
\end{minipage}
\end{table}

\begin{table}[H]
\caption{Pass-Through and Relationship Heterogeneity --- Consumption Goods}
\label{tab:PT_consumption}
\centering
{\small
\renewcommand{\arraystretch}{1.2}
\arrayrulecolor{gray!50}
\begin{tabular}{
  >{\raggedright\arraybackslash}p{5.2cm}
  >{\centering\arraybackslash}p{1.8cm}
  >{\centering\arraybackslash}p{1.8cm}
  >{\centering\arraybackslash}p{1.8cm}
  >{\centering\arraybackslash}p{1.8cm}
}
\toprule
Dependent variable: & \multicolumn{4}{c}{$\Delta\ln p_{ijht}$} \\
\cmidrule(lr){2-5}
& (1) & (2) & (3) & (4) \\
\midrule
$\Delta\ln(1+\tau_{cht})$ & -0.477  & -0.625  & -0.452  & -0.622  \\
                          & (0.320) & (0.344) & (0.322) & (0.349) \\
$\Delta\ln(1+\tau_{cht}) \cdot s_{ijht-1}$
                          &         & 0.371   &         & 0.453   \\
                          &         & (0.181) &         & (0.190) \\
$\Delta\ln(1+\tau_{cht}) \cdot x_{ijht-1}$
                          &         &         & 0.092  & 0.063  \\
                          &         &         & (0.116) & (0.112) \\
\midrule
$\text{FE}_{ht} + \text{FE}_{cs}$
                          & Yes     & Yes     & Yes     & Yes     \\
\midrule
Observations              & 224,000 & 224,000 & 224,000 & 224,000 \\
R-squared                 & 0.04    & 0.04    & 0.04    & 0.04    \\
\bottomrule
\end{tabular}
}
\begin{minipage}{14.5cm}\vspace{0.5em}
\scriptsize  \textit{Notes:} This table reports estimates of the pass-through of statutory tariffs, $\Delta \ln (1 + \tau_{cht})$, to CIF duty-exclusive prices at the exporter-importer-product-year level, $\Delta \ln p_{ijht}$.  The sample is restricted to consumption goods. Column~(2) interacts tariffs with the lagged supplier share, $s_{ijht-1}$, defined as supplier $i$'s share in buyer $j$'s imports of product $h$. Column~(3) interacts tariffs with the lagged buyer share, $x_{ijht-1}$, defined as buyer $j$'s share in supplier $i$'s exports of product $h$. Column~(4) includes both interactions. All regressions include product-year and exporter country-sector fixed effects ($\text{FE}_{ht} + \text{FE}_{cs}$). Controls include: (i) $\ln \text{longevity}_{ijht}$; (ii) $\Delta \ln q_{i(-j)ht}$, the change in exporter $i$'s total sales of $h$ to US buyers other than $j$; and (iii) $\Delta \ln p_{(-i)jht}$, the weighted average price change charged by other suppliers of $h$ to buyer $j$, using lagged shares as weights. Standard errors are clustered at the HS8 product and exporter-country level. The sample is the consumption-good subset of the ``+ Supplier Multi-Buyer'' sample in Panel B of  Table~\ref{tab:sample_stats_bec0}. Observation counts are rounded to four significant digits per US Census Bureau disclosure guidelines. \textit{Source}: FSRDC Project Number 2109 (CBDRB-FY25-P2109-R12520).
\end{minipage}
\end{table}

\newpage
\subsection{Additional Out-of-Sample Evaluation Results}\label{sec:model_fit_robustness} 

\paragraph{Pass-through and Relationship Heterogeneity: Data vs. Model.} 

Table~\ref{tab:table5_new_stringentFE} repeats the pass-through exercise with more demanding fixed effects. The baseline model matches average pass-through and the negative buyer-share gradient. It predicts a modest positive supplier-share gradient, while the empirical estimate is close to zero and imprecise. The difference is not statistically significant.

\paragraph{IV-Based Goodness-of-Fit Test.}
We formally assess model fit using an IV regression following \citet{adao2023putting}: 
\begin{equation}
\Delta\ln p_{ijht}=\beta \, \widehat{\Delta\ln p_{ijht}}
+ \mathbf{FE}
+ u_{ijht},
\label{eq:model_performance_reg}
\end{equation}
where $\Delta\ln p_{ijht}$ is the observed change in the duty-inclusive price and $\widehat{\Delta\ln p_{ijht}}$ is the corresponding model-implied change. Because observed price changes may reflect shocks unrelated to tariffs, we instrument $\widehat{\Delta\ln p_{ijht}}$ with statutory tariff changes to isolate tariff-driven variation.\footnote{Unlike the pass-through analysis in the main text, which uses duty-exclusive prices to mitigate measurement error in applied duties, the IV strategy isolates variation induced by statutory tariffs and evaluates the model against duty-inclusive price changes.} Under this exclusion restriction and the null that the model correctly captures tariff-induced price movements, the IV coefficient $\hat{\beta}$ equals one.

\begin{figure}[p]
\centering
\caption{IV-Based Goodness-of-Fit Test}
\label{fig:IVtest_fit}
\includegraphics[width=0.8\linewidth]{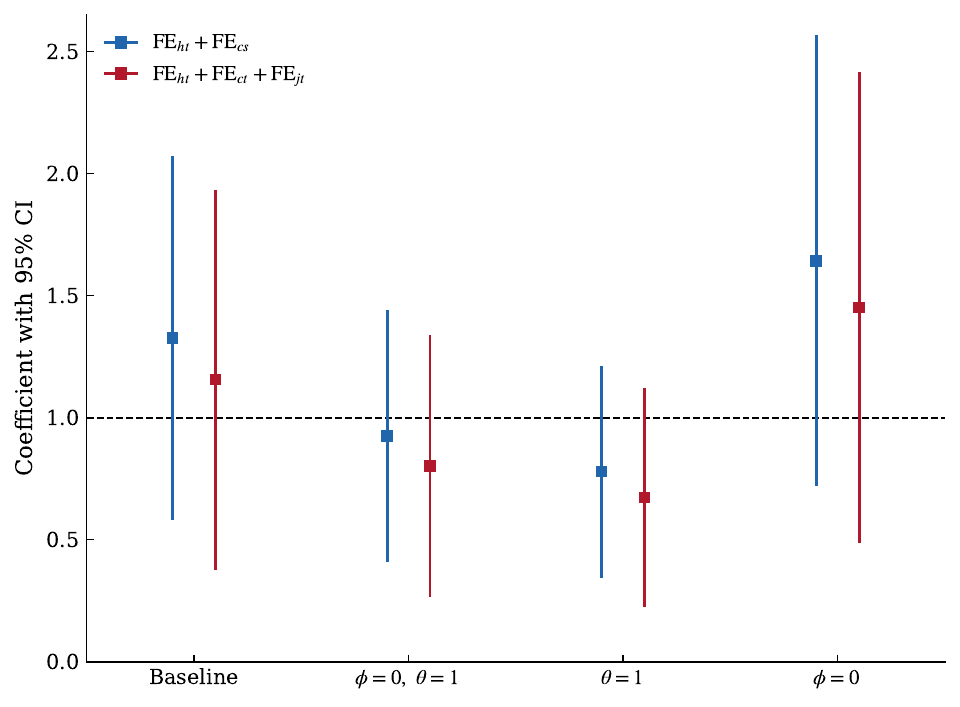}
\begin{minipage}{1\linewidth} \vspace{0.5em}
\footnotesize
\textit{Notes}: Each point reports the coefficient from an IV regression of observed log duty-inclusive price changes on model-predicted changes $\widehat{\Delta \ln p_{ijht}}$, using statutory tariffs as instruments. Lines show 95\% confidence intervals. Blue and red denote regressions with $\mathrm{FE}_{ht} + \mathrm{FE}_{cs}$ and $\mathrm{FE}_{ht} + \mathrm{FE}_{ct} + \mathrm{FE}_{jt}$, respectively. Standard errors combine the sampling variance of the coefficients, clustered by HS8 product and exporter country, with the variance contributed by the estimation of $(\hat{\phi},\hat{\theta})$, computed by a parametric bootstrap with 1,000 replications. Observation counts (249,000) are rounded per Census disclosure guidelines. \textit{Source}: FSRDC Project Number 2109 (CBDRB-FY25-P2109-R12520).
\end{minipage}
\end{figure}

Figure~\ref{fig:IVtest_fit} reports the IV estimates of $\beta$ across the four model variants under the two fixed-effects specifications. Blue markers correspond to regressions with product--time and country\allowbreak--sector fixed effects, while red markers use product--time, country--time, and buyer--time fixed effects. The model-implied price changes are close to those observed in the data. %
The test does not reject the baseline model, but it does not reject the restricted variants either: as noted in the main text, average pass-through has little power to discriminate among them.

\paragraph{Sensitivity of Model Fit to Alternative Parameter Values.} In our baseline estimation we set \(\varrho=1\), \(\nu=4\), and \(\gamma=0.5\). We conduct a robustness exercise where we set \(\varrho=0.5\), close to the returns-to-scale parameter we estimate for the supplier side. 
Further, we also consider a value of \(\rho=5\), the lower end of the range provided in \citet{Anderson2004}.

Table \ref{tab:Modelfit_IV_altparam_panel} repeats the IV test for the alternative calibrations and both sets of fixed effects. No specification is uniformly closest to one. The baseline is closest under the stringent fixed effects; under the baseline fixed effects, some restricted variants are as close or closer. The confidence intervals are too wide to distinguish the models using this average moment.

\paragraph{Out-of-Sample Evaluation: Quantity and Expenditure Responses.} 

Table~\ref{tab:quantity_response_combined} compares observed quantity responses in Columns (1) and (3) with model predictions in Columns (2) and (4). Columns (1)--(2) use baseline fixed effects and Columns (3)--(4) use stringent fixed effects. Panel~A reports main effects; Panel~B adds interactions with supplier and buyer shares.

The model predicts the sign of the average quantity response but overstates its size: the predicted declines are two to three times larger than the estimated ones, and the difference is large relative to the standard errors. The model also predicts positive interactions with both shares, that is, smaller declines for high-share buyers and suppliers, whereas the estimated interactions in the data are negative and imprecise.

Table~\ref{tab:PT_IV_q_sales_combined} reports IV-based tests comparing observed quantity (Panel A) and duty-inclusive expenditure (Panel B) changes to those implied by the model under alternative parameterizations. Quantities respond to the price the buyer pays, which includes the duty, so the model predicts
\begin{equation*}
\widehat{\Delta\ln q_{ijht}}
=
-\varepsilon_{ijht}\,\big(\widehat{\Delta\ln p^*_{ijht}}+\Delta\ln(1+\tau_{c(i)ht})\big)
=
-\varepsilon_{ijht}\,\Phi_{ijht}\,\Delta\ln(1+\tau_{c(i)ht}),
\end{equation*}
where $\varepsilon_{ijht}$ is the match-specific residual demand elasticity evaluated at the lagged (2017) shares and the estimated parameters, $\widehat{\Delta\ln p^*_{ijht}}$ is the duty-exclusive price change in equation~\eqref{eq:implied}, and $\Phi_{ijht}$ is the pass-through elasticity into the duty-inclusive price. The buyer's duty-inclusive expenditure is $e_{ijht}=p_{ijht}q_{ijht}$, where $p_{ijht}=(1+\tau_{c(i)ht})p^*_{ijht}$, so that
\begin{equation*}
\widehat{\Delta\ln e_{ijht}}
=
\widehat{\Delta\ln p_{ijht}}+\widehat{\Delta\ln q_{ijht}}
=
(1-\varepsilon_{ijht})\,\Phi_{ijht}\,\Delta\ln(1+\tau_{c(i)ht}).
\end{equation*}

The IV coefficients lie well below one for the baseline and for most variants, so the model's quantity and expenditure predictions are rejected in most specifications; the specifications with decreasing returns, with and without bargaining, come closest to the data, consistent with the results for prices.

\begin{table}[htbp]
\caption{{Price Responses and Relationship Heterogeneity: Data vs. Model}}
\label{tab:table5_new_stringentFE}
\centering
{\small
\renewcommand{\arraystretch}{1.2}
\arrayrulecolor{gray!50}
\begin{tabular}{
  >{\raggedright\arraybackslash}p{4.6cm}
  >{\centering\arraybackslash}p{1.8cm}
  >{\centering\arraybackslash}p{1.8cm}
  >{\centering\arraybackslash}p{1.8cm}
  >{\centering\arraybackslash}p{1.8cm}
  >{\centering\arraybackslash}p{1.8cm}
}
\toprule \addlinespace[1em]
\multicolumn{6}{c}{{Panel A: Main Effect (Stringent Fixed Effects)}} \\
\midrule
& Data & Baseline & $\phi=0$, $\theta=1$  & $\theta=1$ & $\phi=0$ \\
\cmidrule(lr){2-6}  \\
\addlinespace[-1em]
$\Delta \ln(1+\tau_{cht})$ 
    & -0.184  &	-0.226  &	-0.095	& -0.014	& -0.286 \\
    & (0.107) &	(0.012) & (0.008) &	(0.003)	& (0.012) \\
\midrule
R-squared    & 0.31	& 0.48 &	0.42 &	0.41	& 0.51 \\
\midrule

\addlinespace[1.5em]
\multicolumn{6}{c}{{Panel B: Including Interactions (Stringent Fixed Effects)}} \\
\midrule
& Data & Baseline & $\phi=0$, $\theta=1$ & $\theta=1$ & $\phi=0$ \\
& (1) & (2) & (3) & (4) & (5) \\
\midrule
$\Delta \ln(1+\tau_{cht})$ 
    & -0.114 & -0.137 & -0.030 & -0.002 & -0.200 \\
    & (0.104) & (0.013) & (0.007) & (0.001) & (0.015) \\

$\Delta \ln(1+\tau_{cht}) \cdot s_{ijht-1}$ 
    & 0.004 & 0.116 & -0.120 & -0.022 & 0.073 \\
    & (0.165) & (0.013) & (0.022) & (0.005) & (0.013) \\

$\Delta \ln(1+\tau_{cht}) \cdot x_{ijht-1}$ 
    & -0.280 & -0.433 & -0.078 & -0.013 & -0.372 \\
    & (0.134) & (0.027) & (0.020) & (0.005) & (0.025) \\
\midrule
R-squared 
    & 0.31 & 0.57 & 0.45 & 0.45 & 0.56 \\
  \midrule \addlinespace[1em]

$FE_{ht} + FE_{ct} + FE_{jt}$ & Yes & Yes & Yes & Yes & Yes \\
  \midrule
    Observations & \multicolumn{5}{c}{249,000} \\
\bottomrule
\end{tabular}
}
\begin{minipage}{\textwidth}\vspace{0.75em}
\footnotesize
\textit{Notes}: This table reports the response of duty-exclusive CIF prices to tariff changes at the exporter--importer--product level. Column (1) presents reduced-form estimates using observed price changes in the data. Columns (2)--(5) report the corresponding responses implied by model-predicted price changes under each specification. Panel A reports the main effect of tariff changes, while Panel B allows the response to vary with lagged supplier shares ($s_{ijht-1}$) and buyer shares ($x_{ijht-1}$). All specifications include product--time,  country--time, and buyer-time fixed effects ($FE_{ht} + FE_{ct} + FE_{jt}$).  Standard errors in parentheses combine the sampling variance of the coefficients, clustered by HS8 product and exporter country, with the variance contributed by the estimation of $(\hat{\phi},\hat{\theta})$, computed by a parametric bootstrap with 1,000 replications. In column (1), standard errors are clustered at the HS8 product and exporter-country level only. Observation counts are rounded to four significant digits per US Census Bureau disclosure guidelines. \textit{Source}: FSRDC Project Number 2109 (CBDRB-FY25-P2109-R12520).
\end{minipage}
\end{table}

\clearpage

\pagebreak
\begin{sidewaystable}[p]
\caption{IV-Based Goodness-of-Fit Test, Alternative Parameters}
\label{tab:Modelfit_IV_altparam_panel}
\centering
{\small
\renewcommand{\arraystretch}{1.2}
\arrayrulecolor{gray!50}
\begin{tabular}{
  >{\raggedright\arraybackslash}p{4cm}
  >{\centering\arraybackslash}p{1.5cm}
  >{\centering\arraybackslash}p{1.5cm}
  >{\centering\arraybackslash}p{1.5cm}
  >{\centering\arraybackslash}p{1.5cm}
  >{\centering\arraybackslash}p{1.5cm}
  >{\centering\arraybackslash}p{1.5cm}
  >{\centering\arraybackslash}p{1.5cm}
  >{\centering\arraybackslash}p{1.5cm}
}
\addlinespace[1em]
\multicolumn{9}{c}{\text{Panel A: Baseline Fixed Effects}} \\
\midrule
{Dependent variable:} & \multicolumn{8}{c}{$\Delta \ln p_{ijht}$} \\
\cmidrule(lr){2-5} \cmidrule(lr){6-9} 
& \multicolumn{4}{c}{$\rho=5$} & \multicolumn{4}{c}{$\varrho = 0.5$} \\
\cmidrule(lr){2-5} \cmidrule(lr){6-9} 
 & Baseline & $\phi = 0$, $\theta = 1$ & $\theta = 1$ & $\phi = 0$ & Baseline & $\phi = 0$, $\theta = 1$ & $\theta = 1$ & $\phi = 0$\\
 & (1) & (2) & (3) & (4) & (5) & (6) & (7) & (8) \\
\midrule 
$\widehat{\Delta \ln p_{ijht}}$ 
& 1.158 & 0.889 & 0.771 & 1.391 & 1.291 & 0.936 & 0.776 & 1.619 \\ 
& (0.333) & (0.253) & (0.220) & (0.398) & (0.369) & (0.267) & (0.221) & (0.464)  \\
\midrule
$FE_{ht} + FE_{cs}$ & Yes & Yes & Yes & Yes & Yes & Yes & Yes & Yes \\
\midrule
\addlinespace[1em]
\addlinespace[1em]
\multicolumn{9}{c}{\text{Panel B: Stringent Fixed Effects}} \\
\midrule
{Dependent variable:} & \multicolumn{8}{c}{$\Delta \ln p_{ijht}$} \\
\cmidrule(lr){2-5} \cmidrule(lr){6-9} 
& \multicolumn{4}{c}{$\rho=5$} & \multicolumn{4}{c}{$\varrho = 0.5$} \\
\cmidrule(lr){2-5} \cmidrule(lr){6-9} 
 & Baseline & $\phi = 0$, $\theta = 1$ & $\theta = 1$ & $\phi = 0$ & Baseline & $\phi = 0$, $\theta = 1$ & $\theta = 1$ & $\phi = 0$\\
 & (1) & (2) & (3) & (4) & (5) & (6) & (7) & (8) \\
\midrule 
$\widehat{\Delta \ln p_{ijht}}$ 
& 0.998 & 0.776 & 0.667 & 1.215 & 1.128 & 0.805 & 0.670 & 1.432 \\ 
& (0.342) &	(0.265)	& (0.227) &	(0.416)
 & (0.387) &	(0.275)	& (0.229) &	(0.491)  \\
\midrule
$FE_{ht} + FE_{ct} + FE_{jt}$ & Yes & Yes & Yes & Yes & Yes & Yes & Yes & Yes \\
\midrule
Observations & \multicolumn{8}{c}{249,000} \\
\bottomrule
\end{tabular}
}
\vspace{0.5em}
\begin{minipage}{22cm}\vspace{0.5em}
\footnotesize
\textit{Notes}: Each column reports the coefficient from an IV regression of the observed change in log duty-inclusive price on the corresponding model-predicted change $\widehat{\Delta \ln p_{ijht}}$, using statutory tariff changes as instruments. Columns (1)-(4) use predicted prices using a $\rho=5$ instead of $\rho=10$ as in the baseline. Columns (5)-(8) use predicted prices using a $\varrho=0.5$ instead of $\varrho=1$ as in the baseline. Each case re-estimates the structural parameters $\phi$ and $\theta$, reported in Columns~(4) and~(5) of Table~\ref{tab:robust_model_estimates_d} in the main text. In Panel A, all columns include product-time and country-sector fixed effects ($FE_{ht} + FE_{cs}$). In Panel B, all columns include product-time, country-time, and buyer-time fixed effects ($FE_{ht} + FE_{ct} + FE_{jt}$). Standard errors in parentheses combine the sampling variance of the coefficients, clustered by HS8 product and exporter country, with the variance contributed by the estimation of $(\hat{\phi},\hat{\theta})$, computed by a parametric bootstrap with 1,000 replications. The number of observations is rounded to four significant digits in accordance with US Census Bureau disclosure guidelines. \textit{Source}: FSRDC Project Number 2109 (CBDRB-FY25-P2109-R12520).
\end{minipage}
\end{sidewaystable}

\begin{table}[p]
\caption{\textit{Quantity} Responses and Relationship Heterogeneity: Data vs. Model}
\label{tab:quantity_response_combined}
\centering
{\small
\renewcommand{\arraystretch}{1.2}
\arrayrulecolor{gray!50}
\begin{tabular}{
  >{\raggedright\arraybackslash}p{4cm}
  >{\centering\arraybackslash}p{1.8cm}
  >{\centering\arraybackslash}p{1.8cm}
  >{\centering\arraybackslash}p{1.8cm}
  >{\centering\arraybackslash}p{1.8cm}
}
\toprule
 & \multicolumn{2}{c}{Baseline FE} & \multicolumn{2}{c}{Stringent FE} \\
\cmidrule(lr){2-3}\cmidrule(lr){4-5}
 & Data & Model & Data & Model \\
 & (1) & (2) & (3) & (4) \\
\midrule
\addlinespace[1.5em]
\multicolumn{5}{c}{\textit{Panel A}: Main effects} \\
\midrule

$\Delta\ln(1+\tau_{cht})$ 
& -0.568 
& -1.710 
& -1.021 
& -1.921 \\
& (0.249) 
& (0.088) 
& (0.185) 
& (0.117) \\

\midrule
R-squared 
& 0.06 
& 0.25 
& 0.32 
& 0.40 \\
\midrule
\addlinespace[1.5em]

\multicolumn{5}{c}{\textit{Panel B}: Including Interactions} \\
\midrule

$\Delta\ln(1+\tau_{cht})$ 
& 0.006 
& -3.683 
& -0.761 
& -3.705 \\
& (0.331) 
& (0.216) 
& (0.237) 
& (0.240) \\

$\Delta\ln(1+\tau_{cht}) \cdot s_{ijht-1}$ 
& -0.511 
& 2.755 
& -0.278 
& 2.816 \\
& (0.323) 
& (0.465) 
& (0.276) 
& (0.620) \\

$\Delta\ln(1+\tau_{cht}) \cdot x_{ijht-1}$ 
& -0.201 
& 2.453 
& -0.130 
& 2.703 \\
& (0.229) 
& (0.242) 
& (0.242) 
& (0.349) \\

\midrule
R-squared 
& 0.10 
& 0.36 
& 0.36 
& 0.49 \\

\midrule

$FE_{ht} + FE_{cs}$  & Yes & Yes & No & No \\
$FE_{ht} + FE_{ct} + FE_{jt}$  & No & No & Yes & Yes \\

\midrule
Observations & \multicolumn{4}{c}{249,000} \\
\bottomrule
\end{tabular}
}
\begin{minipage}{\textwidth} \vspace{0.75em}
\footnotesize \textit{Notes}: Panel A reports main effects of tariff changes on quantities. Panel B introduces interactions between tariff changes and lagged supplier share ($s_{ijht-1}$) and lagged buyer share ($x_{ijht-1}$). Columns (1) and (2) use baseline fixed effects ($FE_{ht} + FE_{cs}$), while Columns (3) and (4) use a more demanding specification with product–year, country–year, and buyer–year fixed effects ($FE_{ht} + FE_{ct} + FE_{jt}$). Standard errors in parentheses are clustered by HS8 product and exporter country in columns (1) and (3). In columns (2) and (4), they additionally incorporate the variance contributed by the estimation of $(\hat{\phi},\hat{\theta})$, computed using a parametric bootstrap with 1,000 replications. Observation counts are rounded to four significant digits per US Census Bureau disclosure guidelines. \textit{Source}: FSRDC Project Number 2109 (CBDRB-FY25-P2109-R12520).
\end{minipage}
\end{table}

\pagebreak

\begin{table}[p]
\caption{IV-Based Goodness-of-Fit Test for Quantities and Expenditures}
\label{tab:PT_IV_q_sales_combined}
\centering
{\small
\renewcommand{\arraystretch}{1.2}
\arrayrulecolor{gray!50}
\begin{tabular}{
  >{\raggedright\arraybackslash}p{3.8cm} 
  >{\centering\arraybackslash}p{1.1cm} 
  >{\centering\arraybackslash}p{1.1cm} 
  >{\centering\arraybackslash}p{1.1cm} 
  >{\centering\arraybackslash}p{1.1cm} 
  >{\centering\arraybackslash}p{1.1cm} 
  >{\centering\arraybackslash}p{1.1cm} 
  >{\centering\arraybackslash}p{1.1cm} 
  >{\centering\arraybackslash}p{1.1cm}}
\toprule
\addlinespace[0.8em]

\multicolumn{9}{c}{\textit{Panel A}: Quantities} \\
\midrule
Dependent Variable: & \multicolumn{8}{c}{$\Delta \ln q_{ijht}$} \\
\cmidrule(lr){2-3} \cmidrule(lr){4-5} \cmidrule(lr){6-7} \cmidrule(lr){8-9}
& \multicolumn{2}{c}{Baseline} & \multicolumn{2}{c}{$\phi = 0$, $\theta = 1$} & \multicolumn{2}{c}{$\theta = 1$} & \multicolumn{2}{c}{$\phi = 0$} \\
\cmidrule(lr){2-3} \cmidrule(lr){4-5} \cmidrule(lr){6-7} \cmidrule(lr){8-9}
& (1) & (2) & (3) & (4) & (5) & (6) & (7) & (8) \\
\midrule
$\widehat{\Delta \ln q_{ijht}}$ 
& 0.401	& 0.592	& 0.242	& 0.361	& 0.213	& 0.320	& 0.496	& 0.726 \\
& (0.165)	& (0.163)	& (0.100)	& (0.097)	& (0.087)	& (0.086)	& (0.205)	& (0.198) \\
\addlinespace[0.8em]
\toprule
\addlinespace[0.8em]
\addlinespace[0.8em]

\multicolumn{9}{c}{\textit{Panel B}: Expenditures} \\
\midrule
Dependent Variable: & \multicolumn{8}{c}{$\Delta \ln (p_{ijht} \cdot q_{ijht})$} \\
\cmidrule(lr){2-3} \cmidrule(lr){4-5} \cmidrule(lr){6-7} \cmidrule(lr){8-9}
& \multicolumn{2}{c}{Baseline} & \multicolumn{2}{c}{$\phi = 0$, $\theta = 1$} & \multicolumn{2}{c}{$\theta = 1$} & \multicolumn{2}{c}{$\phi = 0$} \\
\cmidrule(lr){2-3} \cmidrule(lr){4-5} \cmidrule(lr){6-7} \cmidrule(lr){8-9}
& (1) & (2) & (3) & (4) & (5) & (6) & (7) & (8) \\
\midrule
$\widehat{\Delta \ln (p \cdot q)_{ijht}}$ 
& 0.206	& 0.489	& 0.121	& 0.292	& 0.107	& 0.262	& 0.255	& 0.598 \\
& (0.184)	& (0.177)	& (0.108)	& (0.105)	& (0.095)	& (0.094)	& (0.228)	& (0.216) \\
\midrule
\addlinespace[0.8em]
$FE_{ht} + FE_{cs}$ & Yes & No & Yes & No & Yes & No & Yes & No \\
$FE_{ht} + FE_{ct} + FE_{jt}$ & No & Yes & No & Yes & No & Yes & No & Yes \\
\midrule
Observations & \multicolumn{8}{c}{249,000} \\
\bottomrule
\end{tabular}
}
\begin{minipage}{\textwidth} \vspace{0.75em}
\footnotesize
\textit{Notes}: Each column reports the coefficient from an IV regression of the observed change in log quantity (Panel~A) or log duty-inclusive expenditures (Panel~B) on the corresponding model-predicted change, using statutory tariff changes as instruments. Columns (1), (3), (5), and (7) include product–time and country–sector fixed effects ($FE_{ht} + FE_{cs}$), while Columns (2), (4), (6), and (8) include product–time, country–time, and buyer–time fixed effects ($FE_{ht} + FE_{ct} + FE_{jt}$). Standard errors in parentheses combine the sampling variance of the coefficients, clustered by HS8 product and exporter country, with the variance contributed by the estimation of $(\hat{\phi},\hat{\theta})$, computed by a parametric bootstrap with 1,000 replications. Observation counts are rounded to four
significant digits per US Census Bureau disclosure guidelines. \textit{Source}: FSRDC Project Number 2109 (CBDRB-FY25-P2109-R12520).
\end{minipage}
\end{table}

\clearpage

\clearpage{}

\section{Selection into Observed Matches and Identification\label{sec:Estimation Appendix}}
\label{endogenousrelationships}

The estimating equation in Section~\ref{subsec:Estimation-of-parameters-theta-phi} is
\[
\ln p_{ijht}=\ln\mu(\phi,\theta\mid\Omega_{ijht})
+\frac{1-\theta}{\theta}\ln q_{iht}+\ln k_{ijht}.
\]
Here $\ln k_{ijht}$ combines the common exporter--product--year cost shifter with a match-specific pricing residual. This reduced-form residual is not interpreted as a buyer-specific production marginal cost, and its determinants are not modeled. Differencing prices across buyers removes the common component but leaves $\ln k_{ijht}-\ln k_{i\ell ht}$. Identification requires instruments orthogonal to this difference within the sample of observed matches, conditional on controls and fixed effects. This appendix discusses how endogenous match formation could violate that restriction.

Model the value of a potential match between exporter \( i \) and importer \( j \) as
\[
V(X_{ijht}) = f(X_{ijht}) + \Xi_{ijht},
\]
where \( X_{ijht} \) denotes observed match characteristics, \( f(\cdot) \) is a deterministic function, and \( \Xi_{ijht} \) captures unobserved, idiosyncratic match surplus. A match is observed if and only if \( V(X_{ijht}) > 0 \), so that
\begin{equation}\label{eq:selection}
q^{*}_{ijht} = \begin{cases}
0, &\text{if } \Xi_{ijht} \leq -f(X_{ijht}), \\
q_{ijh}(p_{ijht}), &\text{otherwise},
\end{cases}
\end{equation}
where \( q_{ijh}(\cdot) \) is the demand function in equation \eqref{eq:demand function q_ij}. When no match is formed, bilateral prices and quantities are not observed.

Selection can bias the structural estimates if it violates the selected-sample IV moment:
\[
\mathbb E\!\left[\mathbf Z_{ij\ell ht}
(\ln k_{ijht}-\ln k_{i\ell ht})
\mid (i,j,h,t)\text{ and }(i,\ell,h,t)\text{ observed}\right]=\mathbf 0,
\]
after conditioning on the included controls and fixed effects. Correlation of match surplus $\Xi_{ijht}$ with bilateral shares alone does not establish IV bias. Conversely, conditioning on match formation can invalidate a moment that holds in the population of potential matches. Random selection is sufficient under otherwise valid population moments but is not necessary; the relevant requirement is the displayed orthogonality condition. The sign and magnitude of any bias in $\phi$ and $\theta$ depend on the joint distribution of the instruments, pricing residuals, and selection.

Pairwise differencing removes pricing components common to all buyers of an exporter--product--year. It does not, by itself, remove the effects of selection on the remaining match-specific pricing residuals.

The instruments use the numbers of active importers and exporters and lagged leave-pair-out means and medians of the supplier- and buyer-share distributions in each HS10 product--year. These variables describe market structure, but active partner counts and observed shares can themselves reflect match selection. Leaving out the focal pair removes its direct contribution to the share moments; it does not eliminate dependence through selection or common shocks. Their validity therefore rests on the selected-sample exclusion restriction above, rather than following from the leave-pair-out construction alone.

This static selection framework aligns with our theoretical setup. In dynamic formation games, \( k_{ijht} \) may also correlate with the expected continuation value of a match, which would open additional selection channels whose effect on \( \phi \) and \( \theta \) is harder to characterize.

\end{document}